# Corrections to Hawking radiation from asteroid-mass primordial black holes: description of the stochastic charge effect in quantum electrodynamics

Gabriel Vasquez,[1, 2, *] John Kushan,[1, 2, †] Makana Silva,[3, ‡] Emily
Koivu,[1, 2, §] Arijit Das,[1, 2, ¶] and Christopher M. Hirata[1, 2, 4, **]

[1]*Center for Cosmology and Astroparticle Physics, The Ohio State University,
191 West Woodruff Avenue, Columbus OH, 43210, USA*
[2]*Department of Physics, The Ohio State University,
191 West Woodruff Avenue, Columbus OH, 43210, USA*
[3]*Computational Physics and Methods Group (CCS-2),
Los Alamos National Laboratory, Los Alamos NM, 87544, USA*
[4]*Department of Astronomy, The Ohio State University,
140 West 18th Avenue, Columbus OH, 43210, USA*

(Dated: July 12, 2024)

Hawking radiation sets stringent constraints on Primordial Black Holes (PBHs) as a dark matter candidate in the $M \sim 10^{16}$ g regime based on the evaporation products such as photons, electrons, and positrons. This motivates the need for rigorous modeling of the Hawking emission spectrum. Using semi-classical arguments, Page [Phys. Rev. D 16, 2402 (1977)] showed that the emission of electrons and positrons is altered due to the black hole acquiring an equal and opposite charge to the emitted particle. The Poisson fluctuations of emitted particles cause the charge $Z|e|$ to random walk, but since acquisition of charge increases the probability of the black hole emitting another charged particle of the same sign, the walk is biased toward $Z = 0$, and $P(Z)$ approaches an equilibrium probability distribution with finite variance $\langle Z^2 \rangle$. This paper explores how this "stochastic charge" phenomenon arises from quantum electrodynamics (QED) on a Schwarzschild spacetime. We prove that (except for a small Fermi blocking term) the semi-classical variance $\langle Z^2 \rangle$ agrees with the variance of a quantum operator $\langle \hat{Z}^2 \rangle$, where $\hat{Z}$ may be thought of as an "atomic number" that includes the black hole as well as charge near it (weighted by a factor of $2M/r$). In QED, the fluctuations in $\hat{Z}$ do not arise from the black hole itself (whose charge remains fixed), but rather as a collective effect in the Hawking-emitted particles mediated by the long-range electromagnetic interaction. We find the rms charge $\langle Z^2 \rangle^{1/2}$ asymptotes to 3.44 at small PBH masses $M \lesssim 2 \times 10^{16}$ g, declining to 2.42 at $M = 5.2 \times 10^{17}$ g.

## I. INTRODUCTION

The first lines of evidence that most of the matter in the Universe was nonluminous arose from dynamical arguments [1, 2]. Today, a diverse array of astronomical observations point to a cosmic dark matter density $\sim 5$ times that of the baryonic matter and making up $\sim 25\%$ of the present-day energy budget of the Universe, including the cosmic microwave background [3], the Lyman-$\alpha$ forest [4], galaxy clustering and lensing [5], and the constraints on the baryonic matter density from Big Bang Nucleosynthesis [6]. However, the only unambiguous signatures of dark matter originate from its gravitational interactions. The relation between the dark matter and the Standard Model particles studied in terrestrial experiments is one of the major outstanding problems in fundamental physics today. Many physicists have proposed new particles as dark matter candidates that address some long-standing mystery left unsolved by the Standard Model: for example, neutralinos in an attempt to solve the electroweak hierarchy problem with low-energy supersymmetry (see, e.g., the extensive early review [7]); axions as introduced to solve the strong-CP problem [8, 9]; or heavy sterile neutrinos if there are additional states associated with the neutrino mass matrix [10].

It is also possible that the dark matter problem has a solution in the gravitational sector, via the nonlinearity of the Einstein equations. Primordial black holes (PBHs) [11, 12] would act as a form of cold dark matter from a cosmological structure formation point of view. PBHs do not require the introduction of new long-lived particles, although the formation mechanisms that do produce these objects does involve new physics at a very high energy scale since the "standard" inflationary power spectrum does not produce fluctuations large enough to collapse. A wide range of observational constraints have been investigated (see Carr *et al.* [13] for an extensive review), leaving a mass window from $10^{16} - 10^{21}$ g where PBHs can explain the total dark matter abundance.

Many astrophysical probes have been utilized to set stringent constraints based on the mass and formation mechanism of these objects. A few examples of these are optical microlensing constraints, gravitational wave detection constraints, accretion constraints, and dynamical constraints. Each mechanism probes its own individual mass window and have been used to set constraints on the current non-detection of these objects. However, an additional constraint may be placed due to phenomena of *Hawking radiation* [14, 15], the prediction that black holes should emit radiation and other particles causing a decrease in size and mass (commonly known as evaporation). The temperature of this radiation is $T_H = \kappa/(2\pi)$ in natural units, where $\kappa$ is the surface gravity, and for a Schwarzschild black hole of mass $M$ is $T_H = 1/(8\pi M)$. Lower mass black holes radiate at higher temperatures; in the $M \lesssim 10^{17}$ g range, black holes are hot enough to emit electrons and positrons. Black holes with $M \lesssim 5 \times 10^{14}$ g would have a lifetime less than the current age of the Universe [16].

Renewed interest in the asteroid-mass window ($10^{16} - 10^{22}$ g) has developed due to previous constraints being relaxed after more detailed investigation [17–19], opening up a region where PBHs may make up all the dark matter content. For a PBH of mass $M \sim 10^{16}$ g or lower, the Hawking temperature would be $T_H > 100$ keV, well within the $\gamma$-ray regime of the electromagnetic spectrum. At these energies, the dominant emission products are photons, electrons/positrons, and neutrinos. Many attempts have been made to constrain the PBH abundance by investigating how the emission products affect astrophysical observations. One set of constraints is based on the gamma ray photons from Hawking radiation, which contribute both to the extragalactic background [20] and the local sky brightness as seen from our location in the Milky Way [21]. Positrons emitted from PBHs can contribute to the 511 keV annihilation line from the Galactic center; this effect has also been used to set constraints on the PBH abundance [22, 23]. The $e^{\pm}$ spectrum measured from Voyager 1 (outside the heliosphere) also sets stringent bounds on $M < 10^{16}$ g black holes, eliminating the possibility of this mass window contributing a significant fraction of the dark matter [24]. Other limits have also been placed by radio observations of the Galactic center and by investigating the ionizing effects of these evaporation products over the course of cosmological history [25, 26]. These studies have motivated the need of rigorous and detailed study of the emission spectrum to set accurate constraints. As shown in Coogan *et al.* [27], current and future MeV telescopes such as the proposed AMEGO observatory are suitable to directly detect Hawking radiation if PBHs compose a considerable fraction of the dark matter content in the $10^{16} - 10^{18}$ g window. If a detection is made, it will have enormous implications towards the development of the field, not only providing direct experimental evidence of Hawking radiation, but also illuminating the identity of dark matter and providing evidence supporting the existence of PBHs.

The emission spectrum of these objects are well-studied and modeled as a thermal "greybody," a blackbody with an additional term known as a greybody factor that constitutes the energy-dependent cross section for a particle to be absorbed into the black hole. These calculations are performed by partial wave decomposition, computing the transmission probability through the angular momentum barrier, and then summing over modes [28]. Today they can be carried out by large codes such as BLACKHAWK [29–31]. The primary (directly emitted via Hawking radiation) and secondary (via decays and final state radiation) spectra have been studied extensively utilizing flat spacetime techniques borrowed from nuclear and particle physics [32, 33].

As a small black hole emits a charged particle, it will develop an equal and opposite charge (as measured by an external observer using Gauss's Law). If the hole is initially charged, it will lose its charge due to preferential Hawking radiation of particles of that sign charge, or via pair production from the hole's electric field (which is described as part of the same equation) [34, 35]. Page [36] studied the implications for Hawking radiation. The charge on the hole ($Z$ in units of the elementary charge) leads to an electric potential that perturbatively alters the emission rate of subsequent charged particles. While the emission of charged particles will follow a random walk, as the probability of emitting a particle with the same sign or opposite sign is nonzero, the emission of charged particles with the same sign as the charge of the black hole is more probable and causes the black hole to discharge as it approaches neutrality. This process results in a *stochastic charge* on the black hole, and Page [36] showed that the emission of charged leptons for PBHs in the $M \sim 10^{16}$ g mass window is reduced compared to the emission rate for neutral particles by 2–5%.

This paper is the second in a series whose ultimate goal is to calculate the full $\mathcal{O}(\alpha)$ corrections to the photon, electron, and positron spectrum emitted for a Schwarzschild black hole, where $\alpha \approx 1/137$ is the fine structure constant. The goal is to compute these corrections using full quantum electrodynamics (QED) on the Schwarzschild spacetime background. In Silva *et al.* [37], we performed the canonical quantization of the electron and photon fields using a generalization of the Coulomb gauge, and derived the formal expressions for the correction to the photon spectra; this includes the final state radiation effects [27, 38]. In this paper, we wish to understand how the stochastic charge effect arises within the framework of perturbative QED. We expect the "stochastic charge" physics to occur outside the horizon, since in any finite period of time $\Delta t$, no charge can enter or leave the horizon: strictly speaking, the charge of the black hole itself is not changing in Page [36]'s random walk. Rather, the sea of particles emitted by Hawking radiation (as well as the virtual particles) forms a plasma and the stochastic charge effect should be understood as a collective response to the ejection of a charged particle. Understanding how this works in detail is important both conceptually (to understand how "charge" is stored near a black hole horizon in the context of curved-spacetime QED) and technically (we need to know which terms in the QED formalism correspond to the stochastic charge so that later we do not double-count them later). These goals can be summarized in two conjectures about the full quantum mechanical treatment on curved spacetime:

1. **The semi-classical variance $\langle Z^2 \rangle$ as predicted by Page [36] agrees with the variance of a quantum mechanical operator $\langle \hat{\mathcal{Z}}^2 \rangle$ in the full QED calculation.** One of our tasks will be to identify and interpret the operator $\hat{\mathcal{Z}}$.

2. **There are terms in the QED correction to the emission spectrum that correspond to the semi-classical correction predicted by Page [36].** This requires a full investigation of the scalar potential sector of Coulomb gauge QED, which we did not explore in Paper I.

If both conjectures are proven true, then we may conclude that the concept of stochastic charge originates from the inclusion of quantum field theory on curved spacetime rather than rooted by semi-classical arguments as proposed by Page [36], hinting at the quantum nature of black holes.

## II. OUTLINE OF THIS PAPER

This document is part one of a series of two papers intending to prove stochastic charge as a phenomena originating from quantum field theory on a Schwarzschild background. Part 1 aims to prove the conjecture that the semi-classical variance, as predicted by Page [36], agrees with the variance of a quantum mechanical operator $\langle \hat{\mathcal{Z}}^2 \rangle$ in the full QED calculation. For sake of clarity, we outline the steps of the calculation as shown pictorially in Fig. 1.

III. We discuss and adopt the formalism and conventions of Paper I [37].

IV. We show how the stochastic charge effect originates semi-classically as predicted by Page [36]. Then we derive the semi-classical electron spectra $dN_-/dtdh$ and predicted variance $\langle Z^2 \rangle$, where $Z$ is the charge of the black hole in units of the fundamental charge.

V. We analyze the electrostatic sector of QED by deriving the interaction Hamiltonian mediated by the scalar potential $\Phi$ on a Schwarzschild background. While unnecessary in Paper 1 [37], the emission of charged leptons will cause the black hole to accumulate a charge and an electrostatic potential. Then we decompose this Hamiltonian into separate angular momentum modes $\ell$ and consider the subtle issues that arise in the $\ell = 0$ sector of the theory. Finally, we define a quantum operator $\hat{\mathcal{Z}}$ that counts the charge of the fermions mediated by a factor of $2M/r$ in the mode integrals.

VI. We present a hierarchy of expectation values that will prove crucial towards proving Conjecture # 1 and #2. After a lengthy derivation, we show the explicit form of the electron spectra in terms of these expectation values. Finally, we derive the time evolution of the variance $\langle \hat{\mathcal{Z}}^2 \rangle$ in our quantum formalism.

VII. We derive the time evolution of the expectation values as introduced in Sec. VI A which we use to derive the simplified form of $\langle \hat{\mathcal{Z}}^2 \rangle$ in terms of the fermion phase space densities and transmission and reflection coefficients. Once done so, we compare the quantum mechanical variance $\langle \hat{\mathcal{Z}}^2 \rangle$ we derived with the semi-classical variance $\langle Z^2 \rangle$ as predicted by Page [36].

Using the steps outlined above, it is possible to prove Conjecture #1 providing a hint towards the quantum nature of black holes.

## III. FORMALISM

This paper follows the formalism and conventions of Paper I. We use natural units where $G = c = \hbar = k_B = \epsilon_0 = 1$. We use the $+++-$ signature for the metric. The radial coordinate is the tortoise coordinate $r_\star = r + 2M \ln(r/2M - 1)$, so that the Schwarzschild metric is

$$ds^2 = \left(1 - \frac{2M}{r}\right) \left[ dr_\star^2 + \frac{r^2}{1 - 2M/r} (d\theta^2 + \sin^2\theta \, d\phi^2) - dt^2 \right]. \tag{1}$$

## Hierarchy of correlation functions contributing to the electron spectrum at order $e^2$

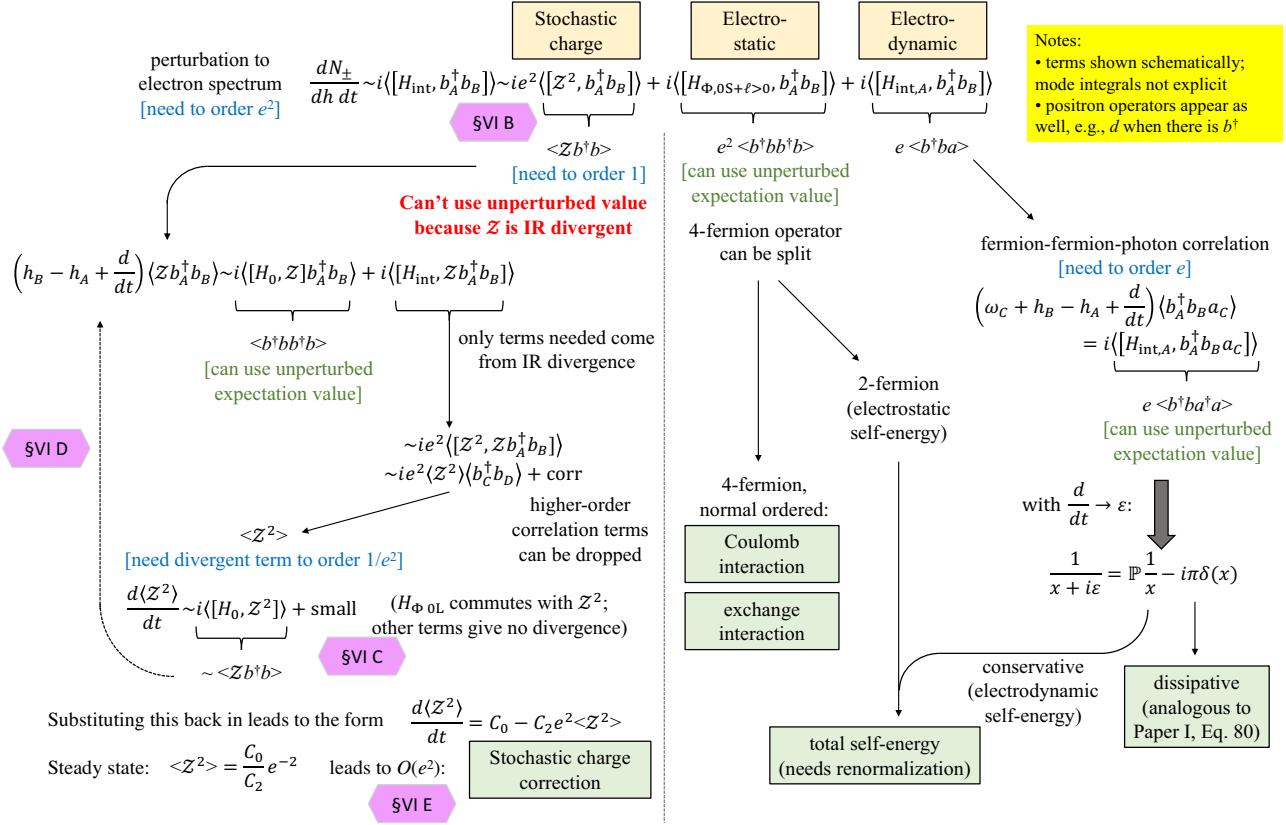

FIG. 1. The flow chart of correlation functions for this project. This paper is concerned with the "stochastic charge" sector; the electrostatic and electrodynamic sectors will be investigated in future papers. The sequence of steps in Section VI is indicated with the section labels.

Note that $r^2/(1-2M/r) \to \infty$ both at the horizon ($r_\star \to -\infty$) and at spatial infinity ($r_\star \to +\infty$).

The quantum numbers for the electrons are $\{X, k, m, h\}$, where $k = \pm 1, \pm 2, \pm 3, \dots$ is the Schrödinger separation constant (which encodes both the total angular momentum $j = |k| - \frac{1}{2}$ and with the sign of $k$ distinguishing between the two parity solutions); $m$ is the projection of the total angular momentum on the $z$-axis; $h$ is the single-particle energy; and $X \in \{\text{in, up}\}$ indexes the scattering solutions (coming "in" from spatial infinity or coming "up" from the horizon). These solutions are related to the basis of final solutions (going "out" to spatial infinity or "down" into the horizon) by the reflection and transmission coefficients, $R_{\frac{1}{2}kh}$ and $T_{\frac{1}{2}kh}$, where the "$\frac{1}{2}$" identifies these as coefficients for fermions, and $k$ and $h$ are the quantum numbers (these coefficients do not depend on $m$ due to spherical symmetry).

The decomposition of the electromagnetic potential $A_\mu$ in the Coulomb-like gauge leads to a dynamical vector potential $A_i$ (restricted by a gauge condition so there are only 2 independent functions of position rather than 3) and a non-dynamical scalar potential $\Phi$. The quantum numbers for the vector are $\{X, \ell, p, m, \omega\}$, where $\ell = 1, 2, 3, \dots$ is the total angular momentum and $m$ is its projection; $p$ is the parity (electric-like or magnetic-like); $\omega$ is the single-particle energy; and again $X \in \{\text{in, up}\}$. The scalar potential can be decomposed in spherical harmonics,

$$\Phi(r_\star, \theta, \phi) = \sum_{\ell=0}^{\infty} \sum_{m=-\ell}^{\ell} \Phi_{\ell m}(r_\star) Y_{\ell m}(\theta, \phi). \tag{2}$$

A key difference from the vector potential sector is that the scalar potential sector includes $\ell = 0$. This special case will feature prominently in this paper because it mediates the stochastic charge effect.

## IV. THE SEMI-CLASSICAL CALCULATION OF STOCHASTIC CHARGE

Here we briefly review the calculation of the stochastic charge effect in Page [36]. Our main objective is to highlight the expressions for $\langle Z^2 \rangle$ and the correction to the emission spectrum $dN/dh\,dt$, so that we can compare these results to the results obtained here in QED. Because of our interest in perturbation theory, we present the Taylor expansion of the machinery in Page [36], rather than quoting specific values of $Z$. The radial mode functions for a charged spin $\frac{1}{2}$ particle with a Dirac mass term when the black hole carries an electric charge are described in Brill and Wheeler [39].

In the semiclassical picture, a black hole has a well-defined charge $Q = Z|e| = -Ze$, and a horizon potential $\Phi_{\mathrm{H}} = -Ze/(8\pi M)$ so that the potential energy of a unit positive charge at the horizon is $|e|\Phi_{\mathrm{H}} = \alpha Z/2M$. We may think of $Z$ as an "atomic number" for the black hole. The emission rate of positrons ($e^+$, upper sign) or electrons ($e^-$, lower sign) is

$$\frac{dN_{e^\pm}}{dh\,dt}(Z) = \frac{1}{2\pi} \sum_k (2j+1) \frac{\mathbb{T}_{\frac{1}{2}kh}(Z \mp 1; \pm)}{e^{8\pi M[h \pm \alpha(Z\mp1)/2M]} + 1} \quad \text{for} \ \ h > \mu, \tag{3}$$

where $\mathbb{T}_{\frac{1}{2}kh}(Z'; \pm) = |T_{\frac{1}{2}kh}(Z'; \pm)|^2$ is the outgoing transmission probability for an $e^\pm$ of energy $h$ and Schrödinger separation constant $k$ in the field of a black hole of charge $Z'$. The use of $Z \mp 1$ is because emission of a positron (electron) from a black hole of atomic number $Z$ leaves behind a residual object of atomic number $Z \mp 1$.

Page [36] argued that the charge on the black hole fluctuates in a Markovian process: if there is an emission rate for electrons $\Gamma(Z \to Z+1)$, and a rate for positrons $\Gamma(Z \to Z-1)$, where the arrows indicate the appropriate remaining charge on the black hole, then the steady-state probability distribution of $Z$ satisfies

$$\frac{P(Z)}{P(Z-1)} = \frac{\Gamma(Z-1 \to Z)}{\Gamma(Z \to Z-1)} \quad \text{or} \quad \ln P(Z) = \sum_{Z'=1}^{Z} \ln \frac{\Gamma(Z'-1 \to Z')}{\Gamma(Z' \to Z'-1)} + \text{const.} \tag{4}$$

The emission rates depend on $e$ and $Z$ only through the combination $\alpha(Z \mp 1)$. Then if the electron emission rate is

$$\Gamma(Z \to Z+1) = \Gamma_{e,0} + \alpha(Z+1)\Gamma_{e,1} + [\alpha(Z+1)]^2\Gamma_{e,2}\ldots, \tag{5}$$

and the positron emission rate is the same except for flipping the sign of $Z$, we find to lowest order in $\alpha$ that

$$\ln P(Z) = \sum_{Z'=1}^{Z} \ln \frac{\Gamma_{e,0} + \alpha Z'\Gamma_{e,1} + \ldots}{\Gamma_{e,0} - \alpha(Z'-1)\Gamma_{e,1} + \ldots} + \text{const} = \alpha\frac{\Gamma_{e,1}}{\Gamma_{e,0}} \sum_{Z'=1}^{Z}(2Z'-1) + \text{const} = \alpha\frac{\Gamma_{e,1}}{\Gamma_{e,0}}Z^2 + \text{const.} \tag{6}$$

It thus follows that to lowest order in $\alpha$, the probability distribution is a Gaussian with variance

$$\langle Z^2 \rangle \approx -\frac{\Gamma_{e,0}}{2\alpha\Gamma_{e,1}}. \tag{7}$$

(Note that $\Gamma_{e,1}$ is negative, since a positively charged hole is less likely to emit an electron.) In the limit of small $\alpha$, this variance is much larger than unity, so the quantization of the black hole's charge can be neglected. We may write this in terms of transmission probabilities by noting that

$$\Gamma_{e,\lambda} = \int_\mu^\infty \frac{dN_{[\lambda]}}{dh\,dt}\,dh, \tag{8}$$

with

$$\frac{dN_{[\lambda]}}{dh\,dt} = \frac{1}{2\pi} \frac{\partial}{\partial Z'^\lambda} \sum_k (2j+1) \frac{\mathbb{T}_{\frac{1}{2}kh}(Z'; -)}{e^{8\pi M[h + \alpha Z'/2M]} + 1}\bigg|_{Z'=0}. \tag{9}$$

The corresponding electron spectrum is

$$\frac{dN_-}{dh\,dt} = \frac{dN_{[0]}}{dh\,dt} + \alpha\frac{dN_{[1]}}{dh\,dt} + \alpha^2(\langle Z^2 \rangle + 1)\frac{dN_{[2]}}{dh\,dt} + \ldots = \frac{dN_{[0]}}{dh\,dt} + \alpha\left(\frac{dN_{[1]}}{dh\,dt} - \frac{\Gamma_{e,0}}{2\Gamma_{e,1}}\frac{dN_{[2]}}{dh\,dt}\right) + \ldots, \tag{10}$$

where the additional terms are of order $\alpha^2$ or higher.

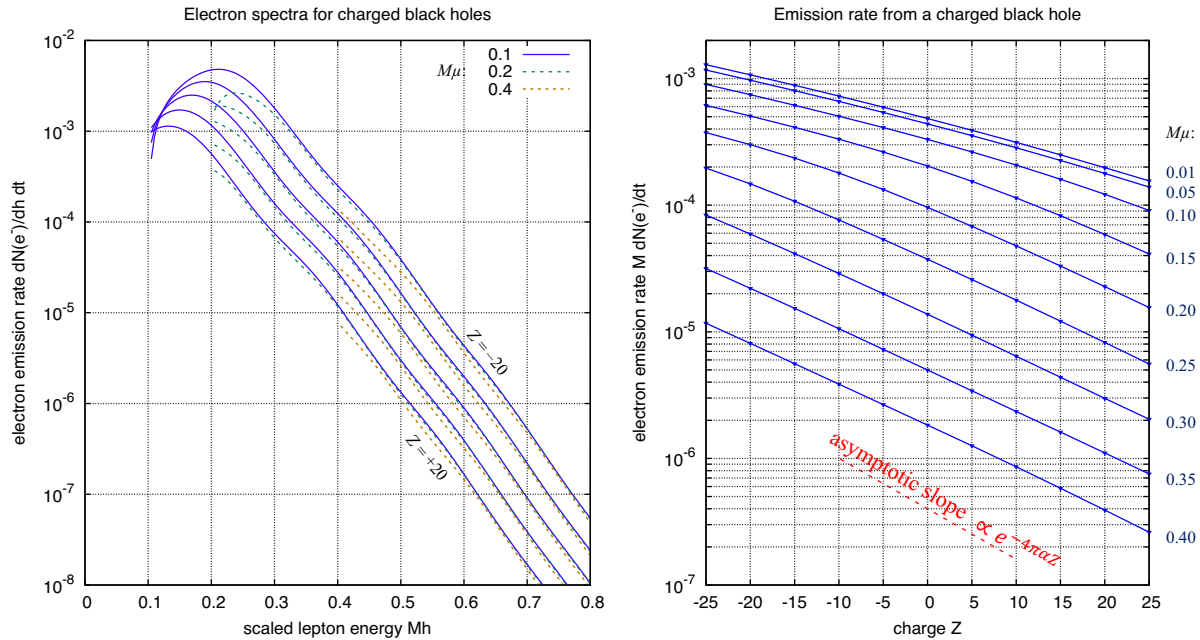

FIG. 2. The emission rate for electrons as a function of black hole mass and charge. The black hole is taken to be nonrotating in all cases. *Left panel*: The electron spectra, for three masses (scaled mass $M\mu = 0.1$, $0.2$, and $0.4$), and with charges $Z$ of $-20$ (top), $-10$, $0$, $+10$, and $+20$ (bottom). *Right panel*: The total electron emission rate as a function of charge $Z$ (and scaled by the black hole mass $M$). The rate decreases as $M\mu$ is increased. The dashed line shows the predicted asymptotic slope, $dN(e^-)/dt \propto e^{-4\pi\alpha Z}$ [36], in the limit of $M\mu \gg 1$.

We may now phrase precisely the two conjectures in the introduction: Conjecture #1 holds that Eq. (7) describes an appropriate expectation value $\langle \hat{\mathcal{Z}}^2 \rangle$ in QED; and Conjecture #2 holds that the contributions to the electron spectrum in Eq. (10) arise via collective effects outside the horizon in QED.

Numerical results are shown in Fig. 2, which are our updates of Figs. 2 and 4 in Page [36]. This calculation used $\ln(r - 2M)$ as an independent variable, and the implicit midpoint method to solve the radial electron wave function.[1] The default settings used for this figure are partial waves up through $j_{\max} = \frac{15}{2}$, and integration through $1.2 \times 10^6$ steps from $r_{\max} = 5000M$ down to $r = (2 + 7.6 \times 10^{-5})M$.

## V. ELECTROSTATIC SECTOR OF THE THEORY

We now turn to the full QED treatment of the stochastic charge problem, building on the formalism of Paper I. Paper I quantized the electromagnetic field in a generalization of the Coulomb gauge [Paper I Eq. (12)]. In this gauge, the propagating photon modes are identified with the vector potential $A_i$; the scalar potential $\Phi$ has no dynamics and mediates an instantaneous electrostatic interaction between charges. Our first step here is to consider the scalar potential sector of QED, which was not needed to compute the $\mathcal{O}(e^2)$ corrections to the photon spectrum. We first follow through the "standard" treatment in Coulomb gauge to derive the interaction Hamiltonian mediated by $\Phi$. Then we consider the subtle issues that arise in the $\ell = 0$ sector, which have some new elements relative to the Minkowski spacetime quantization.

---

[1] Implicit methods are sometimes used to improve stability of stiff solvers, however the key feature that is of interest here is that the implicit midpoint method exactly preserves the Wronskian relation. See, e.g., Iserles [40] for a thorough and accessible discussion of these aspects of ordinary differential equation solvers.

## A. The instantaneous Coulomb interaction

Paper I Eq. (15) included an additional contribution to the Lagrangian:

$$L_{\text{EM},\Phi} = \sum_{\ell m} \int \left\{ \frac{1}{2} \frac{r^2}{1 - 2M/r} |\Phi'_{\ell m}|^2 + \frac{1}{2}\ell(\ell+1)|\Phi_{\ell m}|^2 \right\} dr_\star. \tag{11}$$

There is also an interaction Lagrangian between the electron field and the $\Phi$ sector (following Paper I Eq. 49),

$$\begin{aligned}
L_{\text{int},\Phi} &= -ie \int \sqrt{-g}\, d^3x\, \bar{\psi}(x)\boldsymbol{\gamma}^4\Phi(x)\psi(x) \\
&= -e \int r^2 \sqrt{1 - \frac{2M}{r}}\, dr_\star\, \sin\theta\, d\theta\, d\phi\, \Phi(x)\, \bar{\psi}(x)\boldsymbol{\beta}\psi(x) \\
&= -e \sum_{\ell m} \int dr_\star\, \Phi_{\ell m}(r_\star) D^\dagger_{\ell m}(r_\star),
\end{aligned} \tag{12}$$

where we have defined the operator

$$D_{\ell m}(r_\star) = r^2 \sqrt{1 - \frac{2M}{r}} \int \sin\theta\, d\theta\, d\phi\, Y_{\ell m}(\theta,\phi)\, \bar{\psi}(x)\boldsymbol{\beta}\psi(x). \tag{13}$$

A calculation using the fermion anti-commutation relations (Paper I, Eq. 45) shows that

$$[\bar{\psi}(x)\boldsymbol{\beta}\psi(x), \bar{\psi}(y)\boldsymbol{\beta}\psi(y)] = 0 \tag{14}$$

due to a cancellation of anti-commutation terms.[2] Thus the $D$ operators commute.

A subtle issue is that as written, there is a divergent nonzero charge density associated with the vacuum, even in flat spacetime, since $\bar{\psi}(x)\boldsymbol{\beta}\psi(x) \neq 0$. This can be interpreted as the infinite charge density of electrons associated with the Dirac sea. This problem can be solved by subtracting an infinite 4-current density $J^\mu_{\text{div}}(x)$ (i.e., adding $\int A_\mu(x)J^\mu_{\text{div}}(x)\sqrt{-g}d^4x$ to the action) to cancel this divergence. In flat spacetime QED, we may find the correct divergent term by noting that there is a unique Lorentz-invariant vacuum state for the electrons $|0\rangle$. The Lorentz invariance of the overall theory then requires that $\langle 0|J^\mu|0\rangle = 0$, and we conclude that the correct way to subtract the divergent charge density is then to use normal ordering of the fermion operators, i.e., $:\bar{\psi}(x)\boldsymbol{\beta}\psi(x):$ in place of $\bar{\psi}(x)\boldsymbol{\beta}\psi(x)$. In curved spacetime, there is no such consideration. However, if we require charge conjugation invariance, then we must have $\langle 0_{\text{B}}|J^\mu|0_{\text{B}}\rangle = 0$ for the Boulware vacuum $|0_{\text{B}}\rangle$. So in this case as well, the correct computation is still the replacement

$$\bar{\psi}(x)\boldsymbol{\beta}\psi(x) \quad \rightarrow \quad :\bar{\psi}(x)\boldsymbol{\beta}\psi(x):. \tag{15}$$

In what follows, we make this replacement.[3]

We did not need to consider this part in Paper I, because at order $\mathcal{O}(e^2)$ there are no diagrams by which it can affect the emitted photon spectrum. However, it is important here because it can affect the electrons. Since $\dot{\Phi}$ does not appear in the Lagrangian, there is no dynamics in $\Phi$: we may treat it as a constrained quantity and write $\delta L/\delta\Phi^\dagger_{\ell m}(r_\star) = 0$ (where both $L_{\text{EM},\Phi}$ and $L_{\text{int},\Phi}$ are contributions to $L$). This leads to

$$0 = \frac{\delta L}{\delta\Phi^\dagger_{\ell m}(r_\star)} = -\partial_{r_\star}\left[\frac{r^2}{1 - 2M/r}\partial_{r_\star}\Phi_{\ell m}(r_\star)\right] + \ell(\ell+1)\Phi_{\ell m}(r_\star) - eD_{\ell m}(r_\star), \tag{16}$$

and hence

$$\Phi_{\ell m}(r_\star) = e \int G_\ell(r_\star, r'_\star)D_{\ell m}(r'_\star)\, dr'_\star, \tag{17}$$

---

[2] There is a slight subtlety here because the divergence of $\langle\psi(x)\psi^\dagger(x+\epsilon)\rangle$ as $\epsilon \to 0$ means that some commutators that appear to be zero are actually singular. This includes $[\rho(x), J^i(x)]$ in Minkowski spacetime [41]. Fortunately, the commutator in Eq. (14) is still zero, even in the Schwarzschild geometry. This is most easily seen by regularizing the divergence by placing the fermions on a 3D lattice (leaving time continuous, as one would do if the fermions were sites on a crystal; in Schwarzschild spacetime the lattice does not have discrete translation symmetry, but this presents no issue). In this case, one trivially finds $[\rho(x_a), \rho(x_b)] = 0$, where the $a$ and $b$ indices indicate lattice sites. But current density $J^i = (-g)^{-1/2}\delta L/\delta A_i$ is defined not on the sites themselves but on the links connecting adjacent sites in the $i$-direction, thus spoiling the naive relation $[\rho(x), J^i(x)] = 0$.

[3] A further subtlety is that charge conjugation invariance is a good symmetry of QED, but not of the full Standard Model. Therefore, if we were to expand to all the known particles and fields, we could not make this argument. The issues associated with expanding this work beyond electrodynamics to the full Standard Model on curved spacetime are left for future work.

where the Green's function $G_\ell$ is the solution to

$$-\partial_{r_\star} \left[ \frac{r^2}{1-2M/r} \partial_{r_\star} G_\ell(r_\star, r'_\star) \right] + \ell(\ell+1) G_\ell(r_\star, r'_\star) = \delta(r_\star - r'_\star). \tag{18}$$

Since it is derived from a self-adjoint problem, the Green's function is symmetric in its two arguments. The conclusion is that $\Phi$ mediates an additional contribution to the action,

$$
\begin{aligned}
L_\Phi &= L_{\text{EM},\Phi} + L_{\text{int},\Phi} \\
&= \sum_{\ell m} \int \left\{ -\frac{1}{2} \Phi_{\ell m}(r_\star) \partial_{r_\star} \left[ \frac{r^2}{1-2M/r} \partial_{r_\star} \Phi_{\ell m}^\dagger(r_\star) \right] + \frac{1}{2} \ell(\ell+1) \Phi_{\ell m}^\dagger(r_\star) \Phi_{\ell m}(r_\star) - e \Phi_{\ell m}(r_\star) D_{\ell m}^\dagger(r_\star) \right\} dr_\star \\
&= -\frac{1}{2} e^2 \sum_{\ell m} \int G_\ell(r_\star, r'_\star) D_{\ell m}^\dagger(r_\star) D_{\ell m}(r'_\star) \, dr_\star \, dr'_\star,
\end{aligned}
\tag{19}
$$

where in the first step we integrated by parts in $L_{\text{EM},\Phi}$ and in the second step we substituted the Green's function. We conclude that the $\Phi$ sector mediates a non-local interaction between two fermions,

$$H_\Phi = -L_\Phi = \frac{1}{2} e^2 \sum_{\ell m} \int G_\ell(r_\star, r'_\star) D_{\ell m}^\dagger(r_\star) D_{\ell m}(r'_\star) \, dr_\star \, dr'_\star. \tag{20}$$

Again, this is all analogous to the Coulomb gauge in flat spacetime.

Of special interest to us is the behavior of the Green's function near the horizon and at spatial infinity. Because of the instantaneous nature of the scalar interaction, this is the same Green's function that occurs in electrostatics in the Schwarzschild spacetime [42–47]. The limiting behaviors are

$$G_\ell(r_\star, r'_\star) \sim \begin{cases} \text{constant} \times e^{\ell r_\star / 2M} & r_\star \to -\infty, \quad r'_\star = \text{fixed}, \\ \text{constant} \times r^{-(\ell+1)} & r_\star \to +\infty, \quad r'_\star = \text{fixed}. \end{cases} \tag{21}$$

For $\ell \neq 0$ therefore, the Green's function decays away both at $r_\star \to \pm\infty$, and the electrostatic interaction of two particles decays away as either one approaches the horizon or spatial infinity. The $\ell = 0$ case is different and will require special treatment to avoid divergences associated with the infinite number of particles near the horizon (as expressed in the Boulware Fock space).

### B. The $\ell = 0$ sector

There are two (related) subtleties in the $\ell = 0$ sector. Both are related to the behavior of the $\tilde{\mathcal{H}}_0$ and $\mathcal{H}_0$ operators defined in Paper I Eqs. (13) and (16), which in turn are related to the topology of the quantization domain (with the black hole excised).

Let us start with the gauge fixing in Paper I §IIB. For $\ell = 0$, the gauge choice of Paper I Eq. (12), which eliminated the cross-terms between scalar and vector potentials, allows

$$A_{(r),0,0} = \frac{\xi}{\sqrt{4\pi}} \frac{1-2M/r}{r^2}, \tag{22}$$

where $\xi$ is independent of $r$. We will set $\xi = 0$, so that the vector potential with $\ell = 0$ disappears from the theory entirely and there is only a scalar potential.

The other issue is the Green's function: for $\ell > 0$, $G_\ell(r_\star, r'_\star)$ is well-behaved, and vanishes at both $r_\star \to \pm\infty$. However, a subtlety occurs for the case of $\ell = 0$. Using the fact that $\partial_{r_\star} = (1-2M/r)\partial_r$, one can see that the Green's function is

$$r^2 \partial_r G_0(r_\star, r'_\star) = \begin{cases} C & r_\star < r'_\star \\ C - 1 & r_\star > r'_\star \end{cases}$$

and hence

$$G_0(r_\star, r'_\star) = \begin{cases} B - Cr^{-1} & r_\star < r'_\star \\ B - (C-1)r^{-1} & r_\star > r'_\star \end{cases},$$

where $B$ and $C$ may depend on $r'_\star$. The "simplest" possibility is to take $B = C = 0$, which leads to the Green's function

$$G_0(r_\star, r'_\star) = \frac{1}{\max(r, r')}.\tag{23}$$

However, it is important to consider what influence $B$ and $C$ would have: these correspond to an additional potential of the form

$$\Phi^{\text{ext}}(r_\star) = \Phi_\infty + \frac{Q_\infty}{4\pi r},\tag{24}$$

where $\Phi_\infty$ and $Q_\infty$ are constants. The potential is then

$$\Phi_{\ell m}(r_\star) = e \int G_\ell(r_\star, r'_\star) D_{\ell m}(r'_\star) \, dr'_\star + \sqrt{4\pi} \, \delta_{\ell 0} \delta_{m0} \Phi^{\text{ext}}(r_\star).\tag{25}$$

Following Eq. (25), we also see that the surface term from integration by parts in Eq. (19), which for all $\ell \geq 1$ are zero, is non-trivial for the monopole contribution ($\ell = 0$) and must be accounted for in the final expression for $L_\Phi$.

We begin by writing $L_\Phi$ with the surface term (regardless of multipole moment)

$$\begin{aligned}
L_\Phi &= \sum_{\ell m} \int \left\{ -\frac{1}{2} \Phi_{\ell m}(r_\star) \partial_{r_\star} \left[ \frac{r^2}{1 - 2M/r} \partial_{r_\star} \Phi^\dagger_{\ell m}(r_\star) \right] + \frac{1}{2} \ell(\ell + 1) \Phi^\dagger_{\ell m}(r_\star) \Phi_{\ell m}(r_\star) - e \Phi_{\ell m}(r_\star) D^\dagger_{\ell m}(r_\star) \right\} dr_\star \\
&\quad + \frac{1}{2} \Phi_{\ell m} \frac{r^2}{1 - 2M/r} \partial_{r_\star} \Phi^\dagger_{\ell m} \Big|_{r_\star \to -\infty}^{r_\star \to \infty}, \\
&= -\frac{1}{2} e \sum_{\ell m} \int \left\{ \Phi_{\ell m} D^\dagger_{\ell m}(r_\star) \right\} dr_\star + \frac{1}{2} \Phi_{\ell m} \frac{r^2}{1 - 2M/r} \partial_{r_\star} \Phi^\dagger_{\ell m} \Big|_{r_\star \to -\infty}^{r_\star \to \infty}.
\end{aligned}\tag{26}$$

Focusing on the surface term for now, we can use Eq. (25) for the monopole contribution (since this is the only condition when the surface term does *not* go to zero at the boundaries) to give

$$\begin{aligned}
\frac{1}{2} \Phi_{00} \frac{r^2}{1 - 2M/r} \partial_{r_\star} \Phi^\dagger_{00} \Big|_{r_\star \to -\infty}^{r_\star \to \infty} &= \frac{1}{2} \left( e \int G_0(r_\star, r'_\star) D_{00}(r'_\star) \, dr'_\star + \sqrt{4\pi} \left( \Phi_\infty + \frac{Q_\infty}{4\pi r} \right) \right) \times \\
&\quad r^2 \left( e \int \partial_r G_0(r_\star, r'_\star) D_{00}(r'_\star) \, dr'_\star - \sqrt{4\pi} \frac{Q_\infty}{4\pi r^2} \right) \Big|_{r \to 2M}^{r \to \infty},
\end{aligned}\tag{27}$$

where we used the coordinate conversion between $r_\star$ and the usual radial coordinate to rewrite the derivative and change the boundary conditions. We can evaluate the surface term at boundaries by using the form of the Green's function and its derivatives. Evaluating the surface term at the boundaries gives

$$\lim_{r \to \infty} \frac{1}{2} \Phi_{00} \frac{r^2}{1 - 2M/r} \partial_{r_\star} \Phi^\dagger_{00} = -\frac{1}{2} \left( \sqrt{4\pi} \Phi_\infty \right) \left( e \int D_{00}(r'_\star) \, dr'_\star + \sqrt{4\pi} \frac{Q_\infty}{4\pi} \right),\tag{28}$$

$$\lim_{r \to 2M} \frac{1}{2} \Phi_{00} \frac{r^2}{1 - 2M/r} \partial_{r_\star} \Phi^\dagger_{00} = \frac{1}{2} \left( e \int \frac{1}{r'} D_{00}(r'_\star) \, dr'_\star + \sqrt{4\pi} \left( \Phi_\infty + \frac{Q_\infty}{8\pi M} \right) \right) \left( -\sqrt{4\pi} \frac{Q_\infty}{4\pi} \right).\tag{29}$$

Putting everything together gives the final version of $L_\Phi$ for a general multipole moment is

$$\begin{aligned}
L_\Phi &= -\frac{1}{2} e \sum_{\ell m} \left[ \int \left\{ \Phi_{\ell m} D^\dagger_{\ell m}(r_\star) \right\} dr_\star + \delta_{\ell 0} \delta_{m0} \left( -\frac{1}{2} \Phi_\infty Q_f + \frac{\sqrt{4\pi}}{2} e Q_\infty \int \frac{D_{00}(r_\star)}{4\pi r} dr_\star + \frac{Q_\infty^2}{16\pi M} \right) \right] \\
&= -\frac{1}{2} e^2 \sum_{\ell m} \int G_\ell(r_\star, r'_\star) D^\dagger_{\ell m}(r_\star) D_{\ell m}(r'_\star) \, dr_\star \, dr'_\star - \Phi_\infty Q_f + \frac{Q_\infty^2}{16\pi M},
\end{aligned}\tag{30}$$

where we have defined the operator

$$Q_f = \sqrt{4\pi} \, e \int D_{00}(r_\star) \, dr_\star = e \int r^2 \sqrt{1 - \frac{2M}{r}} \, dr_\star \, \sin\theta \, d\theta \, d\phi \; : \bar\psi(x) \boldsymbol{\beta} \psi(x) :,\tag{31}$$

which we can recognize as the conserved electromagnetic charge associated with the fields. By substituting in the fermion operators from Paper I Eq. (46), and using the orthonormality relations of Paper I Eq. (43), we may express Eq. (31) as

$$Q_{\mathrm{f}} = e \int_0^\infty \frac{dh}{2\pi} \sum_{Xkm} \left( \hat{b}_{X,k,m,h}^\dagger b_{X,k,m,h} - \hat{d}_{X,k,m,h}^\dagger d_{X,k,m,h} \right),\tag{32}$$

where we have dropped the additive constant that is associated with the anti-commutation of the $\hat{d}$ and $\hat{d}^\dagger$ operators. The resulting $Q_{\mathrm{f}}$ operator satisfies $Q_{\mathrm{f}}|\Omega_{\mathrm{B}}\rangle = 0$, where $|\Omega_{\mathrm{B}}\rangle$ is the Boulware vacuum, and it has the simple interpretation in the Boulware Fock space as the total charge of the fermions ($e$ for each electron and $-e$ for each positron). It is easily seen that

$$[H, Q_{\mathrm{f}}] = 0\tag{33}$$

where $H$ is the exact full Hamiltonian for QED, i.e., that $Q_{\mathrm{f}}$ is conserved.

Since the time derivatives of $\Phi_\infty$ and $Q_\infty$ do not appear in the Lagrangian, they also satisfy a constraint, and we should have $\partial L/\partial \Phi_\infty = 0$ and $\partial L/\partial Q_\infty = 0$. So far, the only terms we have included in the Lagrangian that depend on $\Phi_\infty$ and $Q_\infty$ are those in Eq. (30), and these would imply $Q_{\mathrm{f}} = 0$ and $Q_\infty = 0$, respectively. However, the electromagnetic gauge symmetry also allows additional terms in the Lagrangian that depend on the boundaries at $\pm\infty$. Specifically, we know that a charge $q$ placed at a fixed position contributes to the action according to $S = -q \int_{t_i}^{t_f} \Phi(\mathbf{x}, t)\, dt$; and so it is possible to place charges $\mathcal{Q}_+$ at $r_\star = +\infty$ (i.e., at spatial infinity) and $\mathcal{Q}_-$ at $r_\star = -\infty$ (i.e., at the horizon). So adding charges at the boundary then gives us a boundary term in the Lagrangian:

$$L_{\mathrm{bdy}} = -\frac{1}{\sqrt{4\pi}} [\mathcal{Q}_+ \Phi_{00}(r_\star = +\infty) + \mathcal{Q}_- \Phi_{00}(r_\star = -\infty)] = -\mathcal{Q}_+ \Phi_\infty - \mathcal{Q}_- \left[ \Phi_\infty + \frac{Q_\infty}{8\pi M} + \frac{1}{\sqrt{4\pi}} e \int \frac{1}{r'} D_{00}(r_\star') \, dr_\star' \right].\tag{34}$$

The inclusion of these boundary terms leads to the constraints

$$\frac{\partial L}{\partial \Phi_\infty} = 0 \;\to\; Q_{\mathrm{f}} + \mathcal{Q}_+ + \mathcal{Q}_- = 0 \;\text{ and }\; \frac{\partial L}{\partial Q_\infty} = 0 \;\to\; Q_\infty = \mathcal{Q}_-.\tag{35}$$

It can therefore be seen that once the $\mathcal{Q}_\pm$ are specified, only states with $Q_{\mathrm{f}} = -(\mathcal{Q}_+ + \mathcal{Q}_-)$ are allowed. (We may of course work with the full Fock space, but this is a superspace of the Hilbert space of allowed states.) Moreover, $Q_\infty$ satisfies a simple constraint. The resulting total Lagrangian is then

$$L_\Phi + L_{\mathrm{bdy}} = -\frac{1}{2} e^2 \sum_{\ell m} \int G_\ell(r_\star, r_\star') D_{\ell m}^\dagger(r_\star) D_{\ell m}(r_\star') \, dr_\star \, dr_\star' - \frac{\mathcal{Q}_-^2}{16\pi M} - \mathcal{Q}_- \frac{1}{\sqrt{4\pi}} e \int \frac{1}{r'} D_{00}(r_\star') \, dr_\star'.\tag{36}$$

This implies a contribution to the Hamiltonian,

$$H_{\Phi + \mathrm{bdy}} = \frac{1}{2} e^2 \sum_{\ell m} \int G_\ell(r_\star, r_\star') D_{\ell m}^\dagger(r_\star) D_{\ell m}(r_\star') \, dr_\star \, dr_\star' + \frac{\mathcal{Q}_-^2}{16\pi M} + \mathcal{Q}_- \frac{1}{\sqrt{4\pi}} e \int \frac{1}{r'} D_{00}(r_\star') \, dr_\star'.\tag{37}$$

The first term here is the usual Green's function term from the fermions interacting with each other. The second term is a capacitance energy associated with the charge that we have placed on the horizon (recall that a black hole is a capacitor with capacitance $8\pi M$); it is constant and will not concern us. The third term is an electrostatic interaction energy between the hole and the fermions.

Considerations of this type can occur in Minkowski spacetime as well, where one must consider $\Phi_\infty$ and $\mathcal{Q}_+$ (but the inner boundary condition gives no $Q_\infty$ or $\mathcal{Q}_-$). But in the Minkowski case, aside from requiring the conservation of the total charge $Q_{\mathrm{f}}$ of the fields, there are no additional terms in the Hamiltonian analogous to the $\mathcal{Q}_-$ terms in Eq. (37).

### C. The decomposition into short- and long-range components

It is useful to break down the Green's function for $\ell = 0$ into a "long-range" piece that packages the infrared divergence into a single operator, and a "short-range" piece that has no divergence. This decomposition is:

$$G_0(r_\star, r_\star') = G_{0\mathrm{L}}(r_\star, r_\star') + G_{0\mathrm{S}}(r_\star, r_\star'), \quad G_{0\mathrm{L}}(r_\star, r_\star') = \frac{2M}{rr'}, \quad G_{0\mathrm{S}}(r_\star, r_\star') = \left(1 - \frac{2M}{\min(r, r')}\right) \frac{1}{\max(r, r')}.\tag{38}$$

Then $G_{0S}$ is a "short-range" contribution, which converges exponentially to zero at $r_\star \to -\infty$, while $G_{0,L}$ is a "long-range" contribution that goes to a constant near the horizon and should capture all the infrared divergent parts.

Using Eq. (37), we find that

$$H_{\Phi+\text{bdy}} = \sum_{\ell=1}^{\infty} \sum_{m=-\ell}^{\ell} \frac{1}{2}e^2 \int G_\ell(r_\star, r'_\star) D_{\ell m}^\dagger(r_\star) D_{\ell m}(r'_\star) dr_\star dr'_\star + \frac{1}{2}e^2 \int G_{0S}(r_\star, r'_\star) D_{00}(r_\star) D_{00}(r'_\star) dr_\star dr'_\star + H_{\text{int},0L+\text{bdy}},$$
(39)

where we recall that $D_{00}(r_\star)$ is Hermitian. The long range part of the Green's function and the boundary terms are packaged together as

$$H_{\text{int},0L+\text{bdy}} = \frac{1}{2}e^2 \int G_{0L}(r_\star, r'_\star) D_{00}(r_\star) D_{00}(r'_\star) dr_\star dr'_\star + \mathcal{Q}_- - \frac{e}{\sqrt{4\pi}} \int \frac{1}{r'} D_{00}(r'_\star) dr'_\star + \frac{\mathcal{Q}^2}{16\pi M} = \frac{e^2}{16\pi M}\hat{\mathcal{Z}}^2,$$
(40)

where

$$\hat{\mathcal{Z}} = -\sqrt{4\pi} \int_{-\infty}^{\infty} \frac{2M}{r} D_{00}(r_\star) dr_\star - \frac{\mathcal{Q}_-}{e}.$$
(41)

(We put a ^ over $\hat{\mathcal{Z}}$ to emphasize that it is an operator, as distinct from the classical $Z$ in the Page [36] calculation.) The binomial expansion of $\hat{\mathcal{Z}}^2$ gives the three components we need: the square of the first term gives the long-range Green's function ($G_{0L}$); the square of the second term gives the capacitance energy; and the cross-term gives the interaction of the horizon charge and fermion charge in Eq. (37).

We may physically interpret $\hat{\mathcal{Z}}$ by comparison to $Q_f$ from Eq. (31):

$$-e\hat{\mathcal{Z}} = \sqrt{4\pi}e \int_{-\infty}^{\infty} \frac{2M}{r} D_{00}(r_\star) dr_\star + \mathcal{Q}_- \quad \text{versus} \quad Q_f = \sqrt{4\pi}e \int_{-\infty}^{\infty} D_{00}(r_\star) dr_\star.$$
(42)

So while $Q_f$ is the total charge of the fermions, $-e\hat{\mathcal{Z}}$ is the charge of the fermions weighted by $2M/r$: a fermion near the horizon is fully included in the integral, but a fermion near spatial infinity is excluded, and in between the weighting varies smoothly with radius. Also $-e\hat{\mathcal{Z}}$ includes the "charge on the horizon" $\mathcal{Q}_-$. So in this sense, we may think of $\hat{\mathcal{Z}}$ as like the atomic number of the black hole (its charge in units of the proton charge) in the limit that charges are either far away (either not in the hole) or near the horizon (in which case they appear to an external observer to be uniformly spread out over the horizon). In particular, the Hawking-radiated fermions that will emerge in the distant future, or that have fallen back into the hole in the distant past, all contribute to $\hat{\mathcal{Z}}$. However the $2M/r$ weighting implies that $\hat{\mathcal{Z}}$ varies continuously, even though charge is quantized.

### D. Relation of $\hat{\mathcal{Z}}$ to annihilation and creation operators

While the expression for $\hat{\mathcal{Z}}$ in terms of field operators is conceptually useful, for computations we need to write it in terms of the annihilation and creation operators. We write

$$D_{00}(r_\star) = \frac{r^2}{\sqrt{4\pi}} \sqrt{1 - \frac{2M}{r}} \int d\Omega \; : \psi^\dagger(x)\psi(x) :,$$
(43)

where $d\Omega = \sin\theta \, d\theta \, d\phi$ is the differential solid angle, and use the mode expansion of $\psi(x)$ given in Paper I Eq. (46). We expand out $\psi^\dagger(x)\psi(x)$ and use the orthogonality conditions given in Paper I Eq. (33):

$$\int d\Omega \, \Theta_{km}^{(\chi)\dagger}(\theta, \phi) \Theta_{k'm'}^{(\chi')}(\theta, \phi) = \delta_{\chi\chi'} \delta_{kk'} \delta_{mm'},$$
(44)

where $\chi \in \{F, G\}$. This reduces $D_{00}(r_\star)$ to a product of the radial fermion wave functions and the annihilation and creation operators:

$$D_{00}(r_\star) = \frac{1}{\sqrt{4\pi}} \int \frac{dh\,dh'}{(2\pi)^2} \frac{1}{\sqrt{4hh'}} \sum_{XX'} \sum_{km} \Big[ (F_{Xkh}^* F_{X'kh'} + G_{Xkh}^* G_{X'kh'}) \hat{b}_{Xkmh}^\dagger \hat{b}_{X'kmh'}$$
$$+ (F_{Xkh}^* G_{X'-kh'}^* + G_{Xkh}^* F_{X'-kh'}^*) \hat{b}_{Xkmh}^\dagger \hat{d}_{X'kmh'}^\dagger$$
$$+ (G_{X-kh} F_{X'kh'} + F_{X-kh} G_{X'kh'}) \hat{d}_{Xkmh} \hat{b}_{X'kmh'}$$
$$+ (G_{X-kh} G_{X'-kh'}^* + F_{X-kh} F_{X'-kh'}^*) \hat{d}_{Xkmh} \hat{d}_{X'kmh'}^\dagger \Big].$$
(45)

The expression for $\hat{\mathcal{Z}}$ is then given by

$$\hat{\mathcal{Z}} = -\int \frac{dh\,dh'}{(2\pi)^2} \sum_{XX'} \sum_{km} \left[ \mathcal{I}^{(1)}_{XX'k}(h,h')\hat{b}^{\dagger}_{Xkmh}\hat{b}_{X'kmh'} + \mathcal{I}^{(2)}_{XX'k}(h,h')\hat{b}^{\dagger}_{Xkmh}\hat{d}^{\dagger}_{X'kmh'} \right.$$
$$\left. + \mathcal{I}^{(2)*}_{X'Xk}(h',h)\hat{d}_{Xkmh}\hat{b}_{X'kmh'} - \mathcal{I}^{(1)*}_{X'-Xk}(h',h)\hat{d}^{\dagger}_{Xkmh}\hat{d}_{X'kmh'} \right] - \frac{\mathcal{Q}_{\star}}{e}, \tag{46}$$

where we swapped the unprimed and primed variables in the 4th term in order to be consistent with the rest of the expression. Also, we defined the following radial integrals:

$$\mathcal{I}^{(1)}_{XX'k}(h,h') \equiv \frac{1}{\sqrt{4hh'}} \int_{-\infty}^{\infty} dr_{\star} \frac{2M}{r} (F^{*}_{Xkh}F_{X'kh'} + G^{*}_{Xkh}G_{X'kh'}) \quad \text{and}$$
$$\mathcal{I}^{(2)}_{XX'k}(h,h') \equiv \frac{1}{\sqrt{4hh'}} \int_{-\infty}^{\infty} dr_{\star} \frac{2M}{r} (F^{*}_{Xkh}G^{*}_{X'-kh'} + G^{*}_{Xkh}F^{*}_{X'-kh'}), \tag{47}$$

and used the conjugation relations $F_{X,-k,-h} = G^{*}_{X,k,h}$ and $G_{X,-k,-h} = F^{*}_{X,k,h}$ (Paper I, §IVB) to derive

$$\mathcal{I}^{(2)*}_{X'Xk}(h',h) = \mathcal{I}^{(3)}_{XX'k}(h,h') \equiv \frac{1}{\sqrt{4hh'}} \int_{-\infty}^{\infty} dr_{\star} \frac{2M}{r} (G_{X-kh}F_{X'kh'} + F_{X-kh}G_{X'kh'})$$
$$\mathcal{I}^{(1)*}_{X'-Xk}(h',h) = \mathcal{I}^{(4)}_{XX'k}(h,h') \equiv \frac{1}{\sqrt{4hh'}} \int_{-\infty}^{\infty} dr_{\star} \frac{2M}{r} (G_{X'-kh'}G^{*}_{X-kh} + F_{X'-kh'}F^{*}_{X-kh}). \tag{48}$$

The definitions of Eq. (47) imply the hermiticity identities:

$$\mathcal{I}^{(1)}_{XX'k}(h,h') = \mathcal{I}^{(1)*}_{X'Xk}(h',h) \quad \text{and} \quad \mathcal{I}^{(2)}_{XX'k}(h,h') = \mathcal{I}^{(2)*}_{X'X,-k}(h',h). \tag{49}$$

Equation (46) has the familiar $\mathcal{I}^{(1)}$ terms that count a factor $-1$ or $+1$ for a given electron or positron mode, respectively. The integration over $\int dh\,dh'$ ensures that $\langle \hat{\mathcal{Z}} \rangle$ contains interference terms from different electron modes. But there are also the $\mathcal{I}^{(2)}$ terms, which involve interference between a positive-energy electron mode and a negative-energy mode (in the Dirac sea). Thus while $\hat{\mathcal{Z}}$ commutes with the total field charge $Q_{\mathrm{f}}$, it does *not* commute with either the number of electrons, which is $\int_{0}^{\infty} \sum_{Xkm} \hat{b}^{\dagger}_{X,k,m,h}\hat{b}_{X,k,m,h}\,dh/2\pi$, or the number of positrons , which is $\int_{0}^{\infty} \sum_{Xkm} \hat{b}^{\dagger}_{X,k,m,h}\hat{b}_{X,k,m,h}\,dh/2\pi$. This is a manifestation of the well-known effect in QED (even in flat spacetime) that the measurement of the charge *in a particular region, at a particular time* necessarily introduces electric fields and hence can result in pair production or pair annihilation.

## VI. THE HIERARCHY OF EXPECTATION VALUES FOR THE STOCHASTIC CHARGE CONTRIBUTION

We now turn to the hierarchy of expectation values that we need and their evolution equations.

### A. The charge-fermion-fermion correlations

We begin by defining the charge-fermion-fermion correlation functions, defined by

$$\langle \hat{\mathcal{Z}}\hat{b}^{\dagger}_{Xkmh}\hat{b}_{X'k'm'h'} \rangle = \Gamma^{\hat{\mathcal{Z}}b^{\dagger}b(k)}_{XX'}(h,h')\delta_{kk'}\delta_{mm'},$$
$$\langle \hat{\mathcal{Z}}\hat{b}_{Xkmh}\hat{d}_{X'k'm'h'} \rangle = \Gamma^{\hat{\mathcal{Z}}bd(k)}_{XX'}(h,h')\delta_{kk'}\delta_{mm'},$$
$$\langle \hat{\mathcal{Z}}\hat{d}^{\dagger}_{Xkmh}\hat{b}^{\dagger}_{X'k'm'h'} \rangle = \Gamma^{\hat{\mathcal{Z}}d^{\dagger}b^{\dagger}(k)}_{XX'}(h,h')\delta_{kk'}\delta_{mm'}, \quad \text{and}$$
$$\langle \hat{\mathcal{Z}}\hat{d}^{\dagger}_{Xkmh}\hat{d}_{X'k'm'h'} \rangle = -\Gamma^{\hat{\mathcal{Z}}b^{\dagger}b(-k)}_{X'Xk}(h,h')\delta_{kk'}\delta_{mm'}, \tag{50}$$

where we have used spherical symmetry to eliminate unnecessary $m$-dependences. Charge conjugation symmetry implies

$$\Gamma^{\hat{\mathcal{Z}}d^{\dagger}d(k)}_{XX'}(h,h') = -\Gamma^{\hat{\mathcal{Z}}b^{\dagger}b(-k)}_{XX'}(h,h'),$$
$$\Gamma^{\hat{\mathcal{Z}}bd(k)}_{XX'}(h,h') = \Gamma^{\hat{\mathcal{Z}}bd(-k)}_{X'X}(h',h), \quad \text{and}$$
$$\Gamma^{\hat{\mathcal{Z}}d^{\dagger}b^{\dagger}(k)}_{XX'}(h,h') = \Gamma^{\hat{\mathcal{Z}}d^{\dagger}b^{\dagger}(-k)}_{X'X}(h',h). \tag{51}$$

Note that expectation values with the fermion operators before $\hat{\mathcal{Z}}$ can be expressed as, e.g.,

$$\langle \mathcal{O}_1...\mathcal{O}_n \hat{\mathcal{Z}} \rangle = \langle \hat{\mathcal{Z}}\mathcal{O}_n^\dagger...\mathcal{O}_1^\dagger \rangle^*, \tag{52}$$

where $\mathcal{O}_1...\mathcal{O}_n$ are any operators, and so these are not independent.

## B. Evolution of the electron spectrum in terms of the $\Gamma$s

We wish to calculate the changes in the emitted electron spectra due to the $H_{\text{int},0\text{L}}$ operator. (The other contributions will be discussed in a future paper.) There will be a divergent term due to the long-range nature of $H_{\text{int},0\text{L}}$ that shall be represented by a $\langle \hat{\mathcal{Z}}^2 \rangle$ term. Using Ehrenfest's theorem, one may write

$$\frac{dN_-}{dtdh}\bigg|_{0\text{L}} = \frac{1}{2\pi}\sum_{km}\frac{d}{dt}\langle \hat{b}_{out,kmh}^\dagger \hat{b}_{out,kmh}\rangle\bigg|_{0\text{L}} = \frac{i}{2\pi}\sum_{km}\langle[H_{int,0\text{L}}, \hat{b}_{out,kmh}^\dagger \hat{b}_{out,kmh}]\rangle. \tag{53}$$

Using the 3-operator commutation identities, Eq. (A1), we write this as the product of various commutators:

$$\begin{aligned}
\frac{dN_-}{dtdh}\bigg|_{0\text{L}} &= \frac{1}{2\pi}\sum_{km}\frac{d}{dt}\langle \hat{b}_{\text{out,kmh}}^\dagger \hat{b}_{\text{out,kmh}}\rangle\bigg|_{0L} = \frac{ie^2}{32\pi^2 M}\sum_{km}\langle[\mathcal{Z}^2, \hat{b}_{\text{out,kmh}}^\dagger \hat{b}_{\text{out,kmh}}]\rangle \\
&= \frac{ie^2}{32\pi^2 M}\sum_{km}\langle\hat{\mathcal{Z}}[\mathcal{Z}, \hat{b}_{\text{out,kmh}}^\dagger \hat{b}_{\text{out,kmh}}] + [\hat{\mathcal{Z}}, \hat{b}_{\text{out,kmh}}^\dagger \hat{b}_{\text{out,kmh}}]\hat{\mathcal{Z}}\rangle \\
&= \frac{ie^2}{32\pi^2 M}\sum_{km}\langle\hat{\mathcal{Z}}\hat{b}_{\text{out,kmh}}^\dagger[\hat{\mathcal{Z}}, \hat{b}_{\text{out,kmh}}] + \hat{\mathcal{Z}}[\mathcal{Z}, \hat{b}_{\text{out,kmh}}^\dagger]\hat{b}_{\text{out,kmh}} + \hat{b}_{\text{out,kmh}}^\dagger[\hat{\mathcal{Z}}, \hat{b}_{\text{out,kmh}}]\hat{\mathcal{Z}} \\
&\qquad + [\hat{\mathcal{Z}}, \hat{b}_{\text{out,kmh}}^\dagger]\hat{b}_{\text{out,kmh}}\hat{\mathcal{Z}}\rangle,
\end{aligned} \tag{54}$$

which combines the first and fourth expectation value and the second and third expectation value, respectively. Splitting each expectation value as a separate term, so we have four expectation values we need to compute:

$$\begin{aligned}
\frac{dN_-}{dtdh}\bigg|_{0\text{L}} &= \frac{ie^2}{32\pi^2 M}\sum_{km}\Big\{\langle\hat{\mathcal{Z}}\hat{b}_{\text{out,kmh}}^\dagger[\hat{\mathcal{Z}}, \hat{b}_{\text{out,kmh}}]\rangle + \langle\hat{\mathcal{Z}}[\mathcal{Z}, \hat{b}_{\text{out,kmh}}^\dagger]\hat{b}_{\text{out,kmh}}\rangle + \langle\hat{b}_{\text{out,kmh}}^\dagger[\hat{\mathcal{Z}}, \hat{b}_{\text{out,kmh}}]\hat{\mathcal{Z}}\rangle \\
&\qquad + \langle[\hat{\mathcal{Z}}, \hat{b}_{\text{out,kmh}}^\dagger]\hat{b}_{\text{out,kmh}}\hat{\mathcal{Z}}\rangle\Big\},
\end{aligned} \tag{55}$$

which we will equate to eight terms composed of $\hat{\mathcal{Z}}$ and two fermion operators once we substitute Eq. (A1). However, one can prove the relations

$$\begin{aligned}
\langle[\hat{\mathcal{Z}}, \hat{b}_{\text{out,kmh}}^\dagger]\hat{b}_{\text{out,kmh}}\hat{\mathcal{Z}}\rangle &= -\langle\hat{\mathcal{Z}}\hat{b}_{\text{out,kmh}}^\dagger[\hat{\mathcal{Z}}, \hat{b}_{\text{out,kmh}}]\rangle^* \quad \text{and} \\
\langle\hat{b}_{\text{out,kmh}}^\dagger[\hat{\mathcal{Z}}, \hat{b}_{\text{out,kmh}}]\hat{\mathcal{Z}}\rangle &= -\langle\hat{\mathcal{Z}}[\mathcal{Z}, \hat{b}_{\text{out,kmh}}^\dagger]\hat{b}_{\text{out,kmh}}\rangle^*
\end{aligned} \tag{56}$$

making use of the fact that $\langle A^\dagger\rangle = \langle A\rangle^*$ for any operator $A$. This substitution allows one to combine the first and fourth terms and the second and third term, respectively, in Eq. (55), simplifying the emission rate to

$$\frac{dN_-}{dtdh}\bigg|_{0L} = -\frac{e^2}{16\pi^2 M}\sum_{km}\Im\Big\{\langle\hat{\mathcal{Z}}\hat{b}_{\text{out,kmh}}^\dagger[\hat{\mathcal{Z}}, \hat{b}_{\text{out,kmh}}]\rangle + \langle\hat{\mathcal{Z}}[\mathcal{Z}, \hat{b}_{\text{out,kmh}}^\dagger]\hat{b}_{\text{out,kmh}}\rangle\Big\}, \tag{57}$$

which is only nonzero when either expectation value contains an imaginary piece.

Substituting both Eqs. (46) and (50), we may express each of the two expectation values in Eq. (57) in terms of $\mathcal{I}^{(1)}$-type mode integrals and the charge-fermion-fermion correlation functions. These terms are listed below:

$$\begin{aligned}
\langle\hat{\mathcal{Z}}\hat{b}_{\text{out,kmh}}^\dagger[\hat{\mathcal{Z}}, \hat{b}_{\text{out,kmh}}]\rangle &= \int\frac{dh'}{2\pi}\sum_{X'}\Big\{\mathcal{I}_{\text{out,X'k}}^{(1)}(h,h')\langle\hat{\mathcal{Z}}\hat{b}_{\text{out,kmh}}^\dagger\hat{b}_{X'kmh'}\rangle + \mathcal{I}_{\text{out,X'k}}^{(2)}(h,h')\langle\hat{\mathcal{Z}}\hat{b}_{\text{out,kmh}}^\dagger\hat{d}_{X'k'm'h'}^\dagger\rangle\Big\} \\
&= \int\frac{dh'}{2\pi}\sum_{X'}\Big\{\mathcal{I}_{\text{out,X'k}}^{(1)}(h,h')\Gamma_{out,X'}^{\hat{\mathcal{Z}}\hat{b}^\dagger b(k)}(h,h') - \mathcal{I}_{\text{out,X'k}}^{(2)}(h,h')\Gamma_{\text{out,X'k}}^{\hat{\mathcal{Z}}\hat{d}^\dagger b^\dagger(-k)}(h,h')\Big\}
\end{aligned} \tag{58}$$

and

$$
\begin{aligned}
\langle \hat{\mathcal{Z}}[\mathcal{Z}, \hat{b}^\dagger_{\text{out,kmh}}] \hat{b}_{\text{out,kmh}} \rangle &= -\int \frac{dh'}{2\pi} \sum_{X'} \left\{ \mathcal{I}^{(1)\star}_{\text{out,X'k}}(h,h') \langle \hat{\mathcal{Z}} \hat{b}^\dagger_{X'kmh'} \hat{b}_{\text{out,kmh}} \rangle + \mathcal{I}^{(2)\star}_{\text{out,X'k}}(h,h') \langle \hat{\mathcal{Z}} \hat{d}_{X'kmh'} \hat{b}_{\text{out,kmh}} \rangle \right\} \\
&= -\int \frac{dh'}{2\pi} \sum_{X'} \left\{ \mathcal{I}^{(1)\star}_{\text{out,X'k}}(h,h') \Gamma^{\hat{\mathcal{Z}} b^\dagger b(k)}_{X',\text{out}}(h',h) - \mathcal{I}^{(2)\star}_{\text{out,X'k}}(h,h') \Gamma^{\hat{\mathcal{Z}} bd(k)}_{\text{out,X'}}(h,h') \right\}.
\end{aligned}
\tag{59}
$$

Finally, if we substitute Eqs. (58) and (59) into Eq. (57), while also obtaining a factor of $2j+1$ due to the integrand not containing any azimuthal dependence, one obtains

$$
\begin{aligned}
\frac{dN_-}{dt dh}\bigg|_{\text{0L}} = -\frac{e^2}{16\pi^2 M} \Im \int \frac{dh'}{2\pi} \sum_{X'k} (2j+1) \Big\{ &\mathcal{I}^{(1)}_{\text{out,X'k}}(h,h') \Gamma^{\hat{\mathcal{Z}} b^\dagger b(k)}_{\text{out,X'}}(h,h') - \mathcal{I}^{(2)}_{\text{out,X'k}}(h,h') \Gamma^{\hat{\mathcal{Z}} d^\dagger b^\dagger(k)}_{X',\text{out}}(h',h) \\
&+ \mathcal{I}^{(2)\star}_{\text{out,X'k}}(h,h') \Gamma^{\hat{\mathcal{Z}} bd(k)}_{\text{out,X'}}(h,h') - \mathcal{I}^{(1)\star}_{\text{out,X'k}}(h,h') \Gamma^{\hat{\mathcal{Z}} b^\dagger b(k)}_{X',\text{out}}(h',h) \Big\}.
\end{aligned}
\tag{60}
$$

which describes the emission rate due to the $H_{\text{int,0L}}$ operator in the out/down basis. Using the arguments discussed in Appendix D, we express Eq. (60) in the in/up basis by transforming both the $\mathcal{I}^{(1)}$-type integrals and the $\Gamma$'s both into the in/up basis. We obtain

$$
\begin{aligned}
\frac{dN_-}{dt dh}\bigg|_{\text{0L}} = -\frac{e^2}{16\pi^2 M} \Im \int \frac{dh'}{2\pi} \sum_{X'k} (2j+1) \bigg\{ &|T_{\frac{1}{2}kh}|^2 \Big[ \mathcal{I}^{(1)}_{\text{up,X'k}}(h,h') \Gamma^{\hat{\mathcal{Z}} b^\dagger b(k)}_{\text{up,X'}}(h,h') - \mathcal{I}^{(2)}_{\text{up,X'k}}(h,h') \Gamma^{\hat{\mathcal{Z}} d^\dagger b^\dagger(k)}_{X',\text{up}}(h',h) \\
&+ \mathcal{I}^{(2)\star}_{\text{up,X'k}}(h,h') \Gamma^{\hat{\mathcal{Z}} bd(k)}_{\text{up,X'}}(h,h') - \mathcal{I}^{(1)\star}_{\text{up,X'k}}(h,h') \Gamma^{\hat{\mathcal{Z}} b^\dagger b(k)}_{X',\text{up}}(h',h) \Big] \\
&+ |R_{\frac{1}{2}kh}|^2 \Big[ \mathcal{I}^{(1)}_{\text{in,X'k}}(h,h') \Gamma^{\hat{\mathcal{Z}} b^\dagger b(k)}_{\text{in,X'}}(h,h') - \mathcal{I}^{(2)}_{\text{in,X'k}}(h,h') \Gamma^{\hat{\mathcal{Z}} d^\dagger b^\dagger(k)}_{X',\text{in}}(h',h) \\
&+ \mathcal{I}^{(2)\star}_{\text{in,X'k}}(h,h') \Gamma^{\hat{\mathcal{Z}} bd(k)}_{\text{in,X'}}(h,h') - \mathcal{I}^{(1)\star}_{\text{in,X'k}}(h,h') \Gamma^{\hat{\mathcal{Z}} b^\dagger b(k)}_{X',\text{in}}(h',h) \Big] \\
&+ T_{\frac{1}{2}kh} R^\star_{\frac{1}{2}kh} \Big[ \mathcal{I}^{(1)}_{\text{up,X'k}}(h,h') \Gamma^{\hat{\mathcal{Z}} b^\dagger b(k)}_{\text{in,X'}}(h,h') - \mathcal{I}^{(2)}_{\text{up,X'k}}(h,h') \Gamma^{\hat{\mathcal{Z}} d^\dagger b^\dagger(k)}_{X',\text{in}}(h',h) \\
&+ \mathcal{I}^{(2)\star}_{\text{in,X'k}}(h,h') \Gamma^{\hat{\mathcal{Z}} bd(k)}_{\text{up,X'}}(h,h') - \mathcal{I}^{(1)\star}_{\text{in,X'k}}(h,h') \Gamma^{\hat{\mathcal{Z}} b^\dagger b(k)}_{X',\text{up}}(h',h) \Big] \\
&+ T^\star_{\frac{1}{2}kh} R_{\frac{1}{2}kh} \Big[ \mathcal{I}^{(1)}_{\text{in,X'k}}(h,h') \Gamma^{\hat{\mathcal{Z}} b^\dagger b(k)}_{\text{up,X'}}(h,h') - \mathcal{I}^{(2)}_{\text{in,X'k}}(h,h') \Gamma^{\hat{\mathcal{Z}} d^\dagger b^\dagger(k)}_{X',\text{up}}(h',h) \\
&+ \mathcal{I}^{(2)\star}_{\text{up,X'k}}(h,h') \Gamma^{\hat{\mathcal{Z}} bd(k)}_{\text{in,X'}}(h,h') - \mathcal{I}^{(1)\star}_{\text{up,X'k}}(h,h') \Gamma^{\hat{\mathcal{Z}} b^\dagger b(k)}_{X',\text{in}}(h',h) \Big] \bigg\},
\end{aligned}
\tag{61}
$$

which describes the predicted $\mathcal{O}(\alpha)$ spectra due to the long-range interaction as discussed in Sec. V.

### C.  Time Evolution of $\langle \hat{\mathcal{Z}}^2 \rangle$

The time evolution of the variance $\langle \hat{\mathcal{Z}}^2 \rangle$ is given by Ehrenfest's Theorem:

$$
\frac{d}{dt} \langle \hat{\mathcal{Z}}^2 \rangle = i \langle [H_0, \hat{\mathcal{Z}}^2] \rangle = -2\Im \langle \hat{\mathcal{Z}}[H_0, \hat{\mathcal{Z}}] \rangle,
\tag{62}
$$

where we note that $[H_{\Phi+\text{bdy}}, \hat{\mathcal{Z}}] = 0$, and the commutator $[H_{\text{int},A}, \hat{\mathcal{Z}}]$ must be formally 2 orders higher in $e$ than the $H_0$ contribution (since $H_{\text{int},A}$ contains a power of $e$, and then the resulting commutator has a correlation of a photon with the electron operators, which must therefore be of at least order $e$). Thus, only the free-field contributes towards the time evolution where the free-field Hamiltonian is given by

$$H_0 \equiv H_{\text{Dirac}} = \int \frac{dh}{2\pi} h \sum_{Xkm} \left( \hat{b}^\dagger_{Xkmh} \hat{b}_{Xkmh} + \hat{d}^\dagger_{Xkmh} \hat{d}_{Xkmh} \right). \tag{63}$$

From Eq. (46) and the commutation relations of the free-field Hamiltonian (Appendix A 2), one can prove

$$[H_0, \hat{\mathcal{Z}}] = -\int \frac{dh dh'}{(2\pi)^2} \sum_{XX'} \sum_{km} \left\{ (h - h') \mathcal{I}^{(1)}_{XX'k}(h, h') \hat{b}^\dagger_{Xkmh} \hat{b}_{X'kmh'} + (h + h') \mathcal{I}^{(2)}_{XX'k}(h, h') \hat{b}^\dagger_{Xkmh} \hat{d}^\dagger_{X'kmh'} \right.$$
$$\left. - (h + h') \mathcal{I}^{(2)\star}_{X'Xk}(h', h) \hat{d}_{Xkmh} \hat{b}_{X'kmh'} - (h - h') \mathcal{I}^{(1)\star}_{X'X(-k)}(h', h) \hat{d}^\dagger_{Xkmh} \hat{d}_{X'kmh'} \right\}. \tag{64}$$

Finally, we can compute the expectation value $\langle \hat{\mathcal{Z}}[H_0, \hat{\mathcal{Z}}] \rangle$ making use of our definitions of the charge-fermion-fermion expectation values as given in Eq. (50). Note that we are neglecting any additional Kronecker $\delta$ that appear due to redundancy as $k' = k$ and $m' = m$. We find

$$\langle \hat{\mathcal{Z}}[H_0, \hat{\mathcal{Z}}] \rangle = -\int \frac{dh dh'}{(2\pi)^2} \sum_{XX'} \sum_{km} \left\{ (h - h') \mathcal{I}^{(1)}_{XX'k}(h, h') \Gamma^{\hat{\mathcal{Z}} b^\dagger b(k)}_{XX'}(h, h') - (h + h') \mathcal{I}^{(2)}_{XX'k}(h, h') \Gamma^{\hat{\mathcal{Z}} d^\dagger b^\dagger(k)}_{X'X}(h', h) \right.$$
$$\left. + (h + h') \mathcal{I}^{(2)\star}_{X'Xk}(h', h) \Gamma^{\hat{\mathcal{Z}} bd(k)}_{X'X}(h', h) + (h - h') \mathcal{I}^{(1)\star}_{X'X(-k)}(h', h) \Gamma^{\hat{\mathcal{Z}} b^\dagger b(-k)}_{XX'}(h, h') \right\}. \tag{65}$$

Note the sign flips on the $\mathcal{I}^{(2)}$ terms due to reordering of fermion operators, and the sign flip on the last term due to the charge conjugation relation, Eq. (51). Therefore, the time evolution is

$$\frac{d}{dt} \langle \hat{\mathcal{Z}}^2 \rangle = 2 \int \frac{dh dh'}{(2\pi)^2} \sum_{XX'} \sum_{km} \Im \left\{ (h - h') \mathcal{I}^{(1)}_{XX'k}(h, h') \Gamma^{\hat{\mathcal{Z}} b^\dagger b(k)}_{XX'}(h, h') - (h + h') \mathcal{I}^{(2)}_{XX'k}(h, h') \Gamma^{\hat{\mathcal{Z}} d^\dagger b^\dagger(k)}_{X'X}(h', h) \right.$$
$$\left. + (h + h') \mathcal{I}^{(2)\star}_{X'Xk}(h', h) \Gamma^{\hat{\mathcal{Z}} bd(k)}_{X'X}(h', h) + (h - h') \mathcal{I}^{(1)\star}_{X'X(-k)}(h', h) \Gamma^{\hat{\mathcal{Z}} b^\dagger b(-k)}_{XX'}(h, h') \right\}. \tag{66}$$

If we use the hermiticity rule $\mathcal{I}^{(1)\star}_{X'X, -k}(h', h) = \mathcal{I}^{(1)}_{XX', -k}(h, h')$ and relabel $k \to -k$ to enable us to combine the last terms while also swapping $h \leftrightarrow h'$ and $X \leftrightarrow X'$ in the third term so that this may be combined with the second term, we finally obtain

$$\frac{d}{dt} \langle \hat{\mathcal{Z}}^2 \rangle = 2 \Im \int \frac{dh \, dh'}{(2\pi)^2} \sum_{XX'k} (2j + 1) \left\{ 2(h - h') \mathcal{I}^{(1)}_{XX'k}(h, h') \Gamma^{\hat{\mathcal{Z}} b^\dagger b(k)}_{XX'}(h, h') \right.$$
$$\left. - (h + h') \mathcal{I}^{(2)}_{XX'k}(h, h') [\Gamma^{\hat{\mathcal{Z}} d^\dagger b^\dagger(k)}_{X'X}(h', h) + \Gamma^{\hat{\mathcal{Z}} bd(k)*}_{XX'}(h, h')] \right\}. \tag{67}$$

We note that because $\mathcal{I}^{(1)}$ only has a simple pole at $h - h' = 0$ (see Appendix C), the weights $(h - h') \mathcal{I}^{(1)}_{XX'k}(h, h')$ and $(h + h') \mathcal{I}^{(2)}_{XX'k}(h, h')$ appearing in the above integral are finite everywhere in the range of integration. (We will find that there are singularities in the $\Gamma$s.)

### D. Time evolution of the $\Gamma$ correlations

In this section, we will derive the time evolution of every charge-fermion-fermion expectation value (the $\Gamma$'s) in the steady-state limit. This will prove necessary as a means to further simplify Eq. (67) and ultimately calculate $\langle \hat{\mathcal{Z}}^2 \rangle$, as well as being useful in the computation of the electron spectrum correction of Eq. (61), which we will present in a future paper.

### 1. Time evolution of $\langle \hat{\mathcal{Z}} \hat{b}_{Xkmh}^\dagger \hat{b}_{X'k'm'h'} \rangle$

The time evolution of $\langle \hat{\mathcal{Z}} \hat{b}_{Xkmh}^\dagger \hat{b}_{X'k'm'h'} \rangle$ may be written as

$$\frac{d}{dt} \langle \hat{\mathcal{Z}} \hat{b}_{Xkmh}^\dagger \hat{b}_{X'k'm'h'} \rangle = i \langle [H, \hat{\mathcal{Z}} \hat{b}_{Xkmh}^\dagger \hat{b}_{X'k'm'h'}] \rangle \tag{68}$$

where the Hamiltonian $H$ may be decomposed into $H = H_0 + H_{\text{int,0L}} + H_{\text{int,0S}+\ell>0} + H_{\text{int,A}}$ where $H_{\text{int,0L}}$ is the electrostatic long-range Hamiltonian, $H_{\text{int,0S}+\ell>0}$ is the short-range and $\ell > 0$ electrostatic Hamiltonian, and $H_{\text{int,A}}$ is the electrodynamic Hamiltonian and quantifies the interaction between photons and fermions. Our intended goal is to calculate the $\mathcal{O}(1)$ contribution from Eq. (68) which will have two terms: a free-field contribution coming from $H_0$ and a long-range contribution from $H_{\text{int,0L}}$. Since $H_{\text{int,0S}} \sim \mathcal{O}(e^2)$, we need the part of $\langle \hat{\mathcal{Z}}^2 \rangle$ that is $\mathcal{O}(e^{-2})$.

Performing the algebra, and substituting the definition of $H_{\text{int,0L}}$ given in Eq. (40), one obtains

$$\frac{d}{dt} \langle \hat{\mathcal{Z}} \hat{b}_{Xkmh}^\dagger \hat{b}_{X'k'm'h'} \rangle = i \langle \hat{\mathcal{Z}} [H_0, \hat{b}_{Xkmh}^\dagger \hat{b}_{X'k'm'h'}] \rangle + i \langle [H_0, \hat{\mathcal{Z}}] \hat{b}_{Xkmh}^\dagger \hat{b}_{X'k'm'h'} \rangle + \frac{ie^2}{16\pi M} \langle [\hat{\mathcal{Z}}^2, \hat{\mathcal{Z}} \hat{b}_{Xkmh}^\dagger \hat{b}_{X'k'm'h'}] \rangle, \tag{69}$$

where we used the well known commutator relation $[AB, C] = A[B, C] + [A, C]B$ to expand the free-field Hamiltonian. One can show

$$[H_0, \hat{b}_{Xkmh}^\dagger \hat{b}_{X'k'm'h'}] = (h - h') \hat{b}_{Xkmh}^\dagger \hat{b}_{X'k'm'h'}, \tag{70}$$

which allows one to prove

$$\left[ \frac{d}{dt} - i(h - h') \right] \langle \hat{\mathcal{Z}} \hat{b}_{Xkmh}^\dagger \hat{b}_{X'k'm'h'} \rangle = i \langle [H_0, \hat{\mathcal{Z}}] \hat{b}_{Xkmh}^\dagger \hat{b}_{X'k'm'h'} \rangle + \frac{ie^2}{16\pi M} \langle [\hat{\mathcal{Z}}^2, \hat{\mathcal{Z}} \hat{b}_{Xkmh}^\dagger \hat{b}_{X'k'm'h'}] \rangle. \tag{71}$$

Additionally, we may further expand the commutator for the interaction Hamiltonian such that

$$\langle [\hat{\mathcal{Z}}^2, \hat{\mathcal{Z}} \hat{b}_{Xkmh}^\dagger \hat{b}_{X'k'm'h'}] \rangle = \langle \hat{\mathcal{Z}}^2 [\hat{\mathcal{Z}}, \hat{b}_{Xkmh}^\dagger \hat{b}_{X'k'm'h'}] \rangle + \langle \hat{\mathcal{Z}} [\hat{\mathcal{Z}}, \hat{b}_{Xkmh}^\dagger \hat{b}_{X'k'm'h'}] \hat{\mathcal{Z}} \rangle \tag{72}$$

which may be further simplified by splitting these expectation values into a "connected" and "disconnected" piece. The trick is that disconnected piece will give us one copy of $\langle \hat{\mathcal{Z}}^2 \rangle$ which is $\mathcal{O}(e^{-2})$ while the connected piece may be neglected due to being higher order. Since we are purely interested in the $\mathcal{O}(1)$ contribution to $\langle \hat{\mathcal{Z}} \hat{b}_{Xkmh}^\dagger \hat{b}_{X'k'm'h'} \rangle$, one may write

$$\langle [\hat{\mathcal{Z}}^2, \hat{\mathcal{Z}} \hat{b}_{Xkmh}^\dagger \hat{b}_{X'k'm'h'}] \rangle = 2 \langle \hat{\mathcal{Z}}^2 \rangle \langle [\hat{\mathcal{Z}}, b_{Xkmh}^\dagger \hat{b}_{X'k'm'h'}] \rangle + \text{connected terms} \tag{73}$$

and which allows one to write Eq. (71) in the following form:

$$\left[ \frac{d}{dt} - i(h - h') \right] \langle \hat{\mathcal{Z}} \hat{b}_{Xkmh}^\dagger \hat{b}_{X'k'm'h'} \rangle = i \langle [H_0, \hat{\mathcal{Z}}] \hat{b}_{Xkmh}^\dagger \hat{b}_{X'k'm'h'} \rangle + \frac{ie^2 \langle \hat{\mathcal{Z}}^2 \rangle}{8\pi M} \langle [\hat{\mathcal{Z}}, b_{Xkmh}^\dagger \hat{b}_{X'k'm'h'}] \rangle. \tag{74}$$

Substituting Eq. (64), it can be easily shown that

$$\begin{aligned}
\langle [H_0, \hat{\mathcal{Z}}] \hat{b}_{Xkmh}^\dagger \hat{b}_{X'k'm'h'} \rangle = &- \int \frac{dh'' dh'''}{(2\pi)^2} \sum_{X''X'''} \sum_{k''m''} \Big[ (h'' - h''') \mathcal{I}_{X''X'''k''}^{(1)}(h'', h''') \\
&\times \langle \hat{b}_{X''k''m''h''}^\dagger \hat{b}_{X'''k''m''h'''} \hat{b}_{Xkmh}^\dagger \hat{b}_{X'k'm'h'} \rangle \\
&+ (h'' - h''') \mathcal{I}_{X''X'''k''}^{(1)\star}(h''', h'') \langle \hat{d}_{X''k''m''h'''}^\dagger \hat{d}_{X'''k''m''h''} \hat{b}_{Xkmh}^\dagger \hat{b}_{X'k'm'h'} \rangle \Big],
\end{aligned} \tag{75}$$

which describes the free-field contribution and will only be nonzero when $h'' \neq h'''$ because of the factor of $h'' - h'''$ in the numerator. The only contraction that provides a nonzero answer is when $\hat{b}_{X''k''m''h''}$ contracts with $\hat{b}_{Xkmh}^\dagger$ and $\hat{b}_{X'''k''m''h'''}^\dagger$ contracts with $\hat{b}_{X'k'm'h'}$ as this contraction doesn't necessarily imply $h'' = h'''$. Hence, one can obtain

$$\begin{aligned}
\langle [H_0, \hat{\mathcal{Z}}] \hat{b}_{Xkmh}^\dagger \hat{b}_{X'k'm'h'} \rangle &= (h - h') \sum_{X''X'''} \mathcal{I}_{X''X'''k}^{(1)}(h', h) g_{X''Xk}^{e^-}(h) f_{X''X'k}^{e^-}(h') \delta_{kk'} \delta_{mm'} \\
&= (h - h') \mathcal{I}_{X'Xk}^{(1)}(h', h) \Big[ 1 - f_{XXk}^{e^-}(h) \Big] f_{X'X'k}^{e^-}(h') \delta_{kk'} \delta_{mm'}
\end{aligned} \tag{76}$$

which can be thought of the contribution that the long-range Green's function, i.e. $G_{0L}(r_*, r'_*)$, has towards the self-energy of the electron. Note that we substituted the definition of $g^{e^-}_{XX'k}(h)$, i.e. $g^{e^-}_{XX'k}(h) = \delta_{XX'} - f^{e^-\star}_{XX'k}(h)$, while noting these phase space densities are real functions. Additionally, we collapsed the sum in the second line noting the unperturbed $f^{e^-\star}_{XX'k}(h)$ is diagonal.

Our next goal is to further reduce Eq. (74) to an equation as a function of the $\mathcal{I}$-type mode integrals, as defined in Eq. (47). Expanding $\langle [\hat{\mathcal{Z}}, b^{\dagger}_{Xkmh} \hat{b}_{X'k'm'h'}] \rangle$, one may show

$$\langle [\hat{\mathcal{Z}}, b^{\dagger}_{Xkmh} \hat{b}_{X'k'm'h'}] \rangle = \langle b^{\dagger}_{Xkmh}[\hat{\mathcal{Z}}, \hat{b}_{X'k'm'h'}] \rangle + \langle [\hat{\mathcal{Z}}, b^{\dagger}_{Xkmh}] \hat{b}_{X'k'm'h'} \rangle$$
$$= \langle b^{\dagger}_{Xkmh}[\hat{\mathcal{Z}}, \hat{b}_{X'k'm'h'}] \rangle - \langle b^{\dagger}_{X'k'm'h'}[\hat{\mathcal{Z}}, \hat{b}_{Xkmh}] \rangle^* \tag{77}$$

where we used the property that $\langle A^{\dagger} \rangle = \langle A \rangle^*$ for any operator $A$ to obtain the second line from the first line. Substituting in Eq. (A6), one may show that

$$\langle b^{\dagger}_{Xkmh}[\hat{\mathcal{Z}}, \hat{b}_{X'k'm'h'}] \rangle = \mathcal{I}^{(1)}_{X'Xk}(h', h) f^{e^-}_{XXkm}(h) \delta_{kk'} \delta_{mm'}, \tag{78}$$

where we make use of the expectation value of two fermion operators as defined in Silva et al. [37], i.e.

$$\langle \hat{b}^{\dagger}_{Xkmh} \hat{b}_{X'k'm'h'} \rangle = 2\pi f^{e^-}_{XX'km}(h) \delta(h - h') \delta_{XX'} \delta_{kk'} \delta_{mm'}. \tag{79}$$

.Therefore, the final expression for the commutator $\langle [\hat{\mathcal{Z}}, b^{\dagger}_{Xkmh} \hat{b}_{X'k'm'h'}] \rangle$ reduces to

$$\langle [\hat{\mathcal{Z}}, b^{\dagger}_{Xkmh} \hat{b}_{X'k'm'h'}] \rangle = \mathcal{I}^{(1)}_{X'Xk}(h', h) f^{e^-}_{XXkm}(h) \delta_{kk'} \delta_{mm'} - \mathcal{I}^{(1)*}_{XX'k}(h, h') f^{e^-\star}_{X'X'km}(h') \delta_{kk'} \delta_{mm'}$$
$$= \mathcal{I}^{(1)}_{X'Xk}(h', h) \left[ f^{e^-}_{XXkm}(h) - f^{e^-}_{X'X'km}(h') \right] \delta_{kk'} \delta_{mm'} \tag{80}$$

where we make use of the Hermiticity property of the $\mathcal{I}^{(1)}$ integrals, i.e. $\mathcal{I}^{(1)*}_{XX'k}(h, h') = \mathcal{I}^{(1)\star}_{X'Xk}(h', h)$, to go from the first line to the second line while also assuming the phase space densities are real functions.

Substituting each term and solving for the evolution of $\Gamma^{\hat{\mathcal{Z}} b^{\dagger} b(k)}_{XX'}(h, h')$, Eq. (74) may finally be written as

$$\left[ \frac{d}{dt} - i(h - h') \right] \Gamma^{\hat{\mathcal{Z}} b^{\dagger} b(k)}_{XX'}(h, h') = \left\{ i(h - h') \mathcal{I}^{(1)}_{X'Xk}(h', h) \left[ 1 - f^{e^-}_{XXk}(h) \right] f^{e^-}_{X'X'k}(h') \right.$$
$$\left. + \frac{i e^2 \langle \hat{\mathcal{Z}}^2 \rangle}{8\pi M} \mathcal{I}^{(1)}_{X'Xk}(h', h) \left[ f^{e^-}_{XXk}(h) - f^{e^-}_{X'X'k}(h') \right] \right\} \delta_{kk'} \delta_{mm'}. \tag{81}$$

Since we are interested in the steady-state limit, which would normally be done by setting $d/dt \to 0$, one should be able to derive an analytic expression for $\Gamma^{\hat{\mathcal{Z}} b^{\dagger} b}$. However, a singularity at $h = h'$ exists which causes $\Gamma^{\hat{\mathcal{Z}} b^{\dagger} b(k)}_{XX'}(h, h)$ to diverge. We resolve this by setting $d/dt \to +i\eta$ where $\eta$ is some small positive number. The $+$ sign corresponds to the correlation functions being zero (instead of infinite) in the distant past. This allows us to write

$$\Gamma^{\hat{\mathcal{Z}} b^{\dagger} b(k)}_{XX'}(h, h') = -\frac{(h - h') \underline{\mathcal{I}^{(1)}_{X'Xk}}(h', h)}{h - h' + i\eta} \left[ 1 - f^{e^-}_{XXk}(h) \right] f^{e^-}_{X'X'k}(h') - \frac{e^2 \langle \hat{\mathcal{Z}}^2 \rangle}{8\pi M} \frac{\underline{\mathcal{I}^{(1)}_{X'Xk}}(h', h)}{h - h' + i\eta} \left[ f^{e^-}_{XXk}(h) - f^{e^-}_{X'X'k}(h') \right] \tag{82}$$

where we underline the $\mathcal{I}^{(1)}$ to indicate that the singular behavior at $h = h'$ is removed due to the factor of $h - h'$ in the numerator. Succinctly, we may define

$$\mathcal{A}_{XX'k}(h, h') \equiv (h - h') \underline{\mathcal{I}^{(1)}_{XX'k}}(h, h') \tag{83}$$

to isolate the singular behavior of $\mathcal{I}^{(1)}$ when we take the $h = h'$ limit. Note that there is a minus sign when $h$ and $h'$ are switched: $(h - h') \underline{\mathcal{I}^{(1)}_{X'Xk}}(h', h) = -\mathcal{A}_{X'Xk}(h', h)$. The inverse expressions for $\mathcal{I}^{(1)}$ in terms of $\mathcal{A}$, including the correct pole displacements so that one does not divide by zero at $h = h'$, are derived in Appendix C (the key result is Eq. C10). With this substitution, we may write

$$\Gamma^{\hat{\mathcal{Z}} b^{\dagger} b(k)}_{XX'}(h, h') = \mathcal{A}_{X'Xk}(h', h) \left\{ \frac{\left[ 1 - f^{e^-}_{XXk}(h) \right] f^{e^-}_{X'X'k}(h')}{h - h' + i\eta} + \frac{e^2 \langle \hat{\mathcal{Z}}^2 \rangle}{8\pi M} \frac{f^{e^-}_{XXk}(h) - f^{e^-}_{X'X'k}(h')}{(h - h' + i\eta)(h - h' + i\epsilon)} \right\}$$
$$- \frac{e^2 \langle \hat{\mathcal{Z}}^2 \rangle}{8\pi M} \delta_{X,\text{up}} \delta_{X',\text{up}} \frac{2\pi \delta_{\epsilon}(h - h')}{h - h' + i\eta} [f^{e^-}_{\text{up,up},k}(h) - f^{e^-}_{\text{up,up},k}(h')], \tag{84}$$

which has a simple pole for the free-field contribution and second order pole for the $\langle \hat{\mathcal{Z}}^2 \rangle$ term. Our next goal is to understand physically what this term means. The first term corresponds to the occupation number of electrons with an extra factor of $1 - f$ as electrons are fermions. The second term represents the effects of the long-range interaction $H_{\text{int,0L}}$. Let us expand the above expression in the "in/up" basis in order to physically understand what terms represent:

$$
\begin{aligned}
\Gamma_{\text{up,up}}^{\hat{\mathcal{Z}}b^\dagger b(k)}(h,h') &= \mathcal{A}_{\text{up,up,k}}(h',h) \left\{ \frac{[1 - f_{\text{up,up,k}}^{e^-}(h)] f_{\text{up,up,k}}^{e^-}(h')}{h - h' + i\eta} + \frac{e^2 \langle \hat{\mathcal{Z}}^2 \rangle}{8\pi M} \frac{f_{\text{up,up,k}}^{e^-}(h) - f_{\text{up,up,k}}^{e^-}(h')}{(h - h' + i\eta)(h - h' + i\epsilon)} \right\} \\
&\quad - \frac{e^2 \langle \hat{\mathcal{Z}}^2 \rangle}{8\pi M} \frac{2\pi \delta_\epsilon(h - h')}{h - h' + i\eta} [f_{\text{up,up,k}}^{e^-}(h) - f_{\text{up,up,k}}^{e^-}(h')] \\
\Gamma_{\text{in,up}}^{\hat{\mathcal{Z}}b^\dagger b(k)}(h,h') &= \mathcal{A}_{\text{up,in,k}}(h',h) f_{\text{up,up,k}}^{e^-}(h') \left\{ \frac{1}{h - h' + i\eta} - \frac{e^2 \langle \hat{\mathcal{Z}}^2 \rangle}{8\pi M} \frac{1}{(h - h' + i\eta)(h - h' + i\epsilon)} \right\} \\
\Gamma_{\text{up,in}}^{\hat{\mathcal{Z}}b^\dagger b(k)}(h,h') &= -\frac{e^2 \langle \hat{\mathcal{Z}}^2 \rangle}{8\pi M} \mathcal{A}_{\text{up,in,k}}^*(h,h') f_{\text{up,up,k}}^{e^-}(h) \frac{1}{(h - h' + i\eta)(h - h' + i\epsilon)} \\
\Gamma_{\text{in,in}}^{\hat{\mathcal{Z}}b^\dagger b(k)}(h,h') &= 0,
\end{aligned}
$$

(85)

where we make use of the anti-hermiticity of $\mathcal{A}_{XX'k(h,h')}$, i.e. $\mathcal{A}_{XX'k(h,h')} = -\mathcal{A}_{X'Xk}^*(h',h)$, for the $\Gamma_{\text{up,in}}^{\hat{\mathcal{Z}}b^\dagger b(k)}$ term. Physically, the (up,up) may be thought of as analogous to a charge density of the surrounding plasma around the black hole, weighted by $2M/r$, as defined by $\mathcal{I}^{(1)}$ or $\mathcal{A}$. Since electrons are fermions, we have to include an additional factor of $1 - f$ when considering the expected number of electrons at a given energy. Additionally, the unperturbed fermion phase space density in the "in" state is zero, as electrons and positrons are only emitted near the horizon of the black hole, these is no charge density of fermions in the "in" state, and thus, we should expect $\Gamma_{\text{in,in}}^{\hat{\mathcal{Z}}b^\dagger b(k)}(h,h') = 0$. One interesting observation that may be made is that there is a coherence from the free-field for the (in,up) term, but not for the (up,in) term. Physically, this may be thought of that the outgoing electron wave function is some linear superposition of the "in" and "up" basis states as both states interfere with one another. An electron in the "up" state can affect the probability of finding an electron in the "in" state as the wave function may overlap when computing the mode integrals in $\mathcal{I}^{(1)}$ or $\mathcal{A}$. However, the dominant process producing the distribution of electrons and positrons near the black hole is the Hawking emission itself and will dominate. This would be the complete picture except we must also consider the effects of the long-range divergence from $H_{\text{int,0L}}$ as electrons far away may interact with an infinite number of electrons and positrons all the way to the horizon. Because of the long-range nature of this interaction, there is also an additional term related to the variance $\langle \hat{\mathcal{Z}}^2 \rangle$ that must be included in the covariance.

### 2. Time evolution of $\langle \hat{\mathcal{Z}} \hat{b}_{Xkmh} \hat{d}_{X'k'm'h'} \rangle$

In this section, we will derive the time evolution of $\langle \hat{\mathcal{Z}} \hat{b}_{Xkmh} \hat{d}_{X'k'm'h'} \rangle$ starting from first principles. Please refer to Sec. VI D 1 for the discussion of why only the the free-field Hamiltonion $H_0$ and the long range interaction Hamiltonian $H_{\text{int,0L}}$ contribute to lowest order. One can show the time evolution of $\langle \hat{\mathcal{Z}} \hat{b}_{Xkmh} \hat{d}_{X'k'm'h'} \rangle$ will have the following form such that

$$
\frac{d}{dt} \langle \hat{\mathcal{Z}} \hat{b}_{Xkmh} \hat{d}_{X'k'm'h'} \rangle = i \langle [H_0, \hat{\mathcal{Z}}] \hat{b}_{Xkmh} \hat{d}_{X'k'm'h'} \rangle + i \langle \hat{\mathcal{Z}} [H_0, \hat{b}_{Xkmh} \hat{d}_{X'k'm'h'}] \rangle + \frac{ie^2}{16\pi M} \langle [\hat{\mathcal{Z}}^2, \hat{\mathcal{Z}} \hat{b}_{Xkmh} \hat{d}_{X'k'm'h'}] \rangle \quad (86)
$$

where there is a free-field contribution from the first two terms and a contribution from $H_{\text{int,0L}}$ from the third term. Very conveniently, one can show

$$
[H_0, \hat{b}_{Xkmh} \hat{d}_{X'k'm'h'}] = -(h + h') \hat{b}_{Xkmh} \hat{d}_{X'k'm'h'} \quad (87)
$$

which allows us to write the time evolution given as

$$
\left[ \frac{d}{dt} + i(h + h') \right] \langle \hat{\mathcal{Z}} \hat{b}_{Xkmh} \hat{d}_{X'k'm'h'} \rangle = i \langle [H_0, \hat{\mathcal{Z}}] \hat{b}_{Xkmh} \hat{d}_{X'k'm'h'} \rangle + \frac{ie^2}{16\pi M} \langle [\hat{\mathcal{Z}}^2, \hat{\mathcal{Z}} \hat{b}_{Xkmh} \hat{d}_{X'k'm'h'}] \rangle. \quad (88)
$$

Additionally, we may further expand the commutator for the interaction Hamiltonian such that

$$
\begin{aligned}
\langle [\hat{\mathcal{Z}}^2, \hat{\mathcal{Z}} \hat{b}_{Xkmh} \hat{d}_{X'k'm'h'}] \rangle &= \langle \hat{\mathcal{Z}}^2 [\hat{\mathcal{Z}}, \hat{b}_{Xkmh} \hat{d}_{X'k'm'h'}] \rangle + \langle \hat{\mathcal{Z}} [\hat{\mathcal{Z}}, \hat{b}_{Xkmh} \hat{d}_{X'k'm'h'}] \hat{\mathcal{Z}} \rangle \\
&= 2 \langle \hat{\mathcal{Z}}^2 \rangle \langle [\hat{\mathcal{Z}}, b_{Xkmh} \hat{d}_{X'k'm'h'}] \rangle + \text{connected terms}
\end{aligned}
$$

(89)

where we expanded the commutator in the first line noting that $[AB, C] = A[B, C] + [A, C]B$ while in the second line we decomposed our expectation value into a "connected" and "disconnected" piece. The trick is that disconnected piece will give us one copy of $\langle \hat{\mathcal{Z}}^2 \rangle$ where the $\mathcal{O}(e^{-2})$ term while the connected terms will be higher order. Therefore,

$$\left[\frac{d}{dt} + i(h+h')\right]\langle \hat{\mathcal{Z}}\hat{b}_{Xkmh}\hat{d}_{X'k'm'h'}\rangle = i\langle [H_0, \hat{\mathcal{Z}}]\hat{b}_{Xkmh}\hat{d}_{X'k'm'h'}\rangle + \frac{ie^2\langle \hat{\mathcal{Z}}^2\rangle}{8\pi M}\langle [\hat{\mathcal{Z}}, \hat{b}_{Xkmh}\hat{d}_{X'k'm'h'}]\rangle. \tag{90}$$

which has both a free-field contribution and an interaction term. The free-field contribution gives

$$\langle [H_0, \hat{\mathcal{Z}}]\hat{b}_{Xkmh}\hat{d}_{X'k'm'h'}\rangle = (h+h')\mathcal{I}^{(2)}_{XX'k}(h,h')f^{e^+}_{XXk}(h)f^{e^-}_{X'X'k}(h')\delta_{kk'}\delta_{mm'} \tag{91}$$

where the only nonzero contribution comes from the contraction of the $\langle \hat{b}^\dagger \hat{d}^\dagger \hat{b}\hat{d}\rangle$ term when performing the algebra.

We now focus on the derivation of the interaction term. Substituting Eq. (A6) and Eq. (A8), one may show

$$\langle [\hat{\mathcal{Z}}, \hat{b}_{Xkmh}\hat{d}_{X'k'm'h'}]\rangle = \sum_{X''}\left\{\mathcal{I}^{(2)}_{XX''k}(h,h')f^{e^+}_{X''X'k}(h') - \mathcal{I}^{(2)}_{X''X'k}(h,h')g^{e^-}_{XX''k}(h)\right\}\delta_{kk'}\delta_{mm'} \tag{92}$$

which has a contribution from both the phase space density of electrons and the phase space density of positrons. Since we are purely interested in the lowest order contribution, the phase space densities will only be nonzero when $X = X''$ or $X' = X''$. Therefore, it is possible to contract the sum across $X''$ such that

$$\begin{aligned}\langle [\hat{\mathcal{Z}}, \hat{b}_{Xkmh}\hat{d}_{X'k'm'h'}]\rangle &= \mathcal{I}^{(2)}_{XX'k}(h,h')\left[f^{e^+}_{X'X'k}(h') - g^{e^-}_{XXk}(h)\right]\delta_{kk'}\delta_{mm'}\\ &= \mathcal{I}^{(2)}_{XX'k}(h,h')\left[f^{e^-}_{XXk}(h) + f^{e^+}_{X'X'k}(h') - 1\right]\delta_{kk'}\delta_{mm'}\end{aligned} \tag{93}$$

where in the second line we substituted the definition of $g^{e^-}_{XX'k}(h)$, i.e. $g^{e^-}_{XX'k}(h) = \delta_{XX'} - f^{e^-*}_{XX'k}(h)$, noting the phase space densities are real functions. Substituting this expression into Eq. (90), one can derive the time evolution of $\Gamma^{\hat{\mathcal{Z}}bd(k)}_{XX'}(h,h')$ such that

$$\begin{aligned}\left[\frac{d}{dt} + i(h+h')\right]\Gamma^{\hat{\mathcal{Z}}bd(k)}_{XX'}(h,h') &= i(h+h')\mathcal{I}^{(2)}_{XX'k}(h,h')f^{e^-}_{XXk}(h)f^{e^+}_{X'X'k}(h')\\ &\quad + \frac{ie^2\langle \hat{\mathcal{Z}}^2\rangle}{8\pi M}\mathcal{I}^{(2)}_{XX'k}(h,h')\left[f^{e^-}_{XXk}(h) + f^{e^+}_{X'X'k}(h') - 1\right],\end{aligned} \tag{94}$$

which has an $\mathcal{O}(1)$ contribution from both the free-field term and from $H_{\text{int},0L}$.

Unlike $\Gamma^{\hat{\mathcal{Z}}b^\dagger b(k)}_{XX'}(h,h')$, there are no singularities that need to consider in the steady-state limit since $h$ and $h'$ are defined to be positive, real variables, and hence,, we may neglect introducing a small parameter $d/dt \to +\eta$. Starting from Eq. (94) and assuming steady-state,

$$\Gamma^{\hat{\mathcal{Z}}bd(k)}_{XX'}(h,h') = \mathcal{I}^{(2)}_{XX'k}(h,h')f^{e^-}_{XXk}(h)f^{e^+}_{X'X'k}(h') + \frac{e^2\langle \hat{\mathcal{Z}}^2\rangle}{8\pi M}\frac{\mathcal{I}^{(2)}_{XX'k}(h,h')}{h+h'}\left[f^{e^-}_{XXk}(h) + f^{e^+}_{X'X'k}(h') - 1\right]. \tag{95}$$

### 3. Time evolution of $\langle \hat{\mathcal{Z}}\hat{d}^\dagger_{Xkmh}\hat{b}^\dagger_{X'k'm'h'}\rangle$

In this section, we will derive the time evolution of $\langle \hat{\mathcal{Z}}\hat{d}^\dagger_{Xkmh}\hat{b}^\dagger_{X'k'm'h'}\rangle$. We can perform a very similar derivation as what was shown for Sec. VI D 2 and will not perform every step. However, the reader should wade through the previous section if he or she wishes to understand the logic that is being applied.

The time evolution of $\langle \hat{\mathcal{Z}}\hat{d}^\dagger_{Xkmh}\hat{b}^\dagger_{X'k'm'h'}\rangle$ may be shown to have the following form:

$$\frac{d}{dt}\langle \hat{\mathcal{Z}}\hat{d}^\dagger_{Xkmh}\hat{b}^\dagger_{X'k'm'h'}\rangle = i\langle [H_0, \hat{\mathcal{Z}}]\hat{d}^\dagger_{Xkmh}\hat{b}^\dagger_{X'k'm'h'}\rangle + i\langle \hat{\mathcal{Z}}[H_0, \hat{d}^\dagger_{Xkmh}\hat{b}^\dagger_{X'k'm'h'}]\rangle + \frac{ie^2}{16\pi M}\langle [\hat{\mathcal{Z}}^2, \hat{\mathcal{Z}}\hat{d}^\dagger_{Xkmh}\hat{b}^\dagger_{X'k'm'h'}]\rangle \tag{96}$$

where one can show

$$[H_0, \hat{d}^\dagger_{Xkmh}\hat{b}^\dagger_{X'k'm'h'}] = (h+h')\hat{d}^\dagger_{Xkmh}\hat{b}^\dagger_{X'k'm'h'}. \tag{97}$$

Performing exactly the same arguments as before to derive the time evolution, one can show

$$\left[\frac{d}{dt} - i(h+h')\right]\langle \hat{\mathcal{Z}}\hat{d}^\dagger_{Xkmh}\hat{b}^\dagger_{X'k'm'h'}\rangle = i\langle [H_0, \hat{\mathcal{Z}}]\hat{d}^\dagger_{Xkmh}\hat{b}^\dagger_{X'k'm'h'}\rangle + \frac{ie^2\langle \hat{\mathcal{Z}}^2\rangle}{8\pi M}\langle [\hat{\mathcal{Z}}, \hat{d}^\dagger_{Xkmh}\hat{b}^\dagger_{X'k'm'h'}]\rangle. \tag{98}$$

The free-field contribution can be derived noting that the only nonzero term will be that originating from the $\langle \hat{b}\hat{d}\hat{d}^\dagger \hat{b}^\dagger \rangle$ expectation value when expanding $[\hat{H}_0, \hat{\mathcal{Z}}]$. Thus,

$$
\begin{aligned}
\langle [\hat{H}_0, \hat{\mathcal{Z}}] \hat{d}^\dagger_{Xkmh} \hat{b}^\dagger_{X'k'm'h'} \rangle &= -\sum_{X''X'''} \mathcal{I}^{(2)\star}_{X'''X''k}(h', h) g^{e^-}_{X'''X'k}(h') g^{e^+}_{X''Xk}(h) \delta_{kk'} \delta_{mm'} \\
&= -\mathcal{I}^{(2)\star}_{X'Xk}(h', h) \left[ 1 - f^{e^-}_{X'X'k}(h') \right] \left[ 1 - f^{e^+}_{XXk}(h) \right] \delta_{kk'} \delta_{mm'}
\end{aligned}
\tag{99}
$$

where we collapse the sum across $X''$ and $X'''$ noting that the only nonzero unperturbed density matrix components occur when $X = X''$ and $X' = X'''$. Additionally, we also substitute the definition of $g^{e^\pm}_{XX'k}(h, h') = \delta_{XX'} - f^{e^\pm\star}_{XX'k}(h, h')$.

The interacting term $\langle [\hat{\mathcal{Z}}, \hat{d}^\dagger_{Xkmh} \hat{b}^\dagger_{X'k'm'h'}] \rangle$ may be written in terms of our previously derived expression for $\langle [\hat{\mathcal{Z}}, \hat{b}_{Xkmh} \hat{d}_{X'k'm'h'}] \rangle$ such that

$$
\langle [\hat{\mathcal{Z}}, \hat{d}^\dagger_{Xkmh} \hat{b}^\dagger_{X'k'm'h'}] \rangle = -\langle [\hat{\mathcal{Z}}, \hat{b}_{X'k'm'h'} \hat{d}_{Xkmh}] \rangle^\star = -\mathcal{I}^{(2)\star}_{X'Xk}(h', h) \left[ f^{e^-}_{X'X'k}(h') + f^{e^+}_{XXk}(h) - 1 \right].
\tag{100}
$$

Substituting both the free-field term and the interaction piece into Eq. (98), one can derive the time evolution of $\Gamma^{\hat{\mathcal{Z}} \hat{d}^\dagger \hat{b}^\dagger (k)}_{XX'}(h, h')$:

$$
\begin{aligned}
\left[ \frac{d}{dt} - i(h + h') \right] \Gamma^{\hat{\mathcal{Z}} \hat{d}^\dagger \hat{b}^\dagger (k)}_{XX'}(h, h') = &-i(h + h') \mathcal{I}^{(2)\star}_{X'Xk}(h', h) \left[ 1 - f^{e^-}_{X'X'k}(h') \right] \left[ 1 - f^{e^+}_{XXk}(h) \right] \\
&- \frac{ie^2 \langle \hat{\mathcal{Z}}^2 \rangle}{8\pi M} \mathcal{I}^{(2)\star}_{X'Xk}(h', h) \left[ f^{e^-}_{X'X'k}(h') + f^{e^+}_{XXk}(h) - 1 \right],
\end{aligned}
\tag{101}
$$

which has a $\mathcal{O}(1)$ contribution from both the free-field term and from $H_{\text{int,0L}}$.

Similar to pair annihilation, there are no singularities in the steady-state limit when $h = h'$. In particular, we do not need to include a pole displacement ($\eta$) as we did for $\Gamma^{\hat{\mathcal{Z}} \hat{b}^\dagger \hat{b}(k)}_{XX'}(h, h')$. Thus,

$$
\Gamma^{\hat{\mathcal{Z}} \hat{d}^\dagger \hat{b}^\dagger (k)}_{XX'}(h, h') = \mathcal{I}^{(2)\star}_{X'Xk}(h', h) \left[ 1 - f^{e^+}_{XXk}(h) \right] \left[ 1 - f^{e^-}_{X'X'k}(h') \right] + \frac{e^2 \langle \hat{\mathcal{Z}}^2 \rangle}{8\pi M} \frac{\mathcal{I}^{(2)\star}_{X'Xk}(h', h)}{h + h'} \left[ f^{e^+}_{XXk}(h) + f^{e^-}_{X'X'k}(h') - 1 \right].
\tag{102}
$$

### E. Closure of the evolution equations for $\langle \hat{\mathcal{Z}}^2 \rangle$

The next step is to determine how the variance of the charge $\langle Z^2 \rangle$ evolves with time, and in particular what its equilibrium value is. The key is to substitute the expressions from §VI D (Eqs. (84), Eq. (95), and Eq. (102)) for the $\Gamma$s into Eq. (67) for $\langle \hat{\mathcal{Z}}^2 \rangle$, and then resolve the various singularities. Since the $\Gamma$s each contain a "free-field" term with no dependence on $\langle \hat{\mathcal{Z}}^2 \rangle$, and a term proportional to $\langle \hat{\mathcal{Z}}^2 \rangle$, Eq. (67) will reduce to the form

$$
\frac{d}{dt} \langle Z^2 \rangle = C_0 + C_1 \langle Z^2 \rangle,
\tag{103}
$$

where the free-field contributions have been packaged into $C_0$ and all the contributions containing $\langle Z^2 \rangle$ are in the $C_1$ term.

The trickiest issue is navigating the singularities when $h \approx h'$: as one can see from, e.g., Eq. (84), there are singularities at $h' - h = +i\epsilon$, $-i\epsilon$, and $+i\eta$. The $\epsilon$-singularities come from the fact that the Coulomb interaction continues all the way down to the horizon (so to $r_\star \to -\infty$), and the $\eta$-singularities come from the fact that the interaction has been "on" for a formally infinite duration of time in the past (since $t = -\infty$).

We would like to take both $\eta$ and $\epsilon$ to zero, but as often happens with double singularities, the order matters. Physically, $\epsilon^{-1}$ represents the range of the Coulomb interaction down toward the horizon, i.e., the interaction cuts off at a location $r_\star \sim -\epsilon^{-1}$; whereas $\eta^{-1}$ represents the time scale over which the correlations have been building up. So then there are two choices for how to take these to zero, which correspond to different choices for how the fermions near the horizon experience the potential $-e\hat{\mathcal{Z}}/8\pi M$:

1. We take $\epsilon \to 0$ first, so $\epsilon \ll \eta$. In this case, over the timescale $\eta^{-1}$, all fermions reaching the $r_\star \sim 0$ region have been exposed to the full potential $-e\hat{\mathcal{Z}}/8\pi M$ over their whole history. As $\hat{\mathcal{Z}}$ changes, this potential moves up and down, and as it does so the energy (with respect to $\infty$) of any charged particle near the horizon is raised or lowered.

2. We take $\eta \to 0$ first, so $\eta \ll \epsilon$. In this case, over the timescale $\eta^{-1}$, most fermions have started off in a region of zero potential, and then in the last time interval $\Delta t \sim \epsilon^{-1}$ before they reach $r_\star \sim 0$, they adiabatically descend or ascend the potential "ramp" from 0 to $-e\hat{\mathcal{Z}}/8\pi M$.

Of these choices #1 is the physical one; the "ramp" in #2 corresponds to an electric field from the cutoff, and is clearly unphysical. Therefore, we take $\epsilon \to 0$ first in what follows. We start from Eq. (67), working through the contributions from $\Gamma_{XX'}^{\hat{\mathcal{Z}}b^\dagger b(k)}(h, h')$, $\Gamma_{XX'}^{\hat{\mathcal{Z}}bd(k)}(h, h')$, and $\Gamma_{XX'}^{\hat{\mathcal{Z}}d^\dagger b^\dagger (k)}(h, h')$, respectively. Since the evolution of $\langle \hat{\mathcal{Z}}^2 \rangle$ will be rather long and complicated, we split the calculation into separate sections.

### 1. $\Gamma_{XX'}^{\hat{\mathcal{Z}}b^\dagger b(k)}(h, h')$'s contribution towards $\langle \hat{\mathcal{Z}}^2 \rangle$

In this section, we will derive the terms that contribute towards the time evolution of $\langle \hat{\mathcal{Z}}^2 \rangle$ from $\Gamma_{XX'}^{\hat{\mathcal{Z}}b^\dagger b(k)}(h, h')$. We start from the relevant term of Eq. (67),

$$\frac{d}{dt}\langle \hat{\mathcal{Z}}^2 \rangle \Big|_{b^\dagger b} = 2\Im \int \frac{dh\, dh'}{(2\pi)^2} \sum_{XX'k} (2j+1) 2(h-h') \underline{\mathcal{I}}_{XX'k}^{(1)}(h, h') \Gamma_{XX'}^{\hat{\mathcal{Z}}b^\dagger b(k)}(h, h'); \tag{104}$$

plugging in Eq. (84), we find that

$$\frac{d}{dt}\langle \hat{\mathcal{Z}}^2 \rangle \Big|_{\hat{\mathcal{Z}}b^\dagger b} = C_0^{\hat{\mathcal{Z}}b^\dagger b} + C_1^{\hat{\mathcal{Z}}b^\dagger b} \langle \hat{\mathcal{Z}}^2 \rangle, \tag{105}$$

where

$$C_0^{\hat{\mathcal{Z}}b^\dagger b} = -4\Im \int \frac{dh\, dh'}{(2\pi)^2} \sum_{XX'k} (2j+1) \left\{ \frac{|\mathcal{A}_{XX'k}(h, h')|^2}{h - h' + i\eta} \left[ 1 - f_{XXk}^{e^-}(h) \right] f_{X'X'k}^{e^-}(h') \right\} \tag{106}$$

and

$$C_1^{\hat{\mathcal{Z}}b^\dagger b} = \frac{e^2}{2\pi M} \Im \int \frac{dh\, dh'}{(2\pi)^2} \sum_{XX'k} (2j+1) \left\{ \frac{\mathcal{A}_{XX'k}(h, h') \mathcal{I}_{XX'k}^{(1)\star}(h, h')}{h - h' + i\eta} \left[ f_{XXk}^{e^-}(h) - f_{X'X'k}^{e^-}(h') \right] \right\}$$
$$- \frac{e^2}{2\pi M} \Im \int \frac{dh\, dh'}{(2\pi)^2} \sum_k (2j+1) \mathcal{A}_{\text{up,up},k}(h, h') \frac{2\pi \delta_\epsilon(h - h')}{h - h' + i\eta} [f_{\text{up,up},k}^{e^-}(h) - f_{\text{up,up},k}^{e^-}(h')]. \tag{107}$$

We made use of the fact that $\mathcal{I}_{X'Xk}^{(1)}(h', h) = \mathcal{I}_{XX'k}^{(1)\star}(h, h')$ in both expressions. Notice that there is only one singularity for $C_0^{\hat{\mathcal{Z}}b^\dagger b}$ while $C_1^{\hat{\mathcal{Z}}b^\dagger b}$ has two poles that we need to consider. It was previously mentioned what the physical interpretation of these imaginary components represent the length scale of interactions, $\sim \epsilon^{-1}$, and the timescale of correlations in the approximate value, $\sim \eta^{-1}$. We showed that $\epsilon \to 0$ where $\epsilon \ll \eta$ is the physical case and will be the only case we shall consider. We can solve for both coefficients by performing contour integration noting that since we are interested in the imaginary component of these integrals, both terms will only be nonzero at the poles, as the numerator in both functions are real. Therefore, one can rewrite $C_0^{\hat{\mathcal{Z}}b^\dagger b}$ as

$$C_0^{\hat{\mathcal{Z}}b^\dagger b} = -4\Im \int \frac{dh\, dh'}{(2\pi)^2} \sum_{XX'k} (2j+1) |\mathcal{A}_{XX'k}(h, h')|^2 \left[ 1 - f_{XXk}^{e^-}(h) \right] f_{X'X'k}^{e^-}(h') \left\{ \frac{1}{h - h' + i\eta} \right\}. \tag{108}$$

The integrand is real as $\eta \to 0$ except for the singularity at $h = h'$; using the single pole displacement identity, Eq. (E2), we find

$$C_0^{\hat{\mathcal{Z}}b^\dagger b} = 2 \int \frac{dh}{2\pi} \sum_{XX'k} (2j+1) |\mathcal{A}_{XX'k}(h, h)|^2 \left[ 1 - f_{XXk}^{e^-}(h) \right] f_{X'X'k}^{e^-}(h). \tag{109}$$

We may now expand the sum across $X$ and $X'$ noting the only nonzero unperturbed phase space densities occur when $X = $ up and/or $X' = $ up. Therefore,

$$C_0^{\hat{\mathcal{Z}}b^\dagger b} = 2 \int \frac{dh}{2\pi} \sum_k (2j+1) \left\{ |\mathcal{A}_{\text{up,up},k}(h, h)|^2 \left[ 1 - f_{\text{up,up},k}^{e^-}(h) \right] f_{\text{up,up},k}^{e^-}(h) + |\mathcal{A}_{\text{in,up},k}(h, h)|^2 f_{\text{up,up},k}^{e^-}(h) \right\}. \tag{110}$$

We now wish to derive $C_1^{\hat{Z}b^{\dagger}b}$. We can first expand $C_1^{\hat{Z}b^{\dagger}b}$ to have the following form:

$$
\begin{aligned}
C_1^{\hat{Z}b^{\dagger}b} &= \frac{e^2}{2\pi M}\Im \int \frac{dh\,dh'}{(2\pi)^2} \sum_{XX'k} (2j+1)|\mathcal{A}_{XX'k}(h,h')|^2 \left[f_{XXk}^{e^-}(h) - f_{X'X'k}^{e^-}(h')\right] \frac{1}{(h-h'+i\eta)(h-h'+i\epsilon)} \\
&\quad + \frac{e^2}{2\pi M}\Im \int \frac{dh\,dh'}{(2\pi)^2} \sum_k (2j+1)\mathcal{A}_{\mathrm{up,up},k}(h,h') \frac{2\pi\delta_\epsilon(h-h')}{h-h'-i\eta} [f_{\mathrm{up,up},k}^{e^-}(h) - f_{\mathrm{up,up},k}^{e^-}(h')] \\
&\quad - \frac{e^2}{2\pi M}\Im \int \frac{dh\,dh'}{(2\pi)^2} \sum_k (2j+1)\mathcal{A}_{\mathrm{up,up},k}(h,h') \frac{2\pi\delta_\epsilon(h-h')}{h-h'+i\eta} [f_{\mathrm{up,up},k}^{e^-}(h) - f_{\mathrm{up,up},k}^{e^-}(h')],
\end{aligned} \tag{111}
$$

where the second term arises from writing the $\mathcal{I}^{(1)}$-integral in Eq. (107) in terms of $\mathcal{A}$. The second two terms cancel out, since using the rule $\mathcal{A}_{\mathrm{up,up},k}(h,h') = -\mathcal{A}_{\mathrm{up,up},k}^*(h',h)$, the imaginary part of the integrand of the second term at $(h,h')$ is equal to the imaginary part of the integrand of the third term at $(h',h)$. Using the double pole integral identity, Eq. (E3), we find

$$
\begin{aligned}
C_1^{\hat{Z}b^{\dagger}b} &= -\frac{e^2}{4\pi M} \int \frac{dh}{2\pi} \sum_{XX'k} (2j+1)\Bigg\{ \left(\frac{d}{dh'}|\mathcal{A}_{XX'k}(h,h')|^2\right) [f_{XXk}^{e^-}(h) - f_{X'X'k}^{e^-}(h')] \\
&\quad - |\mathcal{A}_{XX'k}(h,h')|^2 \frac{df_{X'X'k}^{e^-}(h')}{dh'} \Bigg\}\bigg|_{h'=h},
\end{aligned} \tag{112}
$$

where the first term represents the derivative of $(h-h')^2|\underline{\mathcal{I}}_{XX'k}^{(1)}(h,h')|^2$ with respect to $h'$ evaluated at $h'=h$ while the second term represents the derivative of the phase space density with respect to $h'$ evaluated at $h'=h$. We can see that the first term is only nonzero when $X \neq X'$ due to the phase space densities canceling. Therefore, this term only contributes when $(X,X') = (\mathrm{in,up})$ or $(X,X') = (\mathrm{up,in})$. The second term will only contribute when $X' = \mathrm{up}$ since to lowest order the phase space density is zero for fermions in the "in" basis. Therefore, we can expand out the sum across $X$ and $X'$ to obtain

$$
\begin{aligned}
C_1^{\hat{Z}b^{\dagger}b} &= \frac{e^2}{4\pi M} \int \frac{dh}{2\pi} \sum_k (2j+1)\Bigg\{ -\left(\frac{d}{dh'}|\mathcal{A}_{\mathrm{up,in},k}(h,h')|^2 - \frac{d}{dh'}|\mathcal{A}_{\mathrm{in,up},k}(h,h')|^2\right)\bigg|_{h'=h} f_{\mathrm{up,up,k}}^{e^-}(h) \\
&\quad + \left(|\mathcal{A}_{\mathrm{up,up},k}(h,h)|^2 + |\mathcal{A}_{\mathrm{in,up},k}(h,h)|^2\right) \left(\frac{df_{\mathrm{up,up},k}^{e^-}(h')}{dh'}\right)\bigg|_{h'=h} \Bigg\},
\end{aligned} \tag{113}
$$

where we can interpret the terms that are derivatives of $\mathcal{A}$ as describing how the transmission and reflection probabilities change with respect to energy of the fermions while terms that are derivatives of the phase space describe how the population of fermions change with respect to energy.

### 2. $\Gamma_{XX'}^{\hat{Z}bd(k)}(h,h')$'s contribution towards $\langle \hat{Z}^2 \rangle$

As seen in Eq. (95), there is no singularity in $\Gamma_{XX'}^{\hat{Z}bd(k)}(h,h')$. Additionally, there will be no singularity in $\mathcal{I}_{XX'k}^{(2)}(h,h')$ as shown in Appendix C. The contribution that $\Gamma_{XX'}^{\hat{Z}bd(k)}(h,h')$ has towards the time evolution of $\langle \hat{Z}^2 \rangle$ is given by

$$
\frac{d}{dt}\langle \hat{Z}^2 \rangle\bigg|_{\hat{Z}bd} = -2\Im \int \frac{dh\,dh'}{(2\pi)^2} \sum_{XX'k} (2j+1)(h+h')\mathcal{I}_{XX'k}^{(2)}(h,h')\Gamma_{XX'}^{\hat{Z}bd(k)\star}(h,h') \tag{114}
$$

Substituting our definition of $\Gamma_{XX'}^{\hat{Z}bd(k)}(h,h')$, one can show

$$
\begin{aligned}
\frac{d}{dt}\langle \hat{Z}^2 \rangle\bigg|_{\hat{Z}bd} &= -2\Im \int \frac{dh\,dh'}{(2\pi)^2} \sum_{XX'k} (2j+1)\Bigg\{ (h+h')|\mathcal{I}_{XX'k}^{(2)}(h,h')|^2 f_{XXk}^{e^-}(h) f_{X'X'k}^{e^+}(h') \\
&\quad + \frac{e^2\langle \hat{Z}^2 \rangle}{8\pi M}|\mathcal{I}_{XX'k}^{(2)}(h,h')|^2 \left[f_{XXk}^{e^-}(h) + f_{X'X'k}^{e^+}(h') - 1\right] \Bigg\} = 0,
\end{aligned} \tag{115}
$$

where we only care about the imaginary part of the integral. However, the integral is real for all possible values of $h$ and $h'$ unlike what has been shown in the $\Gamma_{XX'}^{\hat{Z}b^{\dagger}b(k)}(h,h')$ as singularities exist that need to be considered when $h = h'$. Therefore, we argue that $\Gamma_{XX'}^{\hat{Z}bd(k)}(h,h')$ will not contribute towards the time evolution of $\langle \hat{Z}^2 \rangle$.

### 3. $\Gamma_{XX'}^{\hat{\mathcal{Z}}d^\dagger b^\dagger(k)}(h,h')$'s contribution towards $\langle \hat{\mathcal{Z}}^2 \rangle$

This case similarly has no singularity. The contribution that $\Gamma_{XX'}^{\hat{\mathcal{Z}}bd(k)}(h,h')$ has towards the time evolution of $\langle \hat{\mathcal{Z}}^2 \rangle$ is given by

$$\frac{d}{dt}\langle \hat{\mathcal{Z}}^2 \rangle \Big|_{\hat{\mathcal{Z}}d^\dagger b^\dagger} = -2\Im \int \frac{dh\,dh'}{(2\pi)^2} \sum_{XX'k} (2j+1)(h+h')\mathcal{I}_{XX'k}^{(2)}(h,h')\Gamma_{X'X}^{\hat{\mathcal{Z}}d^\dagger b^\dagger(k)}(h',h) \tag{116}$$

Substituting our definition of $\Gamma_{XX'}^{\hat{\mathcal{Z}}bd(k)}(h,h')$, one can show

$$\begin{aligned}
\frac{d}{dt}\langle \hat{\mathcal{Z}}^2 \rangle \Big|_{\hat{\mathcal{Z}}d^\dagger b^\dagger} = -2\Im \int \frac{dh\,dh'}{(2\pi)^2} \sum_{XX'k} (2j+1) &\Bigg\{ (h+h')|\mathcal{I}_{XX'k}^{(2)}(h,h')|^2 \left[1 - f_{X'X'k}^{e^-}(h')\right]\left[1 - f_{XXk}^{e^+}(h)\right] \\
&+ \frac{e^2\langle \hat{\mathcal{Z}}^2 \rangle}{8\pi M}|\mathcal{I}_{XX'k}^{(2)}(h,h')|^2 \left[f_{XXk}^{e^-}(h) + f_{X'X'k}^{e^+}(h') - 1\right] \Bigg\} = 0
\end{aligned} \tag{117}$$

where we only care about the imaginary part of the integral. However, the integral is real for all possible values of $h$ and $h'$ unlike what has been shown in the $\Gamma_{XX'}^{\hat{\mathcal{Z}}b^\dagger b(k)}(h,h')$ since there are singularities that need to be considered at $h = h'$. Therefore, we argue that $\Gamma_{XX'}^{\hat{\mathcal{Z}}d^\dagger b^\dagger(k)}(h,h')$ will not contribute towards the time evolution of $\langle \hat{\mathcal{Z}}^2 \rangle$.

### 4. Combining all the pieces together

As was shown above, we proved only $\Gamma_{XX'}^{\hat{\mathcal{Z}}b^\dagger b(k)}(h,h')$ contributes towards the time evolution of $\langle \hat{\mathcal{Z}}^2 \rangle$. Assuming steady-state,

$$\langle \hat{\mathcal{Z}}^2 \rangle = -\frac{C_0^{\hat{\mathcal{Z}}b^\dagger b}}{C_1^{\hat{\mathcal{Z}}b^\dagger b}} \tag{118}$$

which after substituting our derived form of $C_0^{\hat{\mathcal{Z}}b^\dagger b}$ and $C_1^{\hat{\mathcal{Z}}b^\dagger b}$ allows one to obtain

$$\langle \hat{\mathcal{Z}}^2 \rangle = -\frac{8\pi M}{e^2}\frac{1}{\Xi}\int \frac{dh}{2\pi}\sum_k (2j+1)\left\{ |\mathcal{A}_{\mathrm{up,up,k}}(h,h)|^2\left[1 - f_{\mathrm{up,up,k}}^{e^-}(h)\right]f_{\mathrm{up,up,k}}^{e^-}(h) + |\mathcal{A}_{\mathrm{in,up,k}}(h,h)|^2 f_{\mathrm{up,up,k}}^{e^-}(h)\right\} \tag{119}$$

where

$$\begin{aligned}
\Xi = \int \frac{dh}{2\pi}\sum_k (2j+1)\Bigg\{ &-\left(\frac{d}{dh'}|\mathcal{A}_{\mathrm{up,in,k}}(h,h')|^2 - \frac{d}{dh'}|\mathcal{A}_{\mathrm{in,up,k}}(h,h')|^2\right)\Bigg|_{h'=h} f_{\mathrm{up,up,k}}^{e^-}(h) \\
&+ \left(|\mathcal{A}_{\mathrm{up,up,k}}(h,h)|^2 + |\mathcal{A}_{\mathrm{in,up,k}}(h,h)|^2\right)\left(\frac{df_{\mathrm{up,up,k}}^{e^-}(h')}{dh'}\right)\Bigg|_{h'=h}\Bigg\},
\end{aligned} \tag{120}$$

which explicitly is $\mathcal{O}(e^{-2})$ as expected. Our next goal is to calculate these $\mathcal{A}$-type integrals and reduce the above expression to solely an integral across phase space densities and transmission and reflection coefficients. Since we have no explicit dependence on $\mathcal{A}_{\mathrm{in,in},k}$ in the integrand, we will neglect its derivation despite being able to perform a very similar derivation. Let us begin. First, one may show

$$\mathcal{A}_{\mathrm{up,up,k}}(h,h') = i\left[1 - R_{\frac{1}{2}kh}R_{\frac{1}{2}kh'}^* e^{2i\left(\arg T_{\frac{1}{2}kh} - \arg T_{\frac{1}{2}kh}\right)}\right] + (h-h')\Lambda_{\mathrm{up,up,k}}(h,h') \tag{121}$$

where we neglected the $\delta(h-h')$ term due to the factor of $h-h'$ causing the term to be zero always. Additionally, we define $\Lambda_{XX'k}(h,h')$ to correspond with the finite part of $\mathcal{I}_{XX'k}^{(1)}(h,h')$ as defined in Appendix C. In the $h = h'$ limit, one may show

$$\mathcal{A}_{\mathrm{up,up,}k}(h,h) = i|T_{\frac{1}{2}kh}|^2 \tag{122}$$

which corresponds with the electron transmission probability. The cross terms, which corresponds to the interference between the transmitted and reflected electron wavefunctions, are given by

$$\mathcal{A}_{\text{in,up},k}(h, h') = iT^*_{\frac{1}{2}kh}R^*_{\frac{1}{2}kh'}e^{2i\arg T_{\frac{1}{2}kh'}} + (h - h')\Lambda_{\text{in,up},k}(h, h') \tag{123}$$

and

$$\mathcal{A}_{\text{up,in},k}(h, h') = iR_{\frac{1}{2}kh}T_{\frac{1}{2}kh'}e^{-2i\arg T_{\frac{1}{2}kh}} + (h - h')\Lambda_{\text{up,in},k}(h, h'). \tag{124}$$

The derivatives of the absolute value of these quantities may be shown to be

$$\frac{d}{dh'}|\mathcal{A}_{\text{in,up},k}(h, h')|^2\bigg|_{h=h'} = |T_{\frac{1}{2}kh}|^2\frac{d|R_{\frac{1}{2}kh'}|^2}{dh'}\bigg|_{h=h'} - 2\Im\left[T^*_{\frac{1}{2}kh}R_{\frac{1}{2}kh}\Lambda_{\text{in,up},k}(h, h)\right] \tag{125}$$

and

$$\frac{d}{dh'}|\mathcal{A}_{\text{up,in},k}(h, h')|^2\bigg|_{h=h'} = |R_{\frac{1}{2}kh}|^2\frac{d|T_{\frac{1}{2}kh'}|^2}{dh'}\bigg|_{h=h'} + 2\Im\left[T^*_{\frac{1}{2}kh}R_{\frac{1}{2}kh}\Lambda_{\text{in,up},k}(h, h)\right] \tag{126}$$

where we used the hermiticity property for the $\Lambda$ term, i.e $\Lambda_{\text{up,in},k}(h, h') = \Lambda^*_{\text{in,up},k}(h', h)$. Substituting our expressions for $\mathcal{A}$ in terms of reflection and transmission coefficients, one may derive the following:

$$\langle \hat{\mathcal{Z}}^2 \rangle = \frac{1}{\Xi} \times 2 \int \frac{dh}{2\pi} \sum_k (2j+1)|T_{\frac{1}{2}kh}|^2 f^{e^-}_{\text{up,up,k}}(h) \left[1 - |T_{\frac{1}{2}kh}|^2 f^{e^-}_{\text{up,up,k}}(h)\right], \tag{127}$$

where

$$\Xi = \frac{e^2}{2\pi M} \int \frac{dh}{2\pi} \sum_k (2j+1) \left\{ -|T_{\frac{1}{2}kh}|^2 \frac{df^{e^-}_{\text{up,up,k}}(h')}{dh'}\bigg|_{h'=h} + 2\Im\left[T^*_{\frac{1}{2}kh}R_{\frac{1}{2}kh}\Lambda_{\text{in,up},k}(h, h)\right] f^{e^-}_{\text{up,up,k}}(h) \right\}. \tag{128}$$

Note that we performed integration by parts to combine the two derivatives together noting that at $h = 0$ and $h \to \infty$ the combination $|T_{\frac{1}{2}kh}|^2 f^{e^-}_{\text{up,up,k}}(h)$ will be zero. Additionally, we note that the phase, $e^{-2i\arg T_{\frac{1}{2}kh}}$, that appears when performing said calculation for the second term may be absorbed to give us the above result. We show in Appendix F that the imaginary part in $\Xi$ is related to the derivative of the transmission coefficient; using Eq. (F20) and $\alpha = e^2/4\pi$, we have

$$\Xi = 2 \times 2 \int \frac{dh}{2\pi} \sum_k (2j+1) \left\{ -\frac{e^2}{8\pi M}\mathbb{T}_{\frac{1}{2}kh}\frac{df^{e^-}_{\text{up,up,k}}(h')}{dh'}\bigg|_{h'=h} - \frac{\partial\mathbb{T}_{\frac{1}{2}kh}}{\partial Z}f^{e^-}_{\text{up,up,k}}(h) \right\}. \tag{129}$$

Intuitively, the numerator of $\langle \hat{\mathcal{Z}}^2 \rangle$ may be thought of as the shot noise originating from the free field emission of charged particles. Since the emission of electrons and positrons from the black hole is a random process and may *almost* be modeled by a Poisson distribution, we recover a term that originates due to the discrete nature of electric charge. With a total emission rate of charged particles of $2 \int \sum_k (2j+1)\mathbb{T}_{1/2,k,h}f^e_{\text{up}}(h) \, dh/2\pi$, where the 2 comes from inclusion of both electrons and positrons, we expect a corresponding rate in the increase of $\langle Z^2 \rangle$. Since electrons are fermions, we expect an additional term of the form $1 - |T|^2 f$ due to Fermi-Dirac statistics.

The denominator of $\Xi$, as defined in Eq. (129), may be understood intuitively as the summation of all $\mathcal{O}(\alpha)$ changes to the emission rate due to the black hole acquiring charge when a charged lepton is emitted. The first term describes the change to the electron phase space density as the black hole acquires one unit of charge. Mathematically,

$$f^{e^-}_{\text{up,up,k}}(h) = \frac{1}{e^{8\pi M(h+Z\alpha/2M)/T_H} + 1} = \bar{f}^{e^-}_{\text{up,up,k}}(h) + \frac{Z\alpha}{2M}\frac{d\bar{f}^{e^-}_{\text{up,up,k}}(h)}{dh} + ..., \tag{130}$$

where $\bar{f}^{e^-}_{\text{up,up,k}} = 1/(e^{8\pi Mh} + 1)$ is the phase space density emitted from a neutral black hole. Thus, the inclusion of the first term in $\Xi$ describes how the phase space density changes at fixed energy $h$ as the charge fluctuates. The second term in Eq. (129) describes how the single-particle transmission probability changes as a function of the charge on the black hole. Since the potential barrier an electron or positron transverses will be perturbed when introducing an additional Coulomb potential, it is expected the transmission and reflection coefficients we derived will have a first order correction we need to consider. Therefore, by using the full quantum formalism we developed, we derived an expression for $\langle \hat{\mathcal{Z}}^2 \rangle$ that is equal to the semi-classical result (Eq. 7) except for the small Fermi blocking term in the numerator.

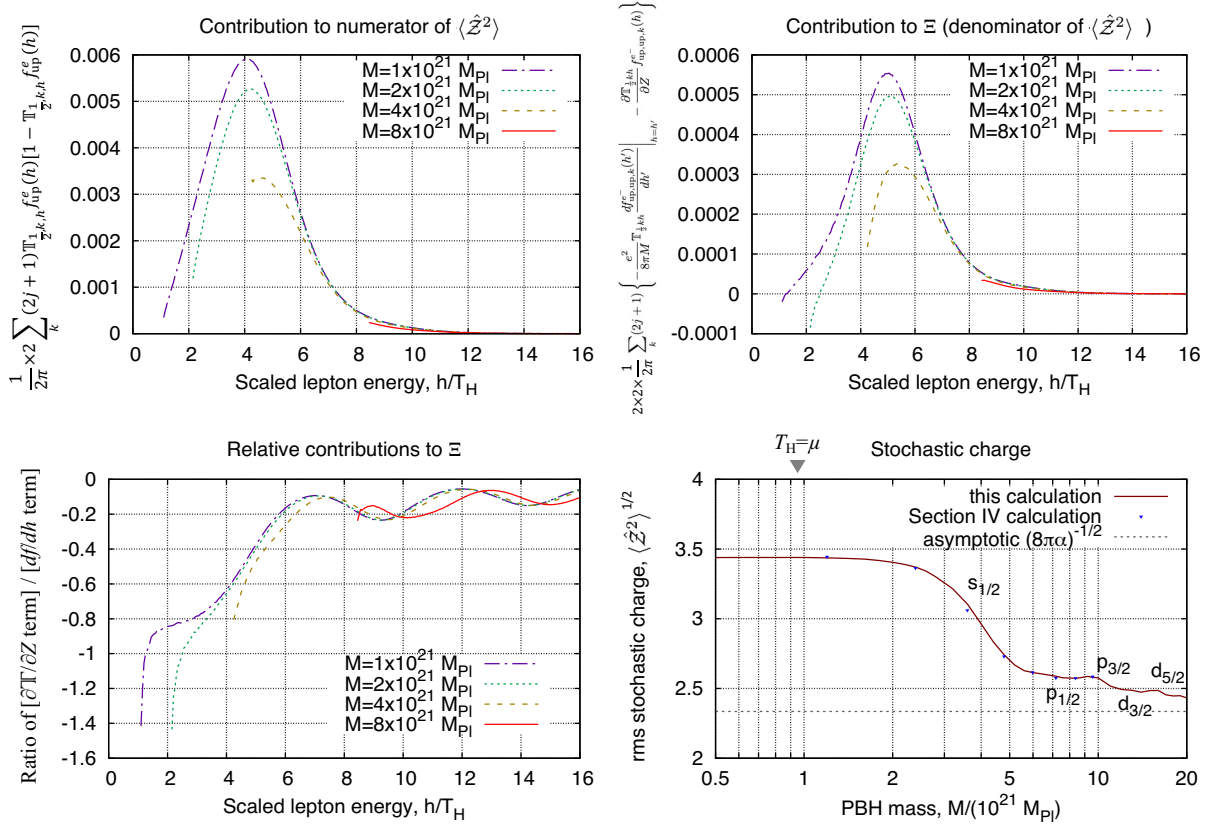

FIG. 3. Numerical results on the stochastic charge, Eq. (127). *Upper-left and upper-right panels*: the numerator (rate of buildup of stochastic charge variance) and denominator (decay rate $\Xi$, Eq. 129) of $\langle \hat{\mathcal{Z}}^2 \rangle$, respectively. These are broken down as a function of the emitted electron energy $h$, for 4 PBH masses: 1, 2, 4, and $8 \times 10^{21} M_{\rm Pl}$. *Lower-left panel*: the relative importance of the two contributions to $\Xi$. The $df/dh$ term acts to neutralize the black hole (positive contribution to $\Xi$), whereas the $\partial \mathbb{T}/\partial Z$ term acts to increase the magnitude of the charge (negative contribution to $\Xi$). At some energies, the integrand in $\Xi$ is negative, but integrated over all energies we have $\Xi > 0$ and the black hole preferentially emits charges of its same sign. *Lower-right panel*: the result for the root-mean-square stochastic charge $\langle \hat{\mathcal{Z}}^2 \rangle^{1/2}$ as a function of PBH mass. The labels indicate the masses above which each partial wave ($s_{1/2}$, $p_{1/2}$, etc.) has $\gtrsim 50\%$ transmission for a slow-moving electron; there are corresponding features in $\langle \hat{\mathcal{Z}}^2 \rangle^{1/2}$. The dashed line shows the asymptotic result $\langle \hat{\mathcal{Z}}^2 \rangle^{1/2} = 1/(8\pi\alpha)$ for large masses ($M \gg \mu^{-1}$, [36]).

## F. Numerical results

We present numerical results for the stochastic charge in Fig. 3. The radial electron wave functions are solved using a 4th-order Runge-Kutta algorithm, integrating with $r_*$ as the independent variable. The matching conditions of Paper I, Eqs. (39, 41) are used to determine the transmission probability $\mathbb{T}$, and 2-sided numerical differentiation with a step of $\Delta Z = \pm 1$ is used to determine $\partial \mathbb{T}/\partial Z$. Default parameters are to cover the range $-2000M \leq r_* \leq 4000M$ with 160,000 steps, with integration step size $\Delta h = 0.05 T_{\rm H}$, and considering values of $|k|$ up to 5. Alternatives such as $\Delta h/T_{\rm H} = 0.04$, 0.06, or 0.1, using 80,000 steps, and using vs. not using the first point with $h > \mu$ in the integration were tried, with resulting changes of $< 1.4\%$ (except for the runs with $\Delta h/T_{\rm H} = 0.1$, which led to changes up to 4.3%). The data are also presented in tabular form in Table I.

We also show, in the points on the right panel of Fig. 3, an alternative computation based on Section IV, Eq. (7). It used computations at $Z = -15, -5, 0, +5, +15$ and a 4th-order polynomial fit of $\ln \mathbb{T}$ vs. $Z$ to determine the derivatives, as well as a different method (implicit midpoint method) to solve the radial wave functions and using the total derivative of the emission rate with respect to $Z$ rather than the formulation with partial derivatives as in Eq. (129). The maximum difference is $\sim 1.5\%$, in line with our expectation based on remaining convergence issues.

In agreement with Page [36], we find that the RMS stochastic charge is approximately $(8\pi\alpha)^{-1/2} \approx 2.335$ elementary charges for large masses where the Hawking temperature is small compared to the electron mass: $T_{\rm H} \ll \mu$. It is helpful

TABLE I. The rms stochastic charge as a function of black hole mass, as plotted in Fig. 3. Columns show the black hole mass in units of $10^{21} M_{\text{Pl}}$; the logarithmic mass in grams; the Hawking temperature in keV; and the rms stochastic charge.

| $M$ $(10^{21} M_{\text{Pl}})$ | $\log_{10} M$ (g) | $T_{\text{H}}$ (keV) | $Z_{\text{rms}}$ | $M$ $(10^{21} M_{\text{Pl}})$ | $\log_{10} M$ (g) | $T_{\text{H}}$ (keV) | $Z_{\text{rms}}$ | $M$ $(10^{21} M_{\text{Pl}})$ | $\log_{10} M$ (g) | $T_{\text{H}}$ (keV) | $Z_{\text{rms}}$ |
|---|---|---|---|---|---|---|---|---|---|---|---|
| 0.50 | 16.037 | 972.9 | 3.439 | 3.20 | 16.843 | 152.0 | 3.215 | 9.20 | 17.302 | 52.9 | 2.587 |
| 0.60 | 16.116 | 810.8 | 3.439 | 3.60 | 16.894 | 135.1 | 3.103 | 9.60 | 17.320 | 50.7 | 2.580 |
| 0.70 | 16.183 | 695.0 | 3.440 | 4.00 | 16.940 | 121.6 | 2.963 | 10.00 | 17.338 | 48.6 | 2.576 |
| 0.80 | 16.241 | 608.1 | 3.440 | 4.40 | 16.981 | 110.6 | 2.833 | 11.00 | 17.379 | 44.2 | 2.517 |
| 0.90 | 16.292 | 540.5 | 3.439 | 4.80 | 17.019 | 101.3 | 2.739 | 12.00 | 17.417 | 40.5 | 2.493 |
| 1.00 | 16.338 | 486.5 | 3.439 | 5.20 | 17.054 | 93.6 | 2.668 | 13.00 | 17.452 | 37.4 | 2.487 |
| 1.20 | 16.417 | 405.4 | 3.436 | 5.60 | 17.086 | 86.9 | 2.627 | 14.00 | 17.484 | 34.7 | 2.473 |
| 1.40 | 16.484 | 347.5 | 3.433 | 6.00 | 17.116 | 81.1 | 2.613 | 15.00 | 17.514 | 32.4 | 2.485 |
| 1.60 | 16.542 | 304.0 | 3.428 | 6.40 | 17.144 | 76.0 | 2.605 | 16.00 | 17.542 | 30.4 | 2.486 |
| 1.80 | 16.593 | 270.3 | 3.418 | 6.80 | 17.170 | 71.5 | 2.596 | 17.00 | 17.568 | 28.6 | 2.454 |
| 2.00 | 16.639 | 243.2 | 3.405 | 7.20 | 17.195 | 67.6 | 2.587 | 18.00 | 17.593 | 27.0 | 2.447 |
| 2.20 | 16.680 | 221.1 | 3.389 | 7.60 | 17.219 | 64.0 | 2.576 | 19.00 | 17.617 | 25.6 | 2.448 |
| 2.40 | 16.718 | 202.7 | 3.370 | 8.00 | 17.241 | 60.8 | 2.575 | 20.00 | 17.639 | 24.3 | 2.433 |
| 2.60 | 16.753 | 187.1 | 3.342 | 8.40 | 17.262 | 57.9 | 2.573 | 22.00 | 17.680 | 22.1 | 2.443 |
| 2.80 | 16.785 | 173.7 | 3.301 | 8.80 | 17.282 | 55.3 | 2.579 | 24.00 | 17.718 | 20.3 | 2.422 |

in reading the figure to note that

$$\frac{T_{\text{H}}}{\mu} = \frac{1}{8\pi M \mu} = \frac{9.52 \times 10^{20} M_{\text{Pl}}}{M} = \frac{2.17 \times 10^{16}\,\text{g}}{M}. \tag{131}$$

At smaller masses, the stochastic charge increases. We find an asymptotic limiting value of 3.44 elementary charges (though this is still $\approx 40\%$ lower than the numerical tables computed in the original paper by Page [36]). The bulk of the transition occurs near $M \sim 4 \times 10^{21} M_{\text{Pl}}$ or $4T_{\text{H}} \sim \mu$; this is to be expected since the peak of the graybody spectrum is near $\sim 4T_{\text{H}}$, and thus this marks the transition between the "massless electron" limit (low $M$) and the case where only the Wien tail of the graybody can escape and even then it consists mainly of non-relativistic electrons (high $M$). We anticipate that further changes in the curve would occur at still lower black hole masses where muons and pions can be produced.

## VII. DISCUSSION

The use of Hawking radiation as a constraint on primordial black holes motivates careful modeling of the electron and positron emission spectra. Page [36] argued that as a small black hole in the $M \sim 10^{16}$ g mass range emits electrons and positrons, the charge of the black hole undergoes a damped random walk. This "stochastic charge" phenomenon was justified by semi-classical arguments, in which the emission of the lepton is treated quantum mechanically but the electric field is classical.

This paper has justified the existence of stochastic charge using interacting quantum field theory on curved spacetime. In Paper 1 [37], we derived the $\mathcal{O}(\alpha)$ change to the photon emission spectra by quantizing the electromagnetic field using a generalization of the Coulomb gauge. This paper completed the formal treatment of the interaction Hamiltonian arising from the scalar potential sector of QED. We showed that in the Schwarzschild spacetime there is an infrared divergence that does not occur in Minkowski spacetime, associated with the fact that a charged particle falling into the horizon continues to produce an electric field measured by an external observer that does not decay as $t \to \infty$. We performed a multipole expansion, and found that this divergent term only occurs in the monopole ($\ell = 0$) component; we packaged this into one term in the interaction Hamiltonian, as shown in Eq. (40):

$$H_{\text{int,0L}} = \frac{e^2 \hat{\mathcal{Z}}^2}{16\pi M}, \tag{132}$$

where we defined the equivalent charge operator $\hat{\mathcal{Z}}$ in Eq. (46) that may be interpreted as an "atomic number" for the black hole. The equivalent charge includes the true charge of the black hole ($\mathcal{Q}_-$, which cannot change in any finite time) as well as the integral of the charge density surrounding the hole with a weighting of $2M/r$ (so that a charge near the horizon counts toward $\hat{\mathcal{Z}}$ with weight 1, a charge far from the hole does not contribute, and the weighting varies smoothly in between). The equivalent charge $\hat{\mathcal{Z}}$ undergoes a random walk, and in the free-field theory its variance diverges as $t \to \infty$. The interaction $H_{\text{int},0L}$ causes the random walk to be "damped" (biased toward zero) and the result is finite. The variance computed by our full quantum treatment, Eq. (127) has the predicted structure of the variance $\langle Z^2 \rangle$ calculated by the semi-classical arguments laid out in Page [36]. Thus, we have proved Conjecture #1 as laid out in the introduction.

Although our calculation using QED in curved spacetime agrees with the semi-classical approach, the interpretation is somewhat different. Page [36] formulated the problem in terms of the charge *on the black hole itself*. In contrast, in our calculation the black hole's charge is fixed. For an initially neutral black hole ($\mathcal{Q}_- = 0$), the charge contributing to $\hat{\mathcal{Z}}$ is located outside the horizon, and its variance arises as a collective effect in the plasma of electrons and positrons (including virtual particles) exterior to the black hole. The true charge $\mathcal{Q}_-$ of the black hole is not relevant to the long-time limiting behavior seen by an external observer, because it only appears in the Hamiltonian through the combination $\hat{\mathcal{Z}}$, whose evolution equations and limiting distribution then make no reference to $\mathcal{Q}_-$.

Our goal, as stated in the Introduction, was to understand whether the full stochastic charge effect is reproduced in QED, and we split the argument between two separate conjectures. Conjecture #1 — that the semi-classical variance $\langle Z^2 \rangle$ [36] agrees with the variance of the quantum mechanical operator $\langle \hat{\mathcal{Z}}^2 \rangle$ when performing the full QED calculation — has been proven here. Conjecture #2 is that the semi-classical corrections to the spectra of radiated electrons and positrons [36] should be replicated when performing the full QED derivation. That calculation involves Eq. (61), which in turn involves the handling of singular integrals of one higher order than handled in this paper (it involves triple instead of double poles at $h - h' \approx 0$). We have therefore deferred testing the validity of Conjecture #2, and hence completion of the program begun here to understand stochastic charge from a QED perspective, to a subsequent manuscript.

## ACKNOWLEDGMENTS

We thank Pierre Heidmann and Oleg Korobkin for useful discussions on electrostatics around a Schwarzschild black hole; and Eric Braaten, Bown Chen, Aditi Fulsundar, and Cara Nel for comments on the draft and presentation. Computations were performed on the Pitzer cluster at the Ohio Supercomputer Center [48]. This project was supported by the David & Lucile Packard Foundation award 2021-72096. C.H. additionally received support from the National Aeronautics and Space Administration, under subaward AWP-10019534 from the Jet Propulsion Laboratory. M.S. received support from the Los Alamos National Laboratory (LANL), operated by Triad National Security, LLC, under the Laboratory Directed Research and Development program of LANL project number 20230863PRD LA - UR: 24-26674.

## Appendix A: Commutator identities

### 1. Commutator identities

In this section, we list different commutators that are essential to future calculations. Using the 3-operator commutation identities

$$[AB, C] = A[B, C] + [A, C]B, \quad [A, BC] = B[A, C] + [A, B]C, \quad \text{and} \quad [AB, C] = A\{B, C\} - \{A, C\}B, \quad \text{(A1)}$$

we can show the following relations:

$$
\begin{aligned}
[\hat{b}^\dagger_{Xkmh}\hat{b}_{Xkmh}, \hat{b}^\dagger_{X'k'm'h'}\hat{b}_{X''k'm'h''}] =& \hat{b}^\dagger_{Xkmh}\hat{b}_{X''k'm'h''} 2\pi\delta(h - h')\delta_{XX'}\delta_{kk'}\delta_{mm'} \\
& - \hat{b}^\dagger_{X'k'm'h'}\hat{b}_{Xkmh} 2\pi\delta(h - h'')\delta_{XX''}\delta_{kk''}\delta_{mm''}, \\
[\hat{b}^\dagger_{Xkmh}\hat{b}_{Xkmh}, \hat{b}^\dagger_{X'k'm'h'}\hat{d}^\dagger_{X''k'm'h''}] =& \hat{b}^\dagger_{Xkmh}\hat{d}^\dagger_{X''k'm'h''} 2\pi\delta(h - h')\delta_{XX'}\delta_{kk'}\delta_{mm'}, \\
[\hat{b}^\dagger_{Xkmh}\hat{b}_{Xkmh}, \hat{d}_{X'k'm'h'}\hat{b}_{X''k'm'h''}] =& \hat{b}_{Xkmh}\hat{d}_{X'k'm'h'} 2\pi\delta(h - h'')\delta_{XX''}\delta_{kk''}\delta_{mm''}, \\
[\hat{d}^\dagger_{Xkmh}\hat{d}_{Xkmh}, \hat{d}^\dagger_{X'k'm'h'}\hat{d}_{X''k'm'h''}] =& \hat{d}^\dagger_{Xkmh}\hat{d}_{X''k'm'h''} 2\pi\delta(h - h')\delta_{XX'}\delta_{kk'}\delta_{mm'} \\
& - \hat{d}^\dagger_{X'k'm'h'}\hat{d}_{Xkmh} 2\pi\delta(h - h'')\delta_{XX''}\delta_{kk''}\delta_{mm''}, \\
[\hat{d}^\dagger_{Xkmh}\hat{d}_{Xkmh}, \hat{b}^\dagger_{X'k'm'h'}\hat{d}^\dagger_{X''k'm'h''}] =& - \hat{d}^\dagger_{Xkmh}\hat{b}^\dagger_{X'k'm'h'} 2\pi\delta(h - h'')\delta_{XX''}\delta_{kk''}\delta_{mm''} \quad \text{and} \\
[\hat{d}^\dagger_{Xkmh}\hat{d}_{Xkmh}, \hat{d}_{X'k'm'h'}\hat{b}_{X''k'm'h''}] =& - \hat{d}_{Xkmh}\hat{b}_{X''k'm'h''} 2\pi\delta(h - h')\delta_{XX'}\delta_{kk'}\delta_{mm'}
\end{aligned}
\quad \text{(A2)}
$$

Note that it is possible to derive the last three expressions by applying charge conjugation to these operators.

### 2. Commutators involving $H_0$

The free-field Hamiltonian has the following commutation relations with raising and lowering operators:

$$[H_0, \hat{b}_{Xkmh}] = -h\hat{b}_{Xkmh}, \quad [H_0, \hat{d}_{Xkmh}] = -h\hat{d}_{Xkmh}, \quad [H_0, \hat{b}^\dagger_{Xkmh}] = h\hat{b}^\dagger_{Xkmh}, \quad \text{and} \quad [H_0, \hat{d}^\dagger_{Xkmh}] = h\hat{d}^\dagger_{Xkmh}. \quad \text{(A3)}$$

These can be combined to give the commutation of $H_0$ with any sequence of raising and lowering operators, e.g.,

$$[H_0, \hat{b}^\dagger_{Xkmh}\hat{b}_{X'k'm'h'}\hat{d}^\dagger_{X''k''m''h''}\hat{d}_{X'''k'''m'''h'''}] = (h - h' + h'' - h''')\hat{b}^\dagger_{Xkmh}\hat{b}_{X'k'm'h'}\hat{d}^\dagger_{X''k''m''h''}\hat{d}_{X'''k'''m'''h'''}; \quad \text{(A4)}$$

in general the rule is that the same operator appears, but with a factor of the energy difference before the application of the raising and lowering operators.

### 3. Commutators involving $\hat{\mathcal{Z}}$

When computing $[\hat{\mathcal{Z}}, \hat{b}_{Xkmh}]$ and $[\hat{\mathcal{Z}}, \hat{d}_{Xkmh}]$, we will only have four sets of commutators that will result in a nonzero value. These are listed as:

$$\begin{aligned}
[\hat{b}^\dagger_{X'k'm'h'}\hat{b}_{X''k''m''h''}, \hat{b}_{Xkmh}] &= -\hat{b}_{X''k''m''h''} \times 2\pi\delta(h - h')\delta_{XX'}\delta_{kk'}\delta_{mm'}, \\
[\hat{b}^\dagger_{X'k'm'h'}\hat{d}^\dagger_{X''k''m''h''}, \hat{b}_{Xkmh}] &= -\hat{d}^\dagger_{X''k''m''h''} \times 2\pi\delta(h - h')\delta_{XX'}\delta_{kk'}\delta_{mm'}, \\
[\hat{b}^\dagger_{X'k'm'h'}\hat{d}^\dagger_{X''k''m''h''}, \hat{d}_{Xkmh}] &= \hat{b}^\dagger_{X'k'm'h'} \times 2\pi\delta(h - h'')\delta_{XX''}\delta_{kk''}\delta_{mm''}, \quad \text{and} \\
[\hat{d}^\dagger_{X'k'm'h'}\hat{d}_{X''k''m''h''}, \hat{d}_{Xkmh}] &= -\hat{d}^\dagger_{X''k''m''h''} \times 2\pi\delta(h - h')\delta_{XX'}\delta_{kk'}\delta_{mm'}.
\end{aligned} \quad \text{(A5)}$$

Note that in our expression for $\hat{\mathcal{Z}}$, as defined in Eq. (46), the primed and double primed variables are dummy indices such that $k' = k''$ and $m' = m''$ which we showed from orthogonality of the angular wave functions. We may relabel every term as a sum of $X'$ as one of the sums over either $X'$ or $X''$ will cancel due to the delta functions in the commutator. Therefore, one can prove

$$[\hat{\mathcal{Z}}, \hat{b}_{Xkmh}] = \int \frac{dh'}{2\pi} \sum_{X'} \left\{ \mathcal{I}^{(1)}_{XX'k}(h, h')\hat{b}_{X'kmh'} + \mathcal{I}^{(2)}_{XX'k}(h, h')\hat{d}^\dagger_{X'kmh'} \right\}, \quad \text{(A6)}$$

$$[\hat{\mathcal{Z}}, \hat{b}^\dagger_{Xkmh}] = -\int \frac{dh'}{2\pi} \sum_{X'} \left\{ \mathcal{I}^{(1)\star}_{XX'k}(h, h')\hat{b}^\dagger_{X'kmh'} + \mathcal{I}^{(2)\star}_{XX'k}(h, h')\hat{d}_{X'kmh'} \right\}, \quad \text{(A7)}$$

and

$$[\hat{\mathcal{Z}}, \hat{d}_{Xkmh}] = -\int \frac{dh'}{2\pi} \sum_{X'} \left\{ \mathcal{I}^{(2)}_{X'Xk}(h', h)\hat{b}^\dagger_{X'kmh'} + \mathcal{I}^{(1)\star}_{XX'-k}(h, h')\hat{d}_{X'kmh'} \right\} \quad \text{(A8)}$$

where we make use of the fact that $\mathcal{I}^{(4)}_{XX'k}(h, h') = \mathcal{I}^{(1)\star}_{X'X-k}(h', h)$. Note that you can easily get $[\hat{\mathcal{Z}}, \hat{b}^\dagger]$ from $[\hat{\mathcal{Z}}, \hat{b}]$ such that

$$[\hat{\mathcal{Z}}, \hat{b}^\dagger_{Xkmh}] = [\hat{b}_{Xkmh}, \hat{\mathcal{Z}}^\dagger]^\dagger = -[\hat{\mathcal{Z}}, \hat{b}_{Xkmh}]^\dagger \quad \text{(A9)}$$

where we make use of the fact that $\hat{\mathcal{Z}}$ is Hermition. You can get $[\hat{\mathcal{Z}}, \hat{d}^\dagger]$ from $[\hat{\mathcal{Z}}, \hat{d}]$ by doing a similar trick.

## Appendix B: Charge conjugation

The goal of this section is to check and verify that our expression for $\hat{\mathcal{Z}}$, given by Eq. (46), behaves correctly under charge conjugation. Since $\hat{\mathcal{Z}}$ is so similar to the charge operator, we should expect that it is $C$-odd, i.e.,

$$C\hat{\mathcal{Z}}C^{-1} = -\hat{\mathcal{Z}}. \quad \text{(B1)}$$

This appendix explicitly shows that this relation holds.

From Appendix E of Paper I [37], each of the fermion operators changes under charge conjugation as listed below:

$$\begin{aligned}
\hat{C}\hat{b}_{Xkmh}\hat{C}^{-1} &= is(-1)^{m+\frac{1}{2}}\hat{d}_{X,-k,-m,h}, \\
\hat{C}\hat{b}^{\dagger}_{Xkmh}\hat{C}^{-1} &= -is(-1)^{m+\frac{1}{2}}\hat{d}^{\dagger}_{X,-k,-m,h}, \\
\hat{C}\hat{d}_{Xkmh}\hat{C}^{-1} &= -is(-1)^{m+\frac{1}{2}}\hat{b}_{X,-k,-m,h}, \qquad \text{and} \\
\hat{C}\hat{d}^{\dagger}_{Xkmh}\hat{C}^{-1} &= is(-1)^{m+\frac{1}{2}}\hat{b}^{\dagger}_{X,-k,-m,h},
\end{aligned}$$
(B2)

where $s = k/|k|$. Enacting the charge conjugation operator on $\hat{\mathcal{Z}}$ given by Eq. (46), we obtain

$$\begin{aligned}
\hat{C}\hat{\mathcal{Z}}\hat{C}^{-1} &= \int \frac{dh\,dh'}{(2\pi)^2} \sum_{XX''} \sum_{km} \Big\{ \mathcal{I}^{(4)}_{XX'k}(h,h')\hat{b}^{\dagger}_{X,-k,-m,h}\hat{b}_{X',-k,-m,h'} + \mathcal{I}^{(2)}_{XX'k}(h,h')\hat{b}^{\dagger}_{X',-k,-m,h'}\hat{d}^{\dagger}_{X,-k,-m,h} \\
&\quad + \mathcal{I}^{(3)}_{XX'k}(h,h')\hat{d}_{X',-k,-m,h'}\hat{b}_{X,-k,-m,h} - \mathcal{I}^{(1)}_{XX'k}(h,h')\hat{d}^{\dagger}_{X,-k,-m,h}\hat{d}_{X',-k,-m,h'} \Big\} - \frac{\hat{C}\mathcal{Q}_-\hat{C}^{-1}}{e} \\
&= \int \frac{dh\,dh'}{(2\pi)^2} \sum_{XX''} \sum_{km} \Big\{ \mathcal{I}^{(1)}_{XX'(-k)}(h,h')\hat{b}^{\dagger}_{X,-k,-m,h}\hat{b}_{X',-k,-m,h'} + \mathcal{I}^{(2)}_{X'X(-k)}(h,h)\hat{b}^{\dagger}_{X',-k,-m,h'}\hat{d}^{\dagger}_{X,-k,-m,h} \\
&\quad + \mathcal{I}^{(3)}_{X'X(-k)}(h',h)\hat{d}_{X',-k,-m,h'}\hat{b}_{X,-k,-m,h} - \mathcal{I}^{(4)}_{XX'(-k)}(h,h')\hat{d}^{\dagger}_{X,-k,-m,h}\hat{d}_{X',-k,-m,h'} \Big\} - \frac{\hat{C}\mathcal{Q}_-\hat{C}^{-1}}{e},
\end{aligned}$$
(B3)

where we use the mappings as defined in Eq. (49) to swap between the $\mathcal{I}$ mode functions. Thus, we acquire a function identical to Eq. (46) when replacing $k \to -k$. Since $k$ is summed over both positive and negative values, the sign of the term is irrelevant and we acquire the same expression for $\hat{\mathcal{Z}}$. In the case where there is a charge on the black hole ($\mathcal{Q}_- \neq 0$) then to complete the chain of logic, we must identify:

$$\hat{C}\mathcal{Q}_-\hat{C}^{-1} = -\mathcal{Q}_-,$$
(B4)

i.e., the charge-conjugation operator must map a state in the subspace with $\mathcal{Q}_-$ to a state in the subspace with $-\mathcal{Q}_-$ (the "charge on the horizon" also needs to be conjugated).

Additionally, we wish to show how $[\hat{\mathcal{Z}},\hat{b}_{Xkhm}]$ and $[\hat{\mathcal{Z}},\hat{d}_{Xkhm}]$ transform under charge conjugation

$$\hat{C}[\hat{\mathcal{Z}},\hat{b}_{Xkmh}]\hat{C}^{-1} = -[\hat{\mathcal{Z}},\hat{C}\hat{b}_{Xkmh}\hat{C}^{-1}] = -is(-1)^{m+\frac{1}{2}}[\hat{\mathcal{Z}},\hat{d}_{X,-k,-m,h}].$$
(B5)

Using the definitions of the commutators in Eq. (A6-A8), one can show the left hand side of Eq. (B5) gives

$$\text{LHS} = \hat{C}[\hat{\mathcal{Z}},\hat{b}_{Xkmh}]\hat{C}^{-1} = is(-1)^{m+\frac{1}{2}}\int\frac{dh'}{2\pi}\sum_{X'}\Big\{\mathcal{I}^{(1)}_{XX'k}(h,h')\hat{d}_{X',-k,-m,h'} + \mathcal{I}^{(2)}_{XX'k}(h,h')\hat{b}^{\dagger}_{X',-k,-m,h'}\Big\}$$
(B6)

while the right hand side may be expanded using the definition of commutator $[\hat{\mathcal{Z}},\hat{d}_{Xkhm}]$ given by Eq. (A8):

$$\begin{aligned}
\text{RHS} &= -is(-1)^{m+\frac{1}{2}}[\hat{\mathcal{Z}},\hat{d}_{X,-k,-m,h}] \\
&= is(-1)^{m+\frac{1}{2}}\int\frac{dh'}{2\pi}\sum_{X'}\Big\{\mathcal{I}^{(2)}_{X'X(-k)}(h',h)\hat{b}^{\dagger}_{X',-k,-m,h'} + \mathcal{I}^{(4)}_{XX'(-k)}(h,h')\hat{d}_{X',-k,-m,h'}\Big\}
\end{aligned}$$
(B7)

which we can relate to the left hand side using the identities derived in Eq. (49) between $\mathcal{I}^{(1)}$ and $\mathcal{I}^{(4)}$ and $\mathcal{I}^{(2)}$ and $\mathcal{I}^{(3)}$, respectively. Therefore, both the left hand side and the right hand side are equivalent as predicted by Eq. (B5).

## Appendix C: Singular behavior of the $\mathcal{I}$-integrals

When $h \approx h'$, the electron mode functions in the $\mathcal{I}$-integrals (Eq. 47) can oscillate at the same frequency. Under this circumstance, the integral of the product of the wave functions may become very large, or even diverge if $h = h'$. This appendix is concerned with the singular behavior of the $\mathcal{I}$-integrals in this case.

We use the limiting forms of Paper I Eqns. (39, 41, 42). We begin with the $\mathcal{I}^{(1)}_{\text{up,up},k}$-type integrals:

$$\mathcal{I}^{(1)}_{\text{up,up},k}(h,h') \sim \frac{1}{\sqrt{4hh'}}\int_{-\infty}^{0} dr_\star\, e^{\epsilon r_\star}(F^*_{Xkh}F_{X'kh'} + G^*_{Xkh}G_{X'k'h'}) + \Lambda_{\text{up,up},k}(h,h'),$$
(C1)

where $\epsilon \to 0^+$ is a regularization parameter that controls how the oscillatory functions are cut off at $r_\star \to -\infty$ and

$$\Lambda_{XX'k}(h, h') = \frac{1}{\sqrt{4hh'}} \int_0^\infty dr_\star \frac{2M}{r} (F^*_{Xkh} F_{X'kh'} + G^*_{Xkh} G_{X'k'h'}) \tag{C2}$$

is the portion of the integral we separated which is finite and extends to spatial infinity. Additionally, $\Lambda_{XX'k}(h, h')$ obeys the same rules of Hermiticity as the $\mathcal{I}^{(1)}$-type integrals, i.e. $\Lambda_{XX'k}(h, h') = \Lambda^*_{X'Xk}(h', h)$. Note that $2M/r \to 1$ at $r_\star \to -\infty$, but this factor naturally cuts off at $r_\star \to +\infty$ causing the integral to converge. Then we have

$$\mathcal{I}^{(1)}_{\text{up,up},k}(h, h') \sim \int_{-\infty}^0 dr_\star \, e^{\epsilon r_\star} \left[ e^{i(h'-h)r_\star} + R_{\frac{1}{2},k,h} R^*_{\frac{1}{2},k,h'} e^{2i(\arg T_{1/2,k,h} - \arg T_{1/2,k,h'})} e^{i(h-h')r_\star} \right] + \Lambda_{\text{up,up},k}(h, h'). \tag{C3}$$

Using the integral $\int_{-\infty}^0 e^{(\epsilon - iy)r_\star} \, dr_\star = i/(y + i\epsilon)$, we conclude that

$$\mathcal{I}^{(1)}_{\text{up,up},k}(h, h') \sim \frac{i}{h - h' + i\epsilon} - |R_{\frac{1}{2},k,h}|^2 \frac{i}{h - h' - i\epsilon} + \Lambda_{\text{up,up},k}(h, h'). \tag{C4}$$

We may perform the same procedure for $\mathcal{I}^{(1)}_{\text{in,in},k}$-type integrals noting that

$$\mathcal{I}^{(1)}_{\text{in,in},k}(h, h') \sim \frac{1}{\sqrt{4hh'}} \int_{-\infty}^0 dr_\star \, e^{\epsilon r_\star} (F^*_{\text{in,kmh}} F_{\text{in},k,h'} + G^*_{\text{in,k,h}} G_{\text{in},k',h'}) + \Lambda_{\text{in,in},k}(h, h'), \tag{C5}$$

where we define $\epsilon$ as the same regularization parameter while noting the upper portion of the integral converges to a finite value. Thus, we have

$$\mathcal{I}^{(1)}_{\text{in,in},k}(h, h') \sim \int_{-\infty}^0 dr_\star \, e^{\epsilon r_\star} T^\star_{\frac{1}{2}kh} T_{\frac{1}{2}kh'} e^{i(h-h')r_\star} + \Lambda_{\text{in,in},k}(h, h'). \tag{C6}$$

If we perform the same substitution for $1/(y + i\epsilon)$, we obtain

$$\mathcal{I}^{(1)}_{\text{in,in},k}(h, h') \sim -|T_{\frac{1}{2}kh}|^2 \frac{i}{h - h' - i\epsilon}, \tag{C7}$$

which has the same singular behavior as $\mathcal{I}^{(1)}_{\text{up,up}}$-type integrals (except that there is only a pole on one side of the real axis, since for $h = h'$ there is only a divergence for the downgoing wave and not the upgoing one). Additionally, we can show $\mathcal{I}^{(1)}_{\text{in,up}}$-type and $\mathcal{I}^{(1)}_{\text{up,in}}$-type integrals are singular and will be functions of the following form:

$$\mathcal{I}^{(1)}_{\text{in,up},k}(h, h') \sim T_{\frac{1}{2}kh} R^*_{\frac{1}{2}kh} \frac{i}{h - h' - i\epsilon} + \Lambda_{\text{in,up},k}(h, h') \tag{C8}$$

and

$$\mathcal{I}^{(1)}_{\text{up,in},k}(h, h') \sim R_{\frac{1}{2}kh} T^*_{\frac{1}{2}kh} \frac{i}{h - h' - i\epsilon} + \Lambda_{\text{up,in},k}(h, h'). \tag{C9}$$

We may express the $\mathcal{I}^{(1)}$-type integrals in terms of the non-singular functions $\mathcal{A}$ (Eqn. (83)):

$$\begin{aligned} \mathcal{I}^{(1)}_{\text{up,up},k}(h, h') &= \frac{\mathcal{A}_{\text{up,up},k}(h, h')}{h - h' - i\epsilon} + 2\pi\delta_\epsilon(h - h'), \\ \mathcal{I}^{(1)}_{\text{up,in},k}(h, h') &= \frac{\mathcal{A}_{\text{up,in},k}(h, h')}{h - h' - i\epsilon}, \\ \mathcal{I}^{(1)}_{\text{in,up},k}(h, h') &= \frac{\mathcal{A}_{\text{in,up},k}(h, h')}{h - h' - i\epsilon}, \quad \text{and} \\ \mathcal{I}^{(1)}_{\text{in,in},k}(h, h') &= \frac{\mathcal{A}_{\text{in,in},k}(h, h')}{h - h' - i\epsilon}, \end{aligned} \tag{C10}$$

where

$$\delta_\epsilon(h - h') = \frac{i}{2\pi} \left( \frac{1}{h - h' + i\epsilon} - \frac{1}{h - h' - i\epsilon} \right) = \frac{\epsilon}{\pi[(h - h')^2 + \epsilon^2]} \tag{C11}$$

is the difference between a pole on each side of the real axis, and accounts for the fact that there is a "$+i\epsilon$" instead of "$-i\epsilon$" pole in Eq. (C4). We note that

$$\int_{-\infty}^\infty \delta_\epsilon(x) \, dx = 1, \tag{C12}$$

i.e., $\delta_\epsilon$ is normalized in the same way as a $\delta$-function.

None of the $\mathcal{I}^{(2)}$'s are singular, since by similar arguments they have $h + h'$ instead of $h - h'$ in the denominator.

## Appendix D: Relating the "in/up" scattering basis to the "out/down" scattering basis for fermions

We wish to perform a similar set of arguments as previously discussed in App. C of Silva *et al.* [37], which only considered how photons transform between the "in/up" and "out/down" basis, to derive how the radial fermion solutions transform. We assume the black hole is in a vacuum and there are no infalling electrons or positrons from an external source. This calculation will prove nearly identitical to what has been done in App. C except with the substitution $T_{1\ell\omega} \to T_{\frac{1}{2}kh}$ and $R_{1\ell\omega} \to R_{\frac{1}{2}kh}$.

As shown in Silva *et al.* [37], the radial evolution of $\chi_{Xkh} = \begin{pmatrix} F_{XKh} \\ G_{XKh} \end{pmatrix}$ is a second order ordinary differential equation with two independent solutions. This allows us to define the asymptotic "out/down" as

$$\begin{pmatrix} \chi_{\text{out,k,h}} \\ \chi_{\text{down,k,h}} \end{pmatrix} = \begin{pmatrix} A & B \\ C & D \end{pmatrix} \begin{pmatrix} \chi_{\text{up,k,h}} \\ \chi_{\text{in,k,h}} \end{pmatrix} \tag{D1}$$

where each $A$, $B$, $C$, and $D$ terms are arbitrary coefficients which we shall derive later on. The asymptotic form in the $r_\star \to \pm\infty$ limit is given by

$$\chi_{(\text{out,down}),k,h}(r_\star) \to \begin{cases} (A,C) \begin{pmatrix} \sqrt{h} \\ -i\sqrt{h} \end{pmatrix} e^{ihr_\star} \\ \quad + \left[ (B,D) T_{\frac{1}{2},k,h} - (A,C) R^*_{\frac{1}{2},k,h} e^{2i\arg T_{\frac{1}{2},k,h}} \right] \begin{pmatrix} \sqrt{h} \\ i\sqrt{h} \end{pmatrix} e^{-i\omega r_\star} & r_\star \to -\infty, \\[2ex] \left[ (A,C) T_{\frac{1}{2},k,h} + (B,D) R_{\frac{1}{2},k,h} \right] v^{-1/2} \begin{pmatrix} \sqrt{h+\mu} \\ -i\sqrt{h-\mu} \end{pmatrix} e^{ix} \\ \quad + (B,D) v^{-1/2} \begin{pmatrix} \sqrt{h+\mu} \\ -i\sqrt{h-\mu} \end{pmatrix} e^{-ix} & r_\star \to \infty, \end{cases} \tag{D2}$$

where

$$v = \frac{\sqrt{h^2 - \mu^2}}{h} \quad \text{and} \quad x = \sqrt{h^2 - \mu^2} r_* + \xi \ln \frac{r_*}{2M} \tag{D3}$$

which describe the velocity and phase, including the logarithmic divergence that occurs with long-range $1/r$ potentials [49], respectively.

In order to maintain the same normalization as in the "in/up" basis, we can constrain the linear combination Eq. (D1) by imposing boundary conditions. For an outgoing wave ($\chi_{\text{out}}$), we wish to have similar boundary conditions to $\chi_{\text{up}}$ except transversing in the opposite direction. This procedure is equivalent to $\chi_{\text{out}}$ matching the same boundary conditions for the complex conjugate of $\chi_{\text{in}}$. Likewise, the boundary conditions for $\chi_{\text{down}}$ will be the same as the boundary conditions for $\chi_{\text{up}}$ except transversing in the opposite direction. Thus, the linear combination that satisfies these normalization conventions are

$$\begin{pmatrix} \chi_{\text{out,k,h}} \\ \chi_{\text{down,k,h}} \end{pmatrix} = \begin{pmatrix} T^*_{\frac{1}{2},k,h} & R^*_{\frac{1}{2},k,h} \\ -R_{\frac{1}{2},k,h} e^{-2i\arg T_{\frac{1}{2},k,h}} & T^*_{\frac{1}{2},k,h} \end{pmatrix} \begin{pmatrix} \chi_{\text{up,k,h}} \\ \chi_{\text{in,k,h}} \end{pmatrix} \tag{D4}$$

where the transformation is unitary, which is expected due to conservation of probability, and has determinant $e^{-2i\arg T_{\frac{1}{2},k,h}}$. Therefore, the procedure of transforming between the "in/up" and "out/down" basis is identical to what was proven for photons, except with their own associated transmission and reflection coefficients.

Since the fermion wave function, $\psi_{Xkhm}$ as defined by Eq. (42) in Silva *et al.* [37], has an explicit dependence on both the radial wave function, $F$'s and $G$'s, along with an explicit dependence on the fermion creation and annihilation operators, we can expand $\psi_{Xkhm}$ into the "out/down" basis using Eq. (60). If we collect operators corresponding to $\psi_{\text{out/down}}$, then it is possible to prove

$$\hat{b}_{\text{out},kmh} = R_{\frac{1}{2}kh} \hat{b}_{\text{in},kmh} + T_{\frac{1}{2}kh} \hat{b}_{\text{up},kmh}$$
$$\hat{d}_{\text{out},kmh} = R_{\frac{1}{2}-kh} \hat{d}_{\text{in},kmh} + T_{\frac{1}{2}-kh} \hat{d}_{\text{up},kmh}$$
$$\hat{b}_{\text{down},kmh} = T_{\frac{1}{2}-kh} \hat{b}_{\text{in},kmh} - R_{\frac{1}{2}-kh} e^{-2i\arg T_{\frac{1}{2}-kh}} \hat{b}_{\text{up},kmh}$$
$$\hat{d}_{\text{down},kmh} = T_{\frac{1}{2}-kh} \hat{d}_{\text{in},kmh} - R_{\frac{1}{2}-kh} e^{-2i\arg T_{\frac{1}{2}-kh}} \hat{d}_{\text{up},kmh}$$
$$\tag{D5}$$

which summarizes how creation and annihilation operators transform between the "in/up" and "out/down" basis.

We wish to now list how the $\mathcal{I}$-type mode integrals, as defined in Eq. (47), and how the charge-fermion-fermion expectation values, as defined in Eq. (50), transform into the "in/up" basis. Because of the implicit dependence on the radial wave functions and the creation and annihilation operators, we may write

$$
\begin{aligned}
\mathcal{I}^{(1)}_{\text{out},X'k}(h,h') &= T_{\frac{1}{2}kh}\mathcal{I}^{(1)}_{\text{up},X'k}(h,h') + R_{\frac{1}{2}kh}\mathcal{I}^{(1)}_{\text{in},X'k}(h,h') \\
\mathcal{I}^{(2)}_{\text{out},X'k}(h,h') &= T_{\frac{1}{2}kh}\mathcal{I}^{(2)}_{\text{up},X'k}(h,h') + R_{\frac{1}{2}kh}\mathcal{I}^{(2)}_{\text{in},X'k}(h,h')
\end{aligned}
\tag{D6}
$$

and

$$
\begin{aligned}
\Gamma^{\hat{\mathcal{Z}}b^{\dagger}b(k)}_{\text{out},X'}(h,h') &= T^{\star}_{\frac{1}{2}kh}\Gamma^{\hat{\mathcal{Z}}b^{\dagger}b(k)}_{\text{up},X'}(h,h') + R^{\star}_{\frac{1}{2}kh}\Gamma^{\hat{\mathcal{Z}}b^{\dagger}b(k)}_{\text{in},X'}(h,h') \\
\Gamma^{\hat{\mathcal{Z}}b^{\dagger}b(k)}_{X',\text{out}}(h',h) &= T_{\frac{1}{2}kh}\Gamma^{\hat{\mathcal{Z}}b^{\dagger}b(k)}_{X',\text{up}}(h',h) + R_{\frac{1}{2}kh}\Gamma^{\hat{\mathcal{Z}}b^{\dagger}b(k)}_{X',\text{in}}(h',h) \\
\Gamma^{\hat{\mathcal{Z}}bd(k)}_{\text{out},X'}(h',h) &= T_{\frac{1}{2}kh}\Gamma^{\hat{\mathcal{Z}}bd(k)}_{\text{up},X'}(h,h') + R_{\frac{1}{2}kh}\Gamma^{\hat{\mathcal{Z}}bd(k)}_{\text{in},X'}(h,h') \\
\Gamma^{\hat{\mathcal{Z}}d^{\dagger}b^{\dagger}(k)}_{X',\text{out}}(h',h) &= T^{\star}_{\frac{1}{2}kh}\Gamma^{\hat{\mathcal{Z}}d^{\dagger}b^{\dagger}(k)}_{X',\text{up}}(h',h) + R^{\star}_{\frac{1}{2}kh}\Gamma^{\hat{\mathcal{Z}}d^{\dagger}b^{\dagger}(k)}_{X',\text{in}}(h',h).
\end{aligned}
\tag{D7}
$$

When deriving the emission spectra, as shown in Eq. (60), terms will appear that are the product of both the $\mathcal{I}$-type integrals and charge-fermion-fermion expectation values. The terms relevant towards deriving Eq. (60) are shown below:

$$
\begin{aligned}
\mathcal{I}^{(1)}_{\text{out},X'k}(h,h')\Gamma^{\hat{\mathcal{Z}}b^{\dagger}b(k)}_{\text{out},X'}(h,h') =& |T_{\frac{1}{2}kh}|^2\mathcal{I}^{(1)}_{\text{up},X'k}(h,h')\Gamma^{\hat{\mathcal{Z}}b^{\dagger}b(k)}_{\text{up},X'}(h,h') + T^{\star}_{\frac{1}{2}kh}R_{\frac{1}{2}kh}\mathcal{I}^{(1)}_{\text{in},X'k}(h,h')\Gamma^{\hat{\mathcal{Z}}b^{\dagger}b(k)}_{\text{up},X'}(h,h') \\
&+ T_{\frac{1}{2}kh}R^{\star}_{\frac{1}{2}kh}\mathcal{I}^{(1)}_{\text{up},X'k}(h,h')\Gamma^{\hat{\mathcal{Z}}b^{\dagger}b(k)}_{\text{in},X'}(h,h') + |R_{\frac{1}{2}kh}|^2\mathcal{I}^{(1)}_{\text{in},X'k}(h,h')\Gamma^{\hat{\mathcal{Z}}b^{\dagger}b(k)}_{\text{in},X'}(h,h')
\end{aligned}
\tag{D8}
$$

$$
\begin{aligned}
\mathcal{I}^{(1)\star}_{\text{out},X',k}(h,h')\Gamma^{\hat{\mathcal{Z}}b^{\dagger}b(k)}_{X',\text{out}}(h',h) =& |T_{\frac{1}{2}kh}|^2\mathcal{I}^{(1)\star}_{\text{up},X',k}(h,h')\Gamma^{\hat{\mathcal{Z}}b^{\dagger}b(k)}_{X',\text{up}}(h',h) + T_{\frac{1}{2}kh}R^{\star}_{\frac{1}{2}kh}\mathcal{I}^{(1)\star}_{\text{in},X',k}(h,h')\Gamma^{\hat{\mathcal{Z}}b^{\dagger}b(k)}_{X',\text{up}}(h',h) \\
&+ T^{\star}_{\frac{1}{2}kh}R_{\frac{1}{2}kh}\mathcal{I}^{(1)\star}_{\text{up},X',k}(h,h')\Gamma^{\hat{\mathcal{Z}}b^{\dagger}b(k)}_{X',\text{in}}(h',h) + |R_{\frac{1}{2}kh}|^2\mathcal{I}^{(1)\star}_{\text{in},X',k}(h',h)\Gamma^{\hat{\mathcal{Z}}b^{\dagger}b(k)}_{X',\text{in}}(h',h)
\end{aligned}
\tag{D9}
$$

$$
\begin{aligned}
\mathcal{I}^{(2)}_{\text{out},X',k}(h',h)\Gamma^{\hat{\mathcal{Z}}d^{\dagger}b^{\dagger}(k)}_{X',\text{out}}(h',h) =& |T_{\frac{1}{2}kh}|^2\mathcal{I}^{(2)}_{\text{up},X'k}(h,h')\Gamma^{\hat{\mathcal{Z}}d^{\dagger}b^{\dagger}(k)}_{X',\text{up}}(h',h) + T_{\frac{1}{2}kh}R^{\star}_{\frac{1}{2}kh}\mathcal{I}^{(2)}_{\text{up},X'k}(h,h')\Gamma^{\hat{\mathcal{Z}}d^{\dagger}b^{\dagger}(k)}_{X',\text{in}}(h',h) \\
&+ T^{\star}_{\frac{1}{2}kh}R_{\frac{1}{2}kh}\mathcal{I}^{(2)}_{\text{in},X'k}(h,h')\Gamma^{\hat{\mathcal{Z}}d^{\dagger}b^{\dagger}(k)}_{X',\text{up}}(h',h) + |R_{\frac{1}{2}kh}|^2\mathcal{I}^{(2)}_{\text{in},X'k}(h',h)\Gamma^{\hat{\mathcal{Z}}d^{\dagger}b^{\dagger}(k)}_{X',\text{in}}(h',h)
\end{aligned}
\tag{D10}
$$

$$
\begin{aligned}
\mathcal{I}^{(2)\star}_{\text{out},X'k}(h',h)\Gamma^{\hat{\mathcal{Z}}bd(k)}_{\text{out},X'}(h,h') =& |T_{\frac{1}{2}kh}|^2\mathcal{I}^{(2)\star}_{\text{up},X'k}(h,h')\Gamma^{\hat{\mathcal{Z}}bd(k)}_{\text{up},X'}(h,h') + T_{\frac{1}{2}kh}R^{\star}_{\frac{1}{2}kh}\mathcal{I}^{(2)\star}_{\text{in},X'k}(h,h')\Gamma^{\hat{\mathcal{Z}}bd(k)}_{\text{up},X'}(h,h') \\
&+ T^{\star}_{\frac{1}{2}kh}R_{\frac{1}{2}kh}\mathcal{I}^{(2)\star}_{\text{up},X'k}(h,h')\Gamma^{\hat{\mathcal{Z}}bd(k)}_{\text{in},X'}(h,h') + |R_{\frac{1}{2}kh}|^2\mathcal{I}^{(2)\star}_{\text{in},X'k}(h,h')\Gamma^{\hat{\mathcal{Z}}bd(k)}_{\text{in},X'}(h,h')
\end{aligned}
\tag{D11}
$$

## Appendix E: Important singular integrals

In this section, we list the various singular integrals that are needed throughout this manuscript. Consider $\epsilon, \eta \in \mathbb{R}^{+}$ and $x_0 \in \mathbb{R}$ such that $\epsilon < \eta$. Furthermore, suppose that $\mathcal{Y}(x)$ is an analytic function of $x$ that is real for real $x$. Then it is possible to show the following integral identities. Note that we explicitly take the limit $\epsilon \to 0$ <u>before</u> the limit $\eta \to 0$. We argued in Sec. VI E that this choice is physically justified.

First, if we Taylor expand $\mathcal{Y}(x)$ around $x_0$:

$$
\mathcal{Y}(x) = y_0 + y_1(x - x_0) + \frac{1}{2}y_2(x - x_0)^2 + ...,
\tag{E1}
$$

so that $y_0 = \mathcal{Y}(x_0)$, $y_1 = d\mathcal{Y}/dx|_{x_0}$, etc., then in the following integrals the $y_2$ and higher terms are all real with no singularities, and can be dropped. We may use partial fractions or contour integration to obtain the relevant imaginary parts:

$$
\lim_{\eta \to 0^{+}} \Im \int_0^{\infty} dx\, \frac{\mathcal{Y}(x)}{x - x_0 + i\eta} = -\pi \mathcal{Y}(x_0).
\tag{E2}
$$

and

$$\lim_{\eta \to 0^+} \lim_{\epsilon \to 0^+} \Im \int_0^\infty dx \, \frac{\mathcal{Y}(x)}{(x - x_0 + i\eta)(x - x_0 \pm i\epsilon)} = -\pi \frac{d\mathcal{Y}}{dx}\bigg|_{x=x_0}. \tag{E3}$$

## Appendix F: Relation of the $\mathcal{I}$-integrals to perturbations of the transmission coefficients

In this appendix, we relate the singular behavior of the $\mathcal{I}$-integrals (appearing in the *quantum* formulation) to the perturbation to transmission coefficients when a *classical* charge is introduced on the black hole. This is the key mathematical relation that we need to compare our results in Section VI E 4 to the Page [36] formulation.

### 1. Setup

We use variation of parameters to describe the perturbation to the electron wave function when a small classical charge is introduced on the black hole. Let us first consider $\vec{\chi} = (F, G)^{\mathrm{T}}$ as the set of radial fermionic wave functions defined in Paper 1 [37]. Note that we are suppressing the subscripts $k$ and $h$ to avoid clutter. Consider the fundamental matrix $\Phi$ such that

$$\Phi = \begin{pmatrix} F_{\mathrm{in}} & F_{\mathrm{up}} \\ G_{\mathrm{in}} & G_{\mathrm{up}} \end{pmatrix} \tag{F1}$$

where each column is an eigenvector for a basis state, i.e. "in/up". As shown in Silva *et al.* [37], there are a set of 1st order differential equations that describe the radial evolution of the fermion wave function

$$\partial_{r_*} \vec{\chi} = (\Upsilon_0 + \delta \Upsilon) \vec{\chi} \tag{F2}$$

where the unperturbed evolution matrix is given by

$$\Upsilon_0 = -\mu \sqrt{1 - \frac{2M}{r}} \hat{\sigma}_1 - ih \hat{\sigma}_2 + \frac{k}{r} \sqrt{1 - \frac{2M}{r}} \hat{\sigma}_3 \tag{F3}$$

with a perturbed term related to the introduction of a Coulomb potential as a charged particle is emitted

$$\delta \Upsilon = -i \frac{Z\alpha}{r} \hat{\sigma}_2 \tag{F4}$$

where $\vec{\sigma} = (\hat{\sigma}_1, \hat{\sigma}_2, \hat{\sigma}_3)$ are the standard Pauli matrices. The introduction of a Coulomb potential acts as a perturbation causing a mixing between the basis states.

Before proceeding, let us explicitly define a few quantities that will prove invaluable. First, the determinant of $\Phi$ will remain constant irrespective of choice of basis:

$$\det \Phi = F_{\mathrm{in}} G_{\mathrm{up}} - F_{\mathrm{up}} G_{\mathrm{in}} = -2ih T_{\frac{1}{2}kh} \sqrt{h^2 - \mu^2}. \tag{F5}$$

Second, we wish to also derive an expression allowing one to transform from $\Phi$ to $\Phi^*$ as a function of the transmission and reflection coefficients. This will prove necessary when attempting to relate how the occupation of the an electron changes and relating said change to the $\mathcal{I}^{(1)}$-type integrals, introduced in Eq. (47). First, we define

$$\Phi^* = \Phi \mathbf{B} \quad \Rightarrow \quad \mathbf{B} = \Phi^{-1} \Phi^* \tag{F6}$$

which, by employing the boundary conditions in the limit $r_* \to -\infty$, can explicitly be derived such that

$$\mathbf{B} = \begin{pmatrix} R_{\frac{1}{2}kh}^* & T_{\frac{1}{2}kh}^* \\ T_{\frac{1}{2}kh}^* & -R_{\frac{1}{2}kh} T_{\frac{1}{2}kh}^* T_{\frac{1}{2}kh}^{-1} \end{pmatrix}. \tag{F7}$$

We now conclude this discussion having the tools available to calculate the 1st order perturbations due to the inclusion of a Coulomb potential.

## 2. 1st order perturbations: $\partial\mathbb{T}/\partial Z$

The radial wave function may be written as a linear combination of the in/up basis

$$\vec{\chi} = \Phi \cdot \vec{c} = \vec{\psi}_{\rm in}c_{in} + \vec{\psi}_{\rm up}c_{up} \tag{F8}$$

where $\vec{c} = \begin{pmatrix} c_{\rm in} \\ c_{\rm up} \end{pmatrix}$. Perturbing these coefficients, such that $\vec{c} = \vec{c}_0 + \delta\vec{c}_1 + \delta\vec{c}_2 + ...$, while also noting that $\partial_{r_\star}\Phi = \Upsilon_0\Phi$ as this is just a statement that these basis states satisfy the general solution, one may show to 1st order

$$\partial_{r_\star}\delta\vec{c}_1 = \Phi^{-1}\delta\Upsilon\Phi \cdot \vec{c}_0. \tag{F9}$$

which, after substituting Eq. (F5), will have a solution of the form

$$\delta\vec{c}_1(r_\star) = \int_{\mathcal{R}}^{r_\star} dr_\star' \mathbf{B}(\Phi^{-1})^* \delta\Upsilon\Phi(r_\star') \cdot \vec{c}_0(r_\star') + \vec{a}_1. \tag{F10}$$

Substituting Eq. (F1), (F4), (F6), and (F7) allows one to write

$$\delta\vec{c}_1(r_\star) = -i\frac{Z\alpha}{2M}\frac{1}{\det\Phi^*}\int_{\mathcal{R}}^{r_\star} dr_\star'\frac{2M}{r}\begin{pmatrix} -R_{\frac{1}{2}kh}^*(F_{up}^\star F_{in} + G_{up}^* G_{in}) + T_{\frac{1}{2}kh}^*(F_{in}^\star F_{in} + G_{in}^\star G_{in}) \\ -T_{\frac{1}{2}kh}^*(F_{up}^\star F_{in} + G_{up}^* G_{in}) - R_{\frac{1}{2}kh}T_{\frac{1}{2}kh}^* T_{\frac{1}{2}kh}^{-1}(F_{in}^\star F_{in} + G_{in}^\star G_{in}) \end{pmatrix} + \vec{a}_1. \tag{F11}$$

If we consider the horizon and spatial infinity as our boundary conditions and substitute the definition of the $\mathcal{I}^{(1)}$-type integrals in Eqn. (47), we finally obtain

$$\delta\vec{c}_1(r_\star) = -i\frac{Z\alpha}{2M}\begin{pmatrix} \mathcal{I}_{\rm in,in,k}^{(1)}(h,h') - \frac{R_{\frac{1}{2}kh}^*}{T_{\frac{1}{2}kh}^*}\mathcal{I}_{\rm up,in,k}^{(1)}(h,h') \\ \mathcal{I}_{\rm up,in,k}^{(1)}(h,h') + \frac{R_{\frac{1}{2}kh}^*}{T_{\frac{1}{2}kh}^*}\mathcal{I}_{\rm in,in,k}^{(1)}(h,h') \end{pmatrix} \tag{F12}$$

which describes the $\mathcal{O}(\alpha)$ change to the radial fermionic wave functions. The result is that the perturbation to the "in" wave function (with $c_{0,\rm in} = 1$ and $c_{0,\rm up} = 0$) is

$$\delta c_{1,\rm in}(r_\star \to \infty) - \delta c_{1,\rm in}(r_\star \to -\infty) = i\frac{Z\alpha}{2M}\left[\frac{R_{\frac{1}{2}kh}^*}{T_{\frac{1}{2}kh}^*}\mathcal{I}_{\rm up,in}^{(1)}(h,h) - \mathcal{I}_{\rm in,in}^{(1)}(h,h)\right]. \tag{F13}$$

If one writes a perturbation to the electron wave function $\boldsymbol{\psi}_{\rm in}$, then at the horizon the perturbation behaves as

$$\begin{aligned}
\delta\boldsymbol{\psi}_{\rm in} &= \delta T_{\frac{1}{2}kh}\begin{pmatrix} \sqrt{h} \\ i\sqrt{h} \end{pmatrix}e^{-ihr_\star} \\
&= \delta c_{1,\rm in}(r_\star \to -\infty)T_{\frac{1}{2}kh}\begin{pmatrix} \sqrt{h} \\ i\sqrt{h} \end{pmatrix}e^{-ihr_\star} \\
&\quad + \delta c_{1,\rm up}(r_\star \to -\infty)\left[\begin{pmatrix} \sqrt{h} \\ -i\sqrt{h} \end{pmatrix}e^{ihr_\star} - R_{\frac{1}{2}kh}^* e^{2i\arg T_{1/2,k,h}}\begin{pmatrix} \sqrt{h} \\ i\sqrt{h} \end{pmatrix}e^{-ihr_\star}\right],
\end{aligned} \tag{F14}$$

where the first equality describes $\delta\boldsymbol{\psi}_{\rm in}$ via a change in the transmission coefficient, while the second uses the perturbation $\delta c_1$. Matching these implies that

$$\delta c_{1,\rm up}(r_\star \to -\infty) = 0 \quad \text{and} \quad \delta c_{1,\rm in}(r_\star \to -\infty) = \frac{\delta T_{\frac{1}{2}kh}}{T_{\frac{1}{2}kh}}. \tag{F15}$$

However, the similar matching of the $e^{-i\sqrt{h^2-\mu^2}\,r_\star}$ wave at $r_\star \to +\infty$ implies $\delta c_{1,\rm in}(r_\star \to \infty) = 0$. This means that

$$\frac{\delta T_{\frac{1}{2}kh}}{T_{\frac{1}{2}kh}} + i\frac{Z\alpha}{2M}\left[\frac{R_{\frac{1}{2}kh}^*}{T_{\frac{1}{2}kh}^*}\mathcal{I}_{\rm up,in}^{(1)}(h,h) - \mathcal{I}_{\rm in,in}^{(1)}(h,h)\right] = 0 \tag{F16}$$

or that the change in power transmission coefficient $\mathbb{T}_{\frac{1}{2}kh} = |T_{\frac{1}{2}kh}|^2$ is

$$\delta \mathbb{T}_{\frac{1}{2}kh} = 2|T_{\frac{1}{2}kh}|^2 \Re \frac{\delta T_{\frac{1}{2}kh}}{T_{\frac{1}{2}kh}} = \frac{Z\alpha}{M} \Im \left[ R^*_{\frac{1}{2}kh} T_{\frac{1}{2}kh} \mathcal{I}^{(1)}_{\mathrm{up,in}}(h,h) - |T_{\frac{1}{2}kh}|^2 \mathcal{I}^{(1)}_{\mathrm{in,in}}(h,h) \right]. \tag{F17}$$

Using the relations for the $\mathcal{I}^{(1)}$-type integrals, we simplify the quantity in brackets:

$$\delta \mathbb{T}_{\frac{1}{2}kh} = \frac{Z\alpha}{M} \Im \left[ R^*_{\frac{1}{2}kh} T_{\frac{1}{2}kh} \left( -\frac{R_{\frac{1}{2}kh} T^*_{\frac{1}{2}kh}}{\epsilon} + \Lambda_{\mathrm{up,in},k}(h,h) \right) - |T_{\frac{1}{2}kh}|^2 \left( \frac{|T_{\frac{1}{2}kh}|^2}{\epsilon} + \Lambda_{\mathrm{in,in},k}(h,h) \right) \right]. \tag{F18}$$

Using the hermiticity property of the $\Lambda$'s, $\Lambda_{XX'k}(h,h') = \Lambda^*_{X'Xk}(h',h)$, this simplifies to

$$\delta \mathbb{T}_{\frac{1}{2}kh} = -\frac{Z\alpha}{M} \Im \left[ R_{\frac{1}{2}kh} T^*_{\frac{1}{2}kh} \Lambda_{\mathrm{in,up},k}(h,h) \right]. \tag{F19}$$

We thus conclude that the first derivative of the transmission coefficient with respect to the charge on the black hole is:

$$\left. \frac{\partial \mathbb{T}_{\frac{1}{2}kh}}{\partial Z} \right|_{Z=0} = -\frac{\alpha}{M} \Im \left[ R_{\frac{1}{2}kh} T^*_{\frac{1}{2}kh} \Lambda_{\mathrm{in,up},k}(h,h) \right]. \tag{F20}$$

---